\documentclass[fleqn,usenatbib]{mnras}
\usepackage{newtxtext,newtxmath}
\usepackage[T1]{fontenc}
\DeclareRobustCommand{\VAN}[3]{#2}
\let\VANthebibliography\thebibliography
\def\thebibliography{\DeclareRobustCommand{\VAN}[3]{##3}\VANthebibliography}

\usepackage{graphicx}	
\usepackage{amsmath}	
\def    \apjl  		{\rm {ApJL}}
\def    \apj  		{\rm {ApJ}}
\def    \mnras  	{\rm {MNRAS}}
\def    \araa  		{\rm {ARA\& A}}
\def    \apjl  		{\rm {ApJL}}

\def   \bOmega  {\,{\boldsymbol{\Omega}}}

\def \bmu {\,{\boldsymbol{\mu}}}

\def	\cm		{\,{\rm {cm}}}

\def	\mum	{\,{\mu \rm{m}}}

\def	\H		{\,{\rm {H}}}

\def    \P     {\,\textbf{P}\,}
\def    \B     {\,\textbf{B}\,}
\def    \k     {\,\textbf{k}\,}
\def    \J     {\,\textbf{J}\,}
\def    \PparaB     {\,\P $\|$ \B}
\def    \PperpB     {\,\P $\perp$ \B}
\def \ahat {\,{\hat{\textbf{a}}}}

\def \bea {\begin{eqnarray}}
\def \ena {\end{eqnarray}}              

\title[Effects of Iron Inclusions on Dust Polarization from Protostellar Cores]{Physical Modeling of Dust Polarization from Magnetically Enhanced Radiative Torque (MRAT) Alignment in Protostellar Cores with POLARIS}

\author[Giang et al.]{Nguyen Chau Giang,$^{1,2}$
Thiem Hoang,$^{1,2}$
Jeong-Gyu Kim,$^{2,3}$
Le Ngoc Tram$^{4}$
\\
$^{1}$ Korea Astronomy and Space Science Institute, Daejeon 34055, Republic of Korea\\
$^{2}$Korea University of Science and Technology, 217 Gajeong-ro, Yuseong-gu, Daejeon, 34113, Republic of Korea\\
$^{3}$ National Astronomical Observatory of Japan, National Institutes of Natural Sciences, 2-21-1 Osawa, Mitaka, Tokyo 181-8588, Japan\\
$^{4}$Max-Planck-Institut f\"ur Radioastronomie, Auf dem H\"ugel 69, 53-121, Bonn, Germany
}

\date{Accepted XXX. Received YYY; in original form ZZZ}

\pubyear{2022}

\begin{document}
\label{firstpage}
\pagerange{\pageref{firstpage}--\pageref{lastpage}}
\maketitle

\begin{abstract}

Magnetic fields (\B) are an important factor that controls the star formation process. The leading method to observe \B orientation is using polarized thermal emission from dust grains aligned with \B. However, in dense environments such as protostellar cores, dust grains may have inefficient alignment due to strong gas randomizations, so that using dust polarization to trace \B is uncertain. Hoang $\&$ Lazarian (2016) demonstrated that the grain alignment by RAdiative Torques is enhanced if dust grains contain embedded iron inclusions. Here we extend POLARIS code to study the effect of iron inclusions on grain alignment and thermal dust polarization toward a protostellar core, assuming uniform \B. We found that paramagnetic grains produce a low polarization degree of $p \sim 1\%$ in the envelope and negligible $p \ll 1\%$ in the central region due to the loss of grain alignment. In contrast, grains with a high level of iron inclusions have perfect alignment and produce high $p \sim 40\%$ in the envelope and low $p \leq 10\%$ in the central region. Grains with a moderate level of iron inclusions induce the polarization flipping from \PparaB at millimeter to \PperpB at submillimeter due to the change in the internal alignment caused by slow internal relaxation. The weak alignment of very large grains of $a \geq 10\mum$ reduces dichroic extinction efficiency at submillimeter. We found a positive correlation between $p$ and the level of iron inclusions, which opens a new window to constrain the abundance of solid iron locked in dust through dust polarimetry.
 
\end{abstract}




\section{Introduction}\label{sec:intro}
Magnetic fields (\B) play an important role in controlling the star and planet formation in the Universe (\citealt{McKee_2007}, \citealt{Crutcher_2012}, \citealt{Pattle_2019}). During the initial stage, they produce magnetic pressure that prevents the gravitational collapse of dense molecular clouds (\citealt{Nakano_1978}, \citealt{Krumholz_2019}), but regulate the gas flow from the cloud inward to form denser structure called filament where stars can start to form (\citealt{Seifried_2015}). In the late phase, \B-fields help to launch outflows and jets due to magnetocentrifugal force, which reduces the angular momentum of protostellar disks and facilitates the growth of protostars (\citealt{Shu_1987}, \citealt{Allen_2003}, \citealt{Frank_2014}). 

Magnetic fields on the plane of sky are usually studied by the polarization of background starlight \citep{Clemens.2020} and polarized thermal emission from dust grains that align with the magnetic field direction (B-RAT) (see e.g., \citealt{Pattle_2019} for a review). The basic idea of such a method relies on the assumption that dust grains are aligned with the longest axis perpendicular to \B. As a result, thermal emission from aligned grains is polarized with the polarization vector (\P) perpendicular with magnetic field directions, i.e., \PperpB (see \citealt{Lazarian_2007,Anderson_2015,Tram_Hoang_2022} for reviews). Then, by rotating \P by $90^{\circ}$, one can get the magnetic field morphology on the plane of sky. Moreover, the observed polarization pattern and polarization fraction contain both information of magnetic fields and dust properties in local environments (e.g., \citealt{Planck_2015}, \citealt{Hensley:2021id}). Therefore, interpreting dust polarization is key to studying the roles of magnetic fields (\citealt{Pattle_2019}, \citealt{Hull_2014}) and dynamics of dust grains during the star and planet formation.

However, the question of how and where dust grains can be aligned with \B is the long-standing question in astrophysics. The process of grain alignment includes (1) the alignment of the axis of maximum moment of inertia ($\ahat_{1}$) with the angular momentum \J, so-called internal alignment and (2) the alignment of \J with a preferred direction in space, so-called external alignment (e.g., \B, see \citealt{Lazarian_2007}). The leading mechanism for grain alignment is based on RAdiative Torques (RATs, see \citealt{LAH.2015} and \citealt{Anderson_2015} for reviews). 
\cite{Dolginov_1976} first suggested that the interaction of an anisotropic radiation with a helical grain can induce RATs due to the differential scattering/absorption of left- and right-handed circularly polarized photons. The numerical calculations of RATs with three irregular shapes by using DDSCAT (\citealt{Draine_Wein_1996}) confirmed the efficient spin-up of dust grains by RATs, but the properties of RATs depend complexly on grain properties (e.g., grain shape, grain size). \cite{Lazarian_Hoang_2007} then introduced an Analytical Model (AMO) of RATs which is based on a helical grain consisting of an oblate spheroid and a weightless mirror. The AMO is shown to reproduce the basic properties of RATs obtained from numerical calculations for irregularly shaped grains \citep{Lazarian_Hoang_2007,Hoang_Lazarian_2008,Herranen_2021}, and enables us to make quantitative predictions of grain alignment by RATs with various conditions \citep{Hoang_Lazarian_2014} and dust compositions \citep{Hoang_Lazarian_2016a}. Many observed optical-NIR polarization features of background stars in the interstellar medium (ISM) and toward molecular clouds (MCs) can be successfully explained by the dichroic extinction of aligned dust grains by RATs (see \citealt{Anderson_2015} for a review).

For polarized thermal emission from aligned dust grains, synthetic modeling of dust polarization based on solid grain alignment physics and radiative transfer is a unique way to accurately interpret observational data and constrain basic alignment theory and dust properties. The first attempts to model polarized thermal dust using the basic RAT theory \citep{Lazarian_Hoang_2007} and radiative transfer are presented in \cite{Bethell_2007} for a clumpy molecular cloud and \cite{Cho_2007} for an accretion disk. Later on, \cite{Reissl_2016} developed the  POLArized RadIation Simulator (POLARIS) code that combines the RAT alignment theory \citep{Hoang_Lazarian_2014} and radiative transfer. POLARIS is based on the improved RAT theory for grains with ordinary paramagnetic material (PM grains) \citep{Hoang_Lazarian_2014}, which assumed that grains have fast internal relaxation due to various effects, including Barnett relaxation and inelastic relaxation \citep{Purcell_1979}. Such assumptions are well-known to be valid for the diffuse ISM and MCs where grains are small of sub-micron sizes and the randomization through gas-grain collisions is not strong enough to significantly affect the internal alignment \citep{Lazarian_2007}. Thus, POLARIS can be used to accurately model the RAT alignment and dust polarization to interpret observed dust polarization in these environments \citep{Reissl_2017,Seifried:2018wx}, especially for interpreting {\it Planck} polarization data at sub-millimeter and millimeter wavelengths (\citealt{Reissl_2020}).

The measurements of dust polarization toward dense environments such as prototostellar cores and protoplanetary disks are significantly improved thanks to the advanced interferometric observations by SMA, VLA, JVLA, and especially ALMA. These observations reveal in detail the complex magnetic field morphology from the core scale of few thousands of au to the disk scale of few hundreds of au (see \citealt{Cox_2018}, \citealt{Pattle_2019}, \citealt{Hull:2019hw} for a review), and show many anomalous polarization features. The first anomalous feature is the detection of high polarization degree $p \sim 20 - 40\%$ in the envelope but low $p \sim 1\%$ in the central 100 au region of low mass Class 0/I young stellar objects (YSOs) by ALMA (\citealt{Hull_2014}; \citealt{Galametz_2018}; \citealt{Cox_2018}). The maximum polarization fraction up to $p \sim 40\%$ detected in protostellar envelopes is twice times higher than the maximum $p \sim 20\%$ observed in the ISM by Planck \citep{2020A&A...641A..12P}. It is inconsistent with the prediction of reduced polarization level in protostellar environments due to the decrease of grain alignment efficiency at higher gas density due to stronger gas randomization \citep{Hoang_et_al_2021}. \cite{Bethell_2007} showed that the high polarization degree in dense environments can be due to the increase of maximum grain size. A theoretical study by \cite{Hirashia_Li_2013} showed that grains can grow to micron-size in dense clouds via the gas accretion and grain-grain collision. A recent study by \cite{Guillet.2020} shows that the grain drift due to ambipolar diffusion and turbulence can increase grain growth and form $\sim 1-10\mum$ grains in protostellar cores. Observations by \cite{Kwon_2009} and \cite{Galametz_2019} reported the evidence for the presence of very large grains (VLGs) of size $a > 10\mum$ in both the protostellar envelopes and protostellar disks around Class 0 YSOs. The modelling with current version of POLARIS by \cite{Valdivia_2019} also showed that grains must grow to $a > 50\mum$ and are efficiently aligned with \B to reproduce the detected level of polarization toward Class 0 YSOs. However, whether large grains can maintain their magnetic alignment in such high gas collision environments is still not well clarified.

The rapid decrease of $p\sim 20-40\%$ in the envelope to $p \sim 1\%$ toward protostars, termed as polarization hole, is  another anomalous feature because it is inconsistent with the prediction of higher polarization degree toward the radiation source in the RAT theory. The origin of the polarization hole is still unclear. \cite{Kataoka_2012b} suggested the polarization hole due to the increase in the toroidal component caused by the collapse of rotating protostellar cores and the effect of inclined magnetic fields.  \cite{Brauer_2016} suggested it is due to the extinction by aligned VLGs. \cite{Hoang_et_al_2021} found that the removal of large grains by RAdiative Torque Disruption (RAT-D) around the protostar also can reduce polarized thermal emision emitting from the central region. However, whether VLGs can efficiently have the magnetic alignment in dense protostellar environments is still not well studied.

Another anomalous feature is the 90 degrees flipping of the polarization pattern with wavelengths detected in two Class 0 YSOs NGC1333-IRAS4A (\citealt{Liu_et_al_2016}, \citealt{Ko_2020}) and OMC-3/MMS 6 (\citealt{Liu_2021}),  and the MC NGC 2071 (\citealt{Lapo_2022}). The origin of the polarization flipping  is unclear. \cite{Ko_2020} and \cite{Liu_2021} assigned the polarization flipping in two dense protostellar cores to the change of the polarization mechanism from dichroic emission to dichroic extinction by aligned dust grains. For the MC NGC 2071, \cite{Lapo_2022} suggested this feature may arise from the change of the grain alignment direction from magnetic fields to radiation fields, i.e., k$-$RAT, or the change of the internal alignment direction due to slow internal relaxation (\citealt{Hoang_et_al_2022}). Understanding the physical mechanisms of the dust polarization signal is crucially important for us to understand whether dust polarization can trace \B-fields in a given astrophysical environments. And if yes, in what conditions we can rotate the thermal dust polarization \P by $90^{\circ}$ to obtain \B-fields.

Recently, \cite{Gouellec_2020} interpreted ALMA polarization data of a dozen Class 0 YSOs (\citealt{Hull_2014}) by post-processing non-ideal magnetohydrodynamic (MHD) simulations of protostellar cores with the current version of POLARIS to perform synthetic observations of dust polarization produced by aligned dust grains. They found that the model of RATs currently implemented in POLARIS produces a much lower polarization degree than values measured by ALMA. However, the observational data can be reproduced by the model with fixed perfect grain alignment, revealing the efficient grain alignment in protostellar cores. But how grains can achieve perfect alignment in such dense environments remains unknown.

Within the RAT paradigm, theoretical studies by \cite{Lazarian_Hoang_2008} and detailed numertical calculations by \cite{Hoang_Lazarian_2016a} showed that grains with embedded iron inclusions, i.e., superparamagnetic material (SPM), can be efficiently aligned with \B due to the combined effect of RATs and enhanced magnetic relaxation, which is termed as Magnetically Enhanced RAdiative Torque (MRAT) alignment. Recently, \cite{Hoang.2022} and \cite{Hoang_et_al_2022} carried out detailed analytical studies on the effect of iron inclusions on grain alignment in protostellar cores and disks. They found that in protostellar cores and disks, only SPM grains can be efficiently aligned with \B due to the enhanced internal alignment (IA) and enhanced Larmor precession by iron inclusions. VLGs can be aligned with \B and can even have perfect alignment in protostellar environments if they contain a very high level of iron inclusions and produce high levels of dust polarization. Such a sensitive correlation between the level of iron inclusions locked inside dust grains and observed polarization fraction provides a new avenue for constraining iron abundance in dust, which is still a mystery now.

The modelling of dust polarization by \cite{Lam_2021} also found that micron-sized grains of $a \sim 1\mum$ must have the magnetic susceptibility enhanced by a factor of $\sim 20$  compared to PM to reproduce the observed $p \sim 1\%$ in the central regions of ten Class 0 YSOs by ALMA (\citealt{Cox_2018}). The modelling with POLARIS by \cite{Valdivia_2019} concluded that grains must grow to $a >  50\mum$ to reproduce observed $p > 1\%$ toward protostellar cores. It indirectly implies that VLGs around Class 0 YSOs must be SPM with a high level of iron inclusions.

However, the study by \cite{Lam_2021} simply considered the perfect alignment for grains with the Larmor precession timescale being shorter than the gas damping timescale (\citealt{Hoang_Lazarian_2016a}), instead of accounting for grain alignment physics. Furthermore, \cite{Lam_2021} and \cite{Valdivia_2019} did not consider the effect of internal alignment, which is crucially important for VLGs in protostellar environments. \cite{Hoang.2022} and \cite{Hoang_et_al_2022} showed that in contrast to the diffuse ISM and MCs where sub-micron grains always have efficient IA by fast internal relaxation, large grains in dense environments tend to have inefficient IA due to slow internal relaxation as a result of the stronger gas randomization. In contrast to grains with fast internal relaxation, with the grain major axis of inertia moment $\ahat_{1}$ always parallel to \J, i.e., $\ahat_{1} \| \J$ (right IA), grains with slow internal relaxation 
do not have
well-defined IA. A numerical study for grains without internal relaxation by \cite{Hoang_Lazarian_2009} found that grains aligned at high-\textit{J} attractors (see Section \ref{sec:external_alignment} of grain alignment theory by RATs for details) still can have right IA due to the spinning torques by RATs, but grains at low-\textit{J} attractors may have right or wrong IA. 
In the latter case, 
grains will align with $\ahat_{1}$ perpendicularly to \B and radiate polarized emission with \PparaB. The conditions that support the internal alignment of grains with slow internal relaxation are not well studied. It thus raises the uncertainty on determining the orientation between polarization vectors arising from the emission of aligned dust grains and the magnetic field direction. 

\cite{Hoang.2022} and \cite{Hoang_et_al_2022} found that the presence of grains with slow internal relaxation is reduced with increasing the fraction of iron locked inside dust grains. It implies the crucial role of iron inclusions in determining both the internal alignment direction and the grain alignment efficiency in dense environments. However, the current version of POLARIS only includes the RAT mechanism for paramagnetic grains. The dependence of grain alignment on the magnetic properties of grains and the presence of grains with slow internal relaxation are not taken into account in POLARIS. Thus, up to date, no detailed numerical modeling of dust polarization that takes the effect of iron inclusions and the co-existence of grains with fast and slow internal relaxation have been performed. This leaves a big gap between  theory and observations and limits the diagnostic power of dust polarimetry.

To connect theory to observations and extend the power of POLARIS on studying magnetic fields and dust physics in dense and dynamic environments around protostars, we include the magnetic susceptibility and the detailed grain alignment physics with the MRAT mechanism into the code. To demonstrate the capability of our implementation, we simulate the polarized thermal emission from a protostellar core. We adopt a simple Bonnor-Ebert sphere and a uniform magnetic field to isolate the effect of iron inclusions on grain alignment and dust polarization by aligned dust grains. In subsequent publications, we will use the updated POLARIS code to perform in detail the synthetic modeling of dust polarization toward protostellar cores and protostellar disks with more realistic \B-fields and gas density distribution from MHD simulations.

Our paper is organized as follows. We first describe the magnetic properties of grains in Section \ref{sec:mag_properties}. Then, we describe the fundamental mechanisms of the internal and external grain alignment, and the main points of the external alignment by RATs for grains with iron inclusions in Section \ref{sec:RAT_theory}. The improvements of POLARIS, the model setup, and the numerical results of grain alignment in the protostellar core are given in Sections \ref{sec:POLARIS} and \ref{sec:grain_alignment}, respectively. We then show the results for the effect of iron inclusions on the polarization pattern in Section \ref{sec:iron_pol_map}, the intensity-dependent polarization degree in Section \ref{sec:iron_P_I}, the efficiency of polarization by dichroic extinction of aligned VLGs in Section \ref{sec:iron_extinction}, and the wavelength-dependent polarization degree in Section \ref{sec:iron_p_lambda}. Implications of our results for observations and the summary of our main findings are presented in Sections \ref{sec:discussion} and \ref{sec:summary}, respectively.
 
\section{Magnetic properties of dust grains} \label{sec:mag_properties}

\subsection{Paramagnetic material} \label{sec:paramagnetic}
Dust grains with diffusely embedded iron atoms (e.g., silicate) are called PM due to the existence of unpaired electrons. The magnetic susceptibility of PM grains of temperature $T_{\rm d}$ in a static magnetic field is given by the Curie's law:
\bea 
\chi_{\rm pm}(0) = \frac{n f_{\rm p} \mu^{2}}{3 k_{\rm B} T_{\rm d}} = 0.006 n_{23} f_{\rm p,-1} \hat{p}^{2} T_{d,1}^{-1}
\label{eq:chi_0_paramagnetic}
\ena
where $n$ is the atomic density of grain material, $f_{\rm p}$ is the fraction of iron atoms within the grain, $\mu = p \mu_{\rm B} = g_{\rm e} \sqrt{J (J+1)} \mu_{\rm B}$ is the magnetic moment per iron atom with $g_{\rm e} \approx 2$ the electron-Lande factor, $J$ the angular momentum quantum number of electron in the outer partially filled shell, $\mu_{\rm B}$ the Bohr magneton, and $k_{\rm B}$ is the Boltzmann constant and $T_{\rm d}$ the dust temperature (see \citealt{Draine_Wein_1996}). In the numerical form, $n_{23} = n/(10^{23}\cm^{-3})$, $f_{\rm p,-1} = f_{\rm p}/0.1$, $\hat{p} = p/5.5$, and $T_{\rm d, 1} = T_{\rm d} / (10~\rm K)$ (\citealt{Draine_Wein_1996}).  
 
If the external magnetic field varies with time at frequency $\omega$, electron spins inside PM cannot respond immediately to the change in the external field, resulting in the phase lag that dissipates magnetic energy into thermal energy of grains. \textbf{The phase lag of electron spins at frequency $\omega$ and the energy dissipation of electromagnetic waves are characterized by the imaginary part of the magnetic susceptibility of dust grains, $\chi_{2}$}, which is given by (see e.g., \citealt{Hoang_Lazarian_2016a}): 

\bea 
\chi_{2, \rm pm}(\omega) = \frac{\chi_{\rm pm}(0) \omega \tau _{\rm el}}{[1 + (\omega \tau _{\rm el}/2)^{2}]^{2}} ,
\label{eq:chi_omega_parmagnetic}
\ena
where $\tau _{\rm el} \approx 2.9\times10^{-12} (f_{\rm p} n_{23})^{-1} \,{\rm s}$ 
is the spin-spin relaxation timescale of electron spins (\citealt{Draine_1996}).
 
\subsection{Superparamagenetic material}\label{sec:superparamagnetic}
Beside 
being distributed diffusely inside dust grains, iron atoms also can be embedded in the form of clusters (\citealt{Jones_Spitzer_1967}). Each iron cluster can be considered as a giant magnetic moment, and the presence of many iron clusters will enhance the magnetic susceptibility of grains and make them SPM. Assuming that each cluster contains $N_{\rm cl}$ iron atoms, the magnetic moment of each cluster is given by $m = N_{\rm cl}\mu$ with $\mu$ the magnetic moment of each iron atom. The magnetic susceptibility of SPM at zero frequency $\chi_{\rm spm}(0)$ is then given by (\citealt{Hoang_Lazarian_2016a}):
\bea 
\chi_{\rm spm}(0) &=& \frac{n_{\rm cl} m^{2}}{3 k_{\rm B} T_{\rm d}}  = \frac{N_{\rm cl} \phi_{\rm sp} \mu^{2}}{4\pi a_{\rm iron}^{3} k_{\rm B} T_{\rm d}} \nonumber \\
&\approx&  0.52 N_{\rm cl,3} \phi_{\rm sp,-2} \hat{p}^{2} T_{\rm d, 1}^{-1},
\label{eq:chi_0_super_paramagnetic}
\ena
where $n_{\rm cl} = N/V$ is the number of iron cluster per unit volume. 
$N  = (\phi_{\rm sp}/N_{\rm cl}) (a_{\rm eff} / a_{\rm iron})^{3}$ 
is the total number of iron clusters with $\phi_{\rm sp}$ the volume filling factor of iron clusters, $a_{\rm eff}$ the effective radius of grains (see Section \ref{sec:RAT_theory}), 
$a_{\rm iron} = 1.26\times10^{-4}\mum$ the radius of an iron atom
, and $V = (4/3)\pi a_{\rm eff}^{3}$ 
the grain volume. In the numerical form, $N_{\rm cl,3} = N_{\rm cl} / 10^3$, $\phi_{\rm sp,-2} = \phi_{\rm sp}/0.01$.

Similar to PM grains, the imaginary part of the magnetic susceptibility of SPM grains at frequency $\omega$ is given by (\citealt{Hoang_Lazarian_2016a}):
\bea 
\chi_{2,\rm spm}(\omega) = \frac{\chi_{\rm spm}(0) \omega \tau _{\rm sp}  }{[1 + (\omega \tau _{\rm sp}/2)^{2}]^{2}},
\label{eq:chi_omega_super_paramagnetic}
\ena
where $\tau_{\rm sp} = \nu_{0}^{-1} \exp(N_{\rm cl} T_{\rm act} / T_{\rm d})$ with $\nu_{0} = 10^{9}\rm s^{-1}$ and $T_{\rm act} = 0.011$ K is the timescale for SPM to undergo thermally activated remagnetization (\citealt{Morrish_2001}).

The magnetic susceptibility at zero frequency of PM and SPM grains decreases with increasing dust temperature due to increasing thermal fluctuations of electron spins. In addition, the magnetic susceptibility at frequency $\omega$ decreases when the oscillation rate of magnetic field is larger than the inverse of the relaxation timescale of spin systems, i.e., $\omega \gtrsim 2/\tau_{\rm el}$ (for PM grains) and $\omega \gtrsim 2/\tau _{\rm sp}$ (for SPM grains).

\section{Physics of Grain Alignment}\label{sec:RAT_theory}
In this section, we briefly describe the physics of grain alignment that includes the internal and external alignment (see \citealt{LAH.2015} and \citealt{Anderson_2015} for reviews). The internal alignment brings the angular momentum $\J$ and the angular velocity $\bOmega$ to align with the major axis of inertia moment in which grains have the lowest rotational energy level. The external alignment brings $\J$ to align with some preferred direction (i.e., magnetic field, radiation field, or gas flow) that allows grains to radiate polarized thermal emission (see \citealt{Hoang_et_al_2022} for more details). In our paper, we focus on modeling the alignment of grains with iron inclusions with magnetic fields using the MRAT model \citep{Hoang_Lazarian_2016a}. 
We first describe the damping timescale due to the random gas-grain collisions, then the mechanisms and the corresponding timescales for the internal and external alignment by RATs and magnetic relaxation. 

\subsection{Gas damping timescale}

We consider an oblate spheroidal grain with the principal axes $\ahat_{1}, \ahat_{2}, \ahat_{3}$ with the corresponding semi-minor axis of length $c$ and the semi-major axes of length $a$ (see Figure \ref{fig:alignment}). The minor axis $\ahat_{1}$ has the maximum inertia moment of $I_{\parallel} =  I_{1} = 8\pi/15 \rho s a^{5}$ and the major axes $\ahat_{2}$, $\ahat_{3}$ have the minimum inertia moment of $I_{\perp} = I_{2} = I_{3} = 4\pi/15 \rho s (1 + s^{2}) a^{5}$ with $\rho$ the dust mass density and $s = c/a < 1$ the axial ratio. The ratio of the maximum and minimum inertia moment is $h = I_{\parallel} / I_{\perp} = 2/(1+s^{2})$. All notations used in our study are summarized in Table \ref{tab:model_parameter}.

Atoms and molecules with Brownian motion colliding with a grain only increase the grain rotational energy, but do not increase the grain angular momentum due to the averaging effect. Subsequently, their evaporation from the grain surface takes away the grain angular momentum and makes grains rotate slower with time (e.g., \citealt{Draine_Wein_1996}). The gas damping timescale is given by (see e.g., \citealt{Hoang_et_al_2022}):
\bea
\tau _{\rm gas} &=& \frac{3}{4\sqrt{\pi}}\frac{I_{\parallel}}{1.2 m_{\rm H} n_{\rm H} v_{\rm th} a^{4} \Gamma_{\parallel}} \nonumber \\ 
&\simeq& 0.083\hat{\rho}\left(\frac{sa_{-5}}{n_{\H,8}T_{\rm gas,1}^{1/2}\Gamma_{\|}}\right)~{\rm yr},\label{eq:tau _gas}
\ena
where $n_{\rm H}$ is the number density of hydrogen atom, $v_{\rm th} = \sqrt{2 k_{\rm B} T_{\rm gas} / m_{\rm H}}$ is the thermal velocity with $T_{\rm gas}$ the gas temperature, $\hat{\rho} = (\rho / 3\rm g \cm^{-3})$, and $\Gamma_{\parallel}$ is the geometrical factor of unity (\citealt{Roberge_1993}, \citealt{Hoang_Lazarian_2016a}, \citealt{Hoang_et_al_2022}). We obtain $\Gamma_{\parallel} = 0.62$ for an oblate spheroidal with $s = 1/2$. In the above, $a_{-5} = a/(10^{-5}\cm)$,  $n_{\rm H,8} = n_{\rm H} / (10^{8}\cm^{-3})$, and $T_{\rm gas,1} = T_{\rm gas}/(10\,\rm K)$.

\subsection{Internal Alignment by the Barnett Relaxation}\label{sec:internal_alignment}
In the rest state, grains with embedded iron atoms have no magnetic moment due to the random orientation of electron spins. When they rotate at $\bOmega$, the rotational energy can force all electron spins to orient along $\bOmega$ direction, inducing the net magnetic moment (\citealt{Barnett_1915}). The grain magnetic moment due to the Barnett effect is given by:
\bea 
\bmu_{\rm Bar} = \frac{\chi(0) V}{\gamma_{\rm e}} \bOmega,
\label{eq:mu_bar}
\ena
where $\gamma_{\rm e} = -g_{\rm e} \mu_{\rm B}/\hbar$ is the gyromagnetic ratio of electron spin, $V = 4 \pi s a^{3} /3 $ is the grain volume, and $\chi(0)$ is the magnetic susceptibility at zero frequency (Section \ref{sec:mag_properties}).

\begin{figure}
        \includegraphics[width = 0.5\textwidth]{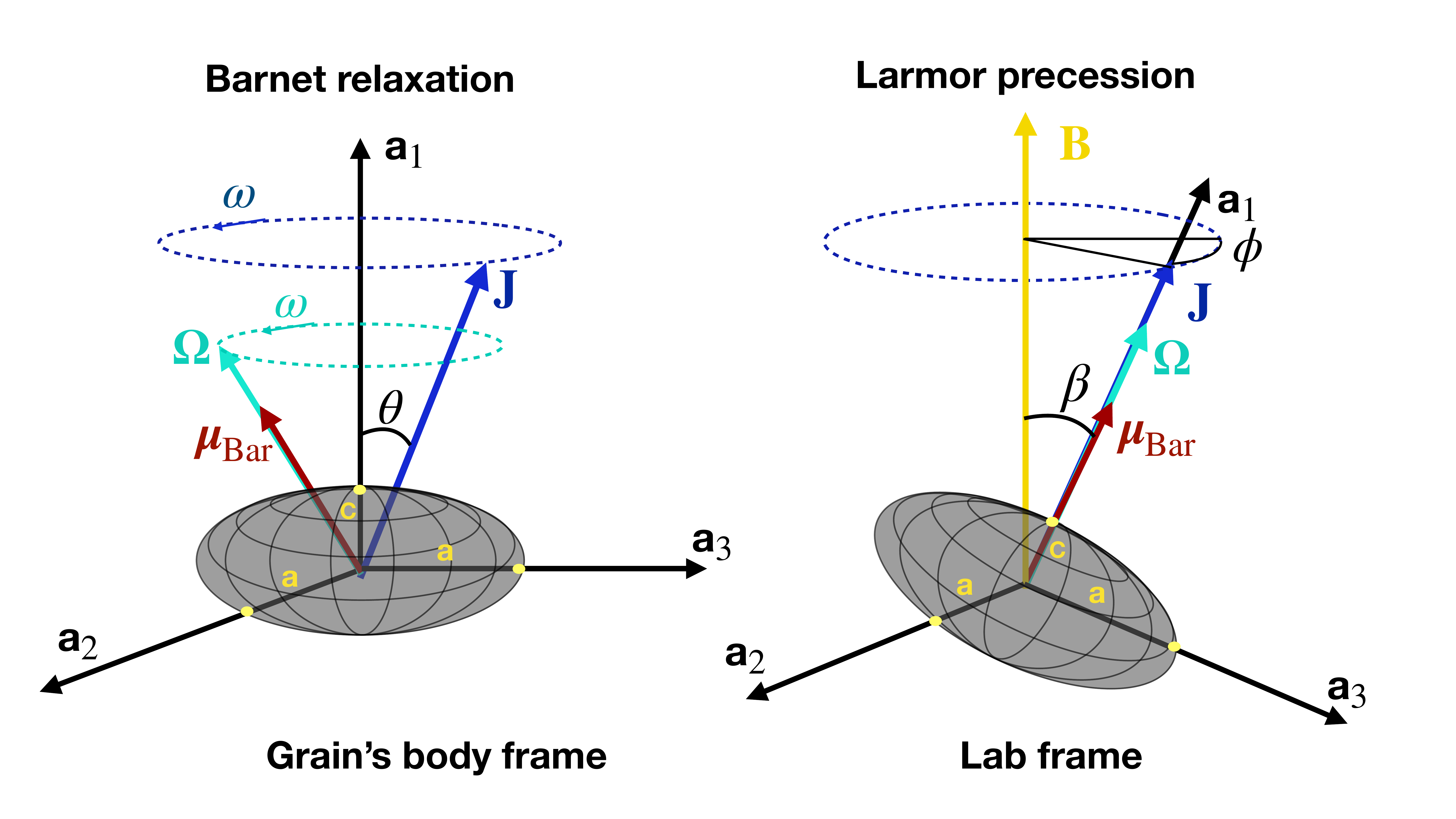}
     \caption{Left panel: Illustration of the internal alignment due to Barnett relaxation viewed from the grain's body frame. The precession of the angular momentum \J and angular velocity $\bOmega$ around the major axis of inertia moment $\ahat_{1}$ at the precession rate $\omega$ induces the precession of the magnetic moment $\bmu_{\rm Bar}$ gained by the Barnett effect around $\ahat_{1}$. The rotating magnetization induces the dissipation of rotational energy, leading to the alignment of \J and $\bOmega$ (and also $\mu_{\rm Bar}$) with $\ahat_{1}$. Right panel: Illustration of Larmor precession due to the magnetic torques caused by the interaction between $\bmu_{\rm Bar}$ and \B. For the external alignment by magnetic relaxation, the rotating magnetic moment induces the dissipation of rotational energy that leads to the alignment between \J and \B .}
 
     \label{fig:alignment}
\end{figure}
 
According to \cite{Purcell_1979}, irregular grains have the high rotational energy due to the misalignment of $\J$ and $\bOmega$ with their major axis of inertia moment $\ahat_{1}$. In the grain's frame of reference, the misalignment of $\J$ and $\bOmega$ causes the precession of $\J$ and $\bOmega$ around  $\ahat_{1}$ at an angular rate $\omega = \Omega (h-1) \cos \theta$ with $\theta$ the angle between $\J$ and $\ahat_{1}$ (Figure \ref{fig:alignment}, left part). For magnetic grains, the rotating $\bOmega$ induces the rotation of magnetic moment $\bmu_{\rm Bar}$ around $\ahat_{1}$ that dissipates the rotational energy into thermal energy. Consequently, grains can stably rotate around $\ahat_{1}$, i.e., $\J$ $\parallel \ahat_{1}$, which corresponds to the state that grains have the minimum rotational energy. This internal alignment due to the rotating magnetization is called the Barnett relaxation (\citealt{Purcell_1979}, \citealt{Roberge_1993})), which occurs on a timescale of (see \citealt{Hoang_Lazarian_2016a}):
\bea 
\tau _{\rm BR} = \frac{\gamma_{\rm e}^{2} I_{\parallel}^{3}}{V K(\omega) h^{2} (h-1) J^{2}}.
\label{eq:tbar}
\ena
where $K(\omega) = \chi_{2}(\omega)/\omega$.  The Barnett relaxation timescale for PM grains is numerically given by:
\bea
\tau _{\rm BR, pm} \simeq 0.5 \hat{\rho}^{2} a_{-5}^{7} f(\hat{s}) \Bigg[1 + \Bigg(\frac{\omega \tau _{\rm el}}{2}\Bigg)^{2} \Bigg]^{2} \Bigg(\frac{J_{\rm d}}{J}\Bigg)^{2}~\rm yr,
\label{eq:tbar_pm}    
\ena
and for SPM grains
(\citealt{Hoang_et_al_2022}):
\bea 
\tau _{\rm BR,spm} &\simeq& 1.6\times10^{-4} \hat{\rho}^{2}f(\hat{s})a_{-5}^{7}\left(\frac{1}{N_{\rm cl,3}\phi_{\rm sp,-2}\hat{p}^{2}}\right)\nonumber\\
&\times& \left(\frac{J_{d}}{J}\right)^{2} \left(\frac{T_{d,1}}{k_{\rm sp}(\omega)}\right) ~ \rm yr,
\label{eq:tbar_spm}    
\ena 
where  $f(\hat{s})=\hat{s}[(1+\hat{s}^{2})/2]^{2}$ with $\hat{s} = s/0.5$, $J_{\rm d} = \sqrt{ k_{\rm B} T_{\rm d} I_{\parallel} / (h-1)}$ is the thermal angular momentum, and $k_{\rm sp}(\omega)$ is given by:
\bea
k_{\rm sp}(\omega)= \exp\left(\frac{N_{\rm cl}T_{\rm act}}{T_{d}}\right)\left[1+\left(\frac{\omega\tau _{\rm sp}}{2}\right)^{2}\right]^{-2}. 
\label{eq:ksp}
\ena

Equations (\ref{eq:tbar_pm}) and (\ref{eq:tbar_spm}) show that large grains experience slower Barnett relaxation. Therefore, in dense environments where the gas damping is significantly shorter (Equation \ref{eq:tau _gas}), large grains could have inefficient IA due to the slow internal relaxation (\citealt{Hoang_et_al_2022}). Following \cite{Hoang_et_al_2022}, we define the  lower and upper cutoff for grain sizes with fast internal relaxation determined by the condition $\tau_{\rm BR} = \tau_{\rm gas}$ as $a_{\rm min,aJ}$ and $a_{\rm max,aJ}$. The range of grains with fast internal relaxation extends with increasing rotational rate and the magnetic susceptibility of grains, i.e., super-Barnett relaxation.

\subsection{External alignment with Magnetic field direction}\label{sec:external_alignment}
\subsubsection{Larmor precession}\label{sec:larmor_precesssion}
The interaction between the grain magnetic moment $\bmu_{\rm Bar}$ and the ambient magnetic field \B produces the magnetic torques that cause the grain angular momentum \J to precess around \B, so-called Larmor precession (Figure \ref{fig:alignment}, right panel). The characteristic timescale of the Larmor precession is given by (see \citealt{Hoang_Lazarian_2016a}):

\bea 
\tau_{\rm Lar} = \frac{2\pi}{d\phi/dt} = \frac{2\pi I_{\parallel} \Omega}{\mu_{\rm Bar} B}, 
\label{eq:tlar}
\ena
where $\phi$ the precession angle of \J with respect to \B (Figure \ref{fig:alignment}, right panel). The typical Larmor precession timescale for PM grains is:

\bea
 \tau _{\rm Lar,pm} \simeq  7\times10^{-4}  \frac{\hat{\rho} T_{\rm d,1} a_{-5}^{2}}{f_{\rm p,-1} \hat{p}^{2} B_{2}}  ~ \rm yr,
\label{eq:tlar_pm}   
\ena
and for SPM grains is:
\bea
 \tau _{\rm Lar, spm}  \simeq 8.1\times 10^{-6}\frac{\hat{\rho} T_{d,1}a_{-5}^{2}}{N_{\rm cl,3}\phi_{\rm sp,-2}\hat{p}^{2}B_{2}} \rm yr,~~~~
\label{eq:tlar_spm}   
\ena
where $B_{2} = B / 100\mu\rm G$.

Dust grains are considered to have the magnetic alignment when grains rapidly precess around \B, so that \J cannot be randomized by gas collisions, which requires $\tau _{\rm Lar}<\tau _{\rm gas}$. Equations (\ref{eq:tlar_pm}) and (\ref{eq:tlar_spm}) show that large grains are more difficultly aligned with \B due to slower Larmor precession. Following \cite{Hoang_et_al_2022}, we define the maximum size that grain still can be aligned with \B is $a_{\rm max,JB}^{\rm Lar}$. More large grains can be aligned with \B, i.e., higher $a_{\rm max,JB}^{\rm Lar}$, if they have higher magnetic susceptibility by iron inclusions (Equation \ref{eq:tlar_spm}).

\subsubsection{Magnetic relaxation}\label{sec:magnetic_relaxation}
A paramagnetic grain rotating with the angular velocity $\bOmega$ misaligned with the ambient field $\B$ experiences the dissipation of the grain rotational energy due to the lag of the grain magnetization with respect to the ambient field, which eventually brings $\bOmega$ (also $\J$) to be aligned with $\B$ that corresponds to the minimum rotational energy (\citealt{Davis_1951}, so-called the Davis-Greenstein mechanism). The timescale for the magnetic relaxation is given by (e.g.,\citealt{Hoang_Lazarian_2016a}):
\bea 
\tau _{\rm mag} = \frac{I_{\parallel}}{V K(\Omega) B^{2}}.
\label{eq:tmag} 
\ena
where $K(\Omega) = \chi_{2}(\Omega)/\Omega$. The timescale of paramagnetic relaxation is:
\bea
 \tau _{\rm mag,pm} \simeq 3440 \frac{\hat{\rho} a_{-5}^{2} T_{\rm d,1}}{\hat{p}B_{2}^{2}} \Bigg(1 + \Bigg(\frac{\Omega \tau _{\rm el}}{2}\Bigg)^{2}\Bigg)^{2} ~\rm yr,
\ena
and the timescale of superparamagnetic relaxation (for SPM) is (\citealt{Hoang_et_al_2022}):
\bea
 \tau _{\rm mag,spm} \simeq 1.5 \frac{\hat{\rho}a_{-5}^{2}}{N_{\rm cl,3}\phi_{\rm sp,-2}\hat{p}^{2}B_{2}^{2}}\frac{T_{d,1}}{k_{\rm sp}(\Omega)} ~ \rm yr,
\label{eq:tmag_spm}   
\ena
where $k_{\rm sp}(\Omega)$ is given by Equation (\ref{eq:ksp}).

To describe the strength of magnetic relaxation on aligning dust grains, we use the magnetic relaxation parameter:
\bea 
\delta_{\rm m} &=& \frac{\tau _{\rm gas}}{\tau_{\rm mag,spm}}\nonumber \\
 &\approx& 0.056 a_{-5}^{-1}\frac{N_{\rm cl,3}\phi_{\rm sp,-2}\hat{p}^{2}B_{2}^{2}}{\hat{\rho} n_{H,8}T_{\rm gas,1}^{1/2}}\frac{k_{\rm sp}(\Omega)}{T_{d,1}}.
\label{eq:delta_m}
\ena
Grains can have some degree of alignment by magnetic relaxation when the magnetic relaxation happens faster than gas randomization, which is given by $\delta_{\rm m} > 1$. Following \cite{Hoang_et_al_2022}, we denote the  maximum size at which the magnetic relaxation is still effective as $a_{\rm max,JB}^{\rm DG}$. SPM grains experience stronger magnetic relaxation due to larger magnetic susceptibility (Equation \ref{eq:tmag_spm}). However, \cite{Hoang_Lazarian_2016a,Hoang_Lazarian_2016b} showed that the grain alignment by magnetic relaxation is still inefficient even for $\delta_{m}\gg 1$ in the absence of grain suprathermal rotation due to internal thermal fluctuations.

\subsection{Radiative Torque Alignment (RAT) Paradigm}\label{sec:RATs}
 
The RAdiative Torque (RAT) mechanism is demonstrated to efficiently drive grain alignment, even for paramagnetic grains (\citealt{Lazarian_Hoang_2007,Hoang_Lazarian_2008}). Moreover, the joint effect between RATs and superparamagnetic relaxation can lead grains to achieve perfect alignment with magnetic fields, which is known as the MRAT mechanism (\citealt{Hoang_Lazarian_2008,Hoang_Lazarian_2016a}). Here, we describe the main properties of the RAT and MRAT alignment mechanisms for reference (see \citealt{Hoang_et_al_2022} for more details). 

\subsubsection{Suprathermal rotation by RATs}\label{sec:spin_up_RAT}

\citet{Dolginov_1976} proposed that irregular grains receive RATs due to differential scattering and absorption of photons with the left and right-hand angular momentum.
Let $a_{\rm eff} = a s^{1/3}$ be an effective radius of a spherical grain that has the same volume as the irregular grain (\citealt{Draine_Wein_1996}; \citealt{Lazarian_Hoang_2007}), the maximum angular speed that the grain achieve by RATs in the constant radiation field is given by (\citealt{Hoang_Lazarian_2014}):
\bea
\Omega_{\rm RAT} = \frac{\Gamma_{\rm RAT}\cos\psi \tau _{\rm damp}}{I_{\parallel}},
\label{eq:omega_rat}
\ena
where $\Gamma_{\rm RAT}$ is the total radiative torques that the grain receives, which is:
\bea 
\Gamma_{\rm RAT} = \int_{\rm \lambda_{\rm min}}^{\lambda_{\rm max}}  \pi a_{\rm eff}^{2} \left(\frac{\lambda}{2\pi}\right)\gamma_{\rm \lambda}  u_{\rm \lambda} Q_{\rm \Gamma} d\lambda,
\label{eq:Gamma_RAT}
\ena
where $u_{\lambda}$ is the radiation energy density at wavelength $\lambda$, $\lambda_{\rm min}$ and $\lambda_{\rm max}$ are the lower and upper boundary of the working region of RAT mechanism (i.e., FUV-FIR), and $\gamma_{\lambda}$ is the anisotropic degree at wavelength $\lambda$, which is defined as (\citealt{Bethell_2007}):
\bea 
\gamma_{\rm \lambda} = \frac{1}{4\pi J_{\lambda}} \Bigg|\int I_{\lambda} \textbf{k} ~ d\Omega\Bigg|,
\label{eq:gamma_lambda}
\ena 
where $I_{\lambda}$ and $J_{\lambda}$ are the specific intensity and the mean intensity at wavelength $\lambda$, and \textbf{k} is the propagation direction of photon. In Equation (\ref{eq:Gamma_RAT}), $Q_{\rm \Gamma}$ is the RAT efficiency, which is a constant for large grains of size $a_{\rm eff} \geq \lambda/1.8$ and decreases with decreasing grain sizes for $a_{\rm eff} < \lambda/1.8$ (see \citealt{Lazarian_Hoang_2007}, \citealt{Hoang_Lazarian_2008}). 

The second term $\psi$ in Equation (\ref{eq:omega_rat}) is the angle between \B and the radiation field direction \k, and the third term $\tau_{\rm damp} = \tau_{\rm gas}/(1 + \rm F_{\rm IR})$ is the total grain damping timescale caused by gas collisions (the term $\tau_{\rm gas}$) and re-emit IR radiation (the dimensionless coefficient $F_{\rm IR}$ describes the rotational damping due to IR emission) (see \citealt{Draine_Lazarian_1998} for a detail).
 
\begin{table*}
  \centering
         \caption{Notations and meanings}
  \begin{tabular} {l|l|l|l}
  \hline 
  Notation & Meaning & Notation & Meaning \\
      \hline
$f_{\rm p}$             & Iron fraction of PM grains                    & a                     & Major axis of oblate spheroidal grain \\
$\tau_{\rm el}$         & Electron spin relaxation timescale        & s                     & Axial ratio \\
$\chi_{\rm pm}(0)$      & Magnetic susceptibility at $\omega = 0$ of PM            & $a_{\rm eff}$         & Effective radius \\
$\chi_{\rm pm}(\omega)$ & Magnetic susceptibility at frequency $\omega$ of PM   & $a_{\rm align}$       & Minimum grain alignment size \\
$N_{\rm cl}$            & Number of iron atoms/cluster                  & $a_{\rm max,JB}^{\rm Lar}$      & Maximum grain alignment size \\ 
$\phi_{\rm sp}$         & Volume filling factor of iron clusters                & $f_{\rm high-J}$      & Fraction of aligned grains at high-\textit{J} \\
$\tau_{\rm sp}$         & Thermally activated remagnetization timescale    & $a_{\rm min,aJ}^{\rm low-J}$   & Minimum size with fast internal relaxation at low-\textit{J} \\ 
$\chi_{\rm spm}(0)$     & Magnetic susceptibility at $\omega = 0$ of SPM  & $a_{\rm max,aJ}^{\rm low-J}$   & Maximum size with fast internal relaxation at low-\textit{J} \\
$\chi_{\rm spm}(\omega)$ & Magnetic susceptibility at frequency $\omega$ of SPM  & $a_{\rm min,aJ}^{\rm high-J}$ & Minimum size with fast internal relaxation at high-\textit{J}\\
$\mu_{\rm Bar}$         & Magnetic moment gained by Barnett effect      & $a_{\rm max,aJ}^{\rm high-J}$  & Maximum size with fast internal relaxation at high-\textit{J} \\
$\tau_{\rm Lar}$        & Larmor precession timescale                   & $a_{\rm max,JB}^{\rm DG,0.5}$  & Maximum grain size has $f_{\rm high-J} = 0.5$ \\
$\tau_{\rm BR}$         & Barnett relaxation timescale                  & $a_{\rm max,JB}^{\rm DG,1}$    & Maximum grain size has $f_{\rm high-J} = 1$ \\
$\tau_{\rm mag}$        & Magnetic relaxation timescale                 & $\Gamma_{\rm RAT}$    & Radiative torque \\
$\delta_{\rm m}$        & Magnetic relaxation parameter                 & $\gamma_{\rm rad}$    & Mean anisotropic degree \\
$\tau_{\rm gas}$        & Gas damping timescale                         & $Q_{\Gamma}$          & RAT efficiency \\
$\theta$                & Angle between $\ahat_{1}$ and \J              & $\Omega_{\rm RAT}$    & Maximum angular velocity gained by RATs \\
$\beta$                 & Angle between \J and  \B                      & $\Omega_{\rm T}$      & Thermal angular velocity \\
$\psi$                  & Angle between \textbf{k} and \B               & $R$                   & Rayleigh reduction factor \\
$Q_{\rm X,low-J}$       & Internal alignment (IA) efficiency at low-\textit{J}               & $Q_{\rm X,high-J}$    & IA efficiency at high-\textit{J}\\
$Q_{\rm J,low-J}$       & External alignment (EA) efficiency at low-\textit{J}               & $Q_{\rm J,high-J}$    & EA efficiency at high-\textit{J}\\
 
  \hline
    \label{tab:model_parameter}
    \end{tabular}

\end{table*}

\subsubsection{Minimum grain size for the RAT alignment}\label{sec:align_RAT}

\cite{Hoang_Lazarian_2008} showed that grains can efficiently align with \J parallel to \B against gas randomization when they are spun up to suprathermal rotation with $\Omega_{\rm RAT} \geq 3\Omega_{\rm th} = 3\sqrt{k_{\rm B}T_{\rm gas}/I_{\parallel}}$. The grain size at which $\Omega_{\rm RAT} = 3 \Omega_{\rm th}$ is called alignment size, $a_{\rm align}$, at which all larger grains will be efficiently aligned with \B (e.g., B-RAT mechanism, \citealt{Hoang_et_al_2022}), provided that their Larmor precession is faster than the gas randomization, i.e., $a < a_{\rm max,JB}^{\rm Lar}$ (Section \ref{sec:larmor_precesssion}).

\subsubsection{A model in the RAT alignment: low-J and high-J attractors}\label{sec:low_highJ_model}

RATs not only can spin up grains to suprathermal rotation but also can align \J with \B by the alignment torque component (\citealt{Lazarian_Hoang_2007}, \citealt{Hoang_Lazarian_2008}, \citealt{Hoang_Lazarian_2016a}). Theoretical and numerical studies of RATs in \cite{Hoang_Lazarian_2008} show that, for aligned grains of $a_{\rm align} < a < a_{\rm max,JB}^{\rm Lar}$, a fraction of grains can stably align with \B at their maximum angular velocity, $\Omega_{\rm RAT}$ (so-called high-\textit{J} attractors). The rest of grains are spun down by RATs and align with \B at thermal rotation $\Omega_{\rm th}$ (so-called low-\textit{J} attractors). The fraction of grains aligning with \B at high-\textit{J} attractors is parameterized by the parameter $f_{\rm high-J}$. The exact value of $f_{\rm high-J}$ depends complexly on the physical properties of grains and the orientation of grains with the ambient radiation and and magnetic fields (\citealt{Hoang_Lazarian_2008,Hoang_Lazarian_2016a,Lazarian_Hoang_2021}). For grains at low-\textit{J} attractors, the gas randomization can gradually transport grains from low-\textit{J} to high-\textit{J} attractors. The value of $f_{\rm high-J}$ for irregular compact grains can vary within $0.25 - 0.7$ (\citealt{Herranen_2021}).

\cite{Hoang_Lazarian_2016a} found that the fraction $f_{\rm high-J}$ is increased for SPM grains due to enhanced magnetic relaxation via the MRAT mechanism. Their numerical calculations of the MRAT alignment showed that SPM grains can have perfect alignment of $f_{\rm high-J} =1$ for $\delta_{\rm m} \geq 10$ (Equation \ref{eq:delta_m}).

\section{Modelling of grain alignment and dust polarization with POLARIS}\label{sec:POLARIS} 
In this section, we will first describe the fundamental components of POLARIS in Section \ref{sec:POLARIS_overview}, then describe how we incorporate the new physical effects induced by iron inclusions on grain alignment in Section \ref{sec:POLARIS+}.
 
\begin{figure*}
\centering
        \includegraphics[width=1.03\textwidth,height=\textheight,keepaspectratio]{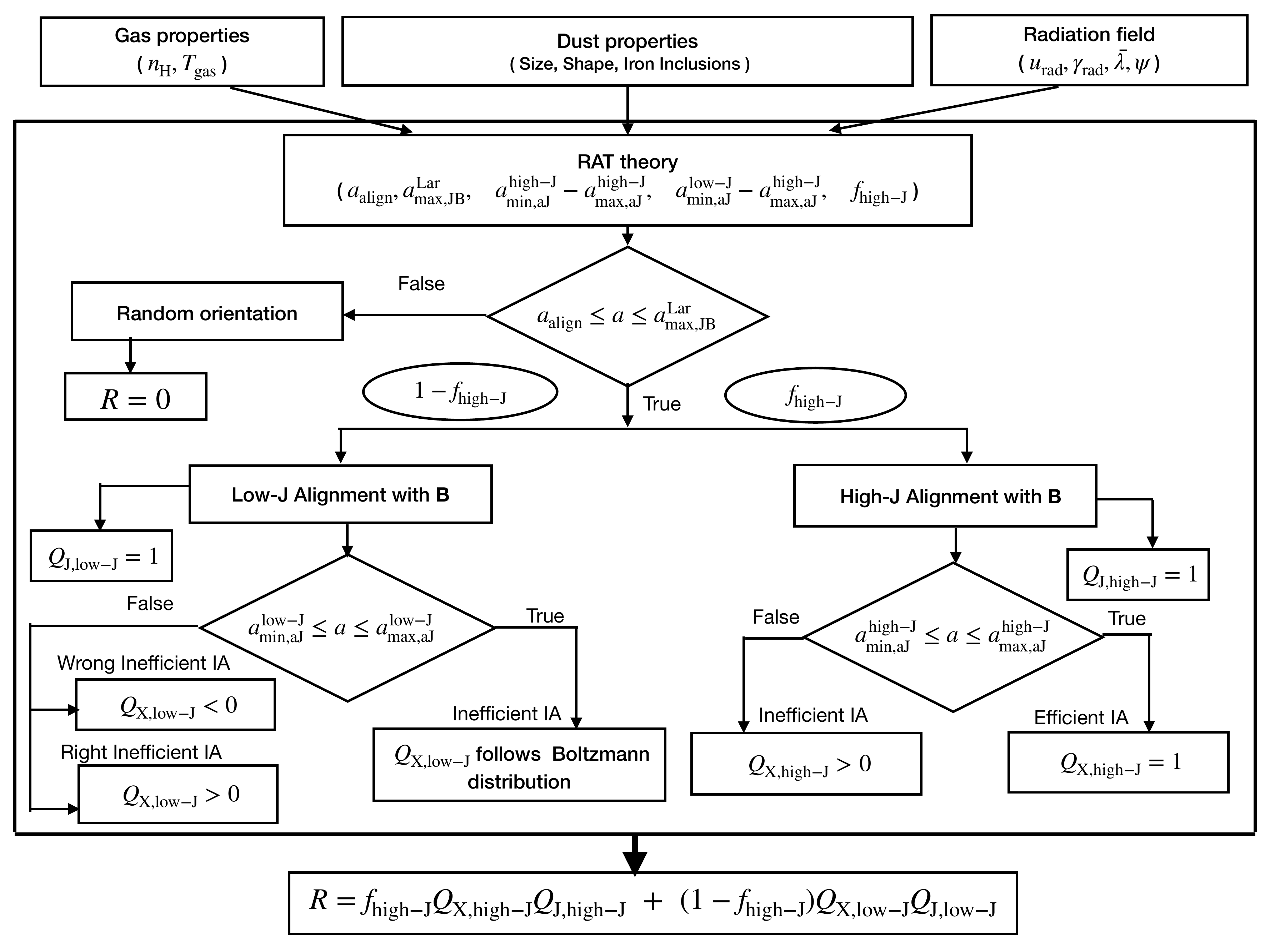}
     \caption{Schematic to determine the alignment degree of grains with magnetic fields in POLARIS. 
     Given the dust model, gas properties, and radiation field information ($u_{\lambda}, ~\gamma_{\lambda}, ~\psi$), POLARIS can determine the typical grain size for grain alignment by RATs ($a_{\rm align} - a_{\rm max,JB}^{\rm Lar}$), the internal alignment degree $Q_{\rm X}$ of grains at low and high-\textit{J} attractors, and the fraction of grains aligned with \B at high-\textit{J} attractors $f_{\rm high-J}$. Finally, one can calculate the Rayleigh reduction factor $R$ that describes the total internal and external alignment degree of grains with \B. }
     \label{fig:Rayleigh}
\end{figure*}

\subsection{An overview of the POLARIS code}\label{sec:POLARIS_overview}
There are two main simulations inside POLARIS: the three-dimensional (3D) radiative transfer simulation using the Monte-Carlo technique that provides the information of radiation field required for calculating dust and gas temperature and radiative torques for grain alignment by RATs. The results from the Monte-Carlo Radiative Transfer (MCRT) simulation are used to solve the polarized radiative transfer of Stokes parameters by the ray-tracing method to obtain the polarization degree $p$ and polarization angle (PA). The detailed description of the working flow in POLARIS is given in detail in \cite{Reissl_2016} and \cite{Reissl_2020}. Here, we just briefly describe the main features of POLARIS. 
 
\subsubsection{Monte Carlo Radiative Transfer}\label{sec:POLARIS_MCMC}
 
For the MCRT simulation, given the 3D gas density distribution of an astronomical object and the dust model, POLARIS simulates the interactions of photons 
emitted from the radiation source with surrounding dust grains by using the MCMC technique introduced by \cite{Lucy_1999}, which allows us to immeadiately correct dust temperatures to simulate the spontaneous thermal emission of dust grains. In detail, when each absorption event between grain of size $a$ and photon happens, POLARIS assumes the temporary thermal equilibrium between the current radiation absorption and dust emission to 
immediately correct the grain temperature $T_{\rm d}(a)$ (see Section \ref{sec:POLARIS_temperature}). One new photon with a wavelength sampled from the Planck function of grains with updated $T_{\rm d}(a)$ is sent to the grid cell immediately to guarantee the energy conservation between the stellar absorption and thermal dust emission. For the radiation field, the mean intensity at wavelength $J_{\lambda}$ inside each cell is calculated by summing energy deposited from all photons entering, being scattering or absorbing by dust grains inside cell, and leaving the cell. The direction of photons inside the cell is saved to calculate the anisotropic degree at wavelengths $\lambda$, $\gamma_{\lambda}$, as defined by Equation (\ref{eq:gamma_lambda}).

\subsubsection{Dust and gas temperature}\label{sec:POLARIS_temperature}
Knowing the mean intensity inside each cell after the MCRT simulation, the temperature of grain of size $a$, $T_{\rm d}(a)$, is calculated following the thermal equilibrium between the radiation absorption and emission, which is (\citealt{Lucy_1999}, \citealt{Reissl_2016}):
\bea 
\int C_{\rm abs, \lambda} J_{\lambda} d\lambda = \int C_{\rm abs,\lambda} B_{\lambda}(T_{\rm d}(a))d\lambda,
\ena 
where $C_{\rm abs}$ is the absorption cross section of grain size $a$ with wavelength $\lambda$ (see Appendix \ref{sec:stoke_parameter}). The average dust temperature in the cell is taken by integrating over the grain size distribution $dn/da$, i.e., $T_{\rm d} = \int_{\rm a_{\rm min}}^{\rm a_{\rm max}} T_{\rm d}(a) (dn/da) da$. The gas temperature is considered to correlate with the dust temperature via a dimensionless coefficient $\alpha$, giving $T_{\rm gas} = \alpha T_{\rm d}$. For dense environments as protostellar cores, gas is mainly heated by colliding with grains, inducing the coupling between gas and dust temperature. For our calculations, we consider the thermal equilibrium between gas and grains with $T_{\rm gas} = T_{\rm d}$ (or $\alpha = 1$).

\subsubsection{Modelling Grain Alignment by RATs}\label{sec:POLARIS_alignment_current}
{\bf (a) Minimum grain size of RAT alignment, $a_{\rm align}$}\\

Figure \ref{fig:Rayleigh} shows the flow to determine the grain alignment degree with \B in POLARIS. From the top of the figure, given the radiation field and gas temperature from the MCRT simulation and the input dust model, POLARIS can calculate the radiative torques $\Gamma_{\rm RAT}$ and the maximum angular speed gained by RATs, $\Omega_{\rm RAT}$, for all grain sizes as described in Section \ref{sec:spin_up_RAT}. The minimum alignment size, $a_{\rm align}$, is calculated by the alignment criteria described in Section \ref{sec:align_RAT}.

{\bf (b) Maximum size of grain alignment with \B by Larmor precession, $a_{\rm max,JB}^{\rm Lar}$}\\

Another criterion for stable grain alignment with \B (i.e., magnetic alignment) is that the Larmor precession must be faster than the grain randomization by gas collisions, which is described by the maximum alignment size $a_{\rm max,JB}^{\rm Lar}$ (Section \ref{sec:larmor_precesssion}). In the current version of POLARIS, PM grains are considered to have the magnetic alignment if $\tau _{\rm Lar} \leq \tau _{\rm gas}$ (\citealt{Hoang_Lazarian_2009,Hoang_et_al_2022}), giving  \footnote{There is a misprint in Equation (31) in \cite{Reissl_2016} where the prefactor of $4.1\times 10^{-9}$ is written instead of the correct one of $4.1\times10^{-19}$, in which $a_{\rm max,JB}^{\rm Lar}$ is given in m, $n_{\rm H}$ is given in $m^{-3}$, and B is given in T. In POLARIS, they adopt the constant of $4.1\times10^{-21}$, two orders of magnitude smaller than the exact values derived for PM grains, which overestimate the value of $a_{\rm max,JB}^{\rm Lar}$ by two orders of magnitude. Although this misprint does not cause any problem with grain alignment in the ISM and MCs with low density, it will overestimate the effect of magnetic alignment for large PM grains in dense cores and protostellar environments of very high density.}:
\bea  
a_{\rm max,JB}^{\rm Lar} \approx 7.589 \frac{ s^{2} B_{2}}{ n_{\rm H,8} T_{\rm d,1} \sqrt{T_{\rm gas,1}}} ~\mum.
\label{eq:amax_JB}
\ena 

{\bf (c) Degree of grain alignment: Rayleigh reduction factor, $R$}\label{sec:R_origin}\\
\cite{Greenberg_1968} introduced the Rayleigh reduction factor $R$ which describes the degree of internal alignment between \J and $\ahat_{1}$ and external alignment between \J and \B, which is defined as $R = \langle Q_{\rm X} Q_{\rm J} \rangle$ where the bracket denotes the averaging over an essembles of grains. The exact calculations of $R$ for dust grains using the RAT theory are rather challenging and require numertical simulations (\citealt{Hoang_Lazarian_2008};\citealt{Hoang_Lazarian_2016a}). Therefore, POLARIS uses the parametric model of the RAT alignment proposed in \cite{Hoang_Lazarian_2014}. This parametric model is based on the fact that, in general, RATs can align a fraction $f_{\rm high-J}$ of grains at high-\textit{J} attractors, and the remaining fraction $(1-f_{\rm high-J})$ of grains are aligned at low-\textit{J} attractors. Therefore, the Rayleigh reduction factor for the RAT alignment can be written as (the bottom of Figure \ref{fig:Rayleigh}):
\bea 
R &=& f_{\rm high-J} Q_{\rm X,high-J} Q_{\rm J, high-J} \nonumber \\
&+&  (1 - f_{\rm high-J}) Q_{\rm X,low-J} Q_{\rm J,low-J},
\label{eq:R_new}
\ena
where $Q_{\rm X,high-J}, Q_{\rm J,high-J}$ are the degrees of internal and external alignment for grains at high-J attractors, and $Q_{\rm X,low-J}, Q_{\rm J,low-J}$ are for grains at low-J attractors.

The degree of internal alignment is defined as $Q_{\rm X} = \langle G(\cos^{2}\theta) \rangle$ with $G(\cos^{2}\theta) =  1/2(3 \cos^{2}\theta - 1)$ and $\theta$ the angle between \J and $\ahat_{1}$ ($\theta \in [0-\pi]$, Figure \ref{fig:alignment}). For grains at high-\textit{J} attractors, we adopt $Q_{\rm X,high-J} = 1$ because grains can have perfect IA due to their suprathermal rotation. In contrast, grains at low-\textit{J} attractors have inefficient IA due to the thermal fluctuations inside grains (\citealt{Purcell_1979}). POLARIS assumes that internal relaxation process is always faster than gas randomization, i.e., fast internal relaxation, i.e., $\tau_{\rm Bar} < \tau_{\rm gas}$ (Section \ref{sec:internal_alignment}), such that there is an efficient energy exchange between the grain rotational energy and the grain vibrational system. The angle $\theta$ between \J and $\ahat_{1}$ for grains rotating thermally then can be described by the local thermal equilibrium (TE) Boltzmann distribution (\citealt{Lazarian_Roberge_1997}):
 \bea 
f_{\rm TE}(\theta) = Z\exp\left(-\frac{J_{\rm th}^{2}}{2 I_{\parallel}k_{\rm B} T_{\rm d}} [1 + (h-1)\sin^{2}(\theta)]\right),
\label{eq:f_zeta}
\ena
where $Z$ is the normalization factor such that $\int_{0}^{\pi} f_{\rm LE}\sin\theta d\theta=1$, and $J_{\rm th}$ is the thermal angular momentum. 

The degree of internal alignment by fast internal relaxation at low-\textit{J} attracrors is calculated as:
\bea
Q_{\rm X,low-J}=\int_{0}^{\pi} G(\cos^{2}\theta) f_{\rm LE}(\theta) \sin\theta d\theta.
\label{eq:QX_lowJ}
\ena

The external alignment degree is described by $Q_{\rm J} = \langle G(\cos^{2}\beta) \rangle$ with $\beta$ the angle between \J and \B (Figure \ref{fig:alignment}). POLARIS considers that grains at both low and high-\textit{J} attractors have perfect external alignment by RATs, giving $Q_{\rm J,low-J} = Q_{\rm J,high-J} = 1$. Grains are assumed to be randomly oriented with $R=0$ for $a<a_{\rm align}$ and $a>a_{\rm max, JB}^{\rm Lar}$.

\subsubsection{Polarized radiative transfer of Stokes vector and the Dust polarization information }\label{sec:POLARIS_stoke_parameter}

After the Monte-Carlo calculation of the radiation field for RAT alignment, 
POLARIS calculates the polarized radiative transfer of the Stokes vector $\vec{S} = [I ~ Q ~ U ~ V]^{\rm T}$ with Stokes $I$ 
representing the total intensity, Stokes $Q$ and $U$ 
the linear polarization, and Stokes $V$ 
It uses the Runge-Kutta method to solve the full polarized radiative transfer equation of $\vec{S}$ along the propagation of photons, given the gas density, grain temperature, and the alignment degree of grains with local magnetic fields, i.e., the Rayleigh reduction factor $R$, in the grid space calculated from the first simulation (Sections \ref{sec:POLARIS_temperature} and \ref{sec:POLARIS_alignment_current}). The detailed description of the polarized radiative transfer of the Stokes vector is given in Appendix \ref{sec:stoke_parameter}.

The degree of linear polarization $p$ is given by:
 
\bea
p = 100 \frac{\sqrt{Q^{2} + U^{2}}}{I} ~\rm (\%),
\ena
and the polarization angle (PA) is:
\bea  
PA = \frac{1}{2}\arctan\Bigg(\frac{Q}{U}\Bigg).
\ena
 
\subsection{Extending POLARIS for dust grains with Iron Inclusions}\label{sec:POLARIS+}
As discussed above, the current version of POLARIS is based on the classical RAT theory for PM grains and assumed aligned dust grains always have fast internal relaxation \citep{Hoang_Lazarian_2014}, which is valid only for sub-micron grains in the diffuse ISM and MCs. In very dense environments such as protostellar cores and disks, both the gas density and grain sizes increase many orders of magnitudes, which requires a detailed treatment of internal and external alignment to 
accurately model the synthetic dust polarization toward this regions \citep{Hoang.2022,Hoang_et_al_2022}. Here, we describe our improvements of POLARIS by taking into account new effects which are important for large grains in protostellar environments, including grains with iron inclusions, effects of iron inclusions on Barnett relaxation and internal alignment, Larmor precession and enhanced external alignment by MRAT alignment.

\subsubsection{Treatment of Enhanced Magnetic Susceptibility by Iron Inclusions}\label{sec:POLARIS_magnetic_susceptibility}
Following Section \ref{sec:mag_properties}, we describe the magnetic susceptibility of PM grains by using the parameter $f_{\rm p}$ (the fraction of iron atoms diffusely distributed inside grains), and SPM grains by using parameter $\phi_{\rm sp}$ (the volume filling factor of iron clusters) and $N_{\rm cl}$ (the number of iron atoms in a single cluster). For given $f_{p},\phi_{\rm sp}$, $N_{\rm cl}$, and the grain temperature calculated in Section \ref{sec:POLARIS_temperature}, we can calculate $\chi_{\rm pm}(0)$ for PM grains using Equation (\ref{eq:chi_0_paramagnetic}) and $\chi_{\rm spm}(0)$ for SPM grains using Equation (\ref{eq:chi_0_super_paramagnetic}). Note that, for SPM grains, even with a fixed $\phi_{\rm sp}$ (i.e., constant volumn of iron clusters locked in dust), the SPM susceptibility can be increased by increasing $N_{\rm cl}$.  
 
\subsubsection{Improved treatment of the Larmor precession and Magnetic alignment}\label{sec:POLARIS_amaxJB}

The first effect that we revisit is the disalignment of grains by gas collisions when the Larmor precession is slower than gas randomization (Section \ref{sec:larmor_precesssion}). As shown in Section \ref{sec:POLARIS_alignment_current}, the current version of POLARIS adopts smaller prefactor (i.e., $4.1\times10^{-21}$) for the maximum alignment size $a_{\rm max,JB}^{\rm Lar}$ of PM grains, that certainly overestimates the range of grain alignment in dense environments. In addition, dust grains are currently considered to be efficiently aligned with \B if they can complete one Larmor precession before being randomized by gas collisions, or $\tau_{\rm Lar} \leq \tau_{\rm gas}$. This condition may be insufficient for grains to be strongly coupled with \B in high gas-grains collision environments (\citealt{Yang_2021}). Here, we assume that grains can only be coupled to \B when the Larmor precession is at least ten times faster than the gas randomization, or in terms of timescales, $\tau _{\rm Lar} \leq \tau _{\rm gas}/10$ (e.g., \citealt{Yang_2021}). 

To calculate $a_{\rm max,JB}^{\rm Lar}$ for PM and SPM grains, we plug $\chi(0)$ calculated in Section \ref{sec:POLARIS_magnetic_susceptibility} in Equation (\ref{eq:mu_bar}) to obtain the grain magnetic moment $\bmu_{\rm Bar}$. Then, we use the value of $\bmu_{\rm Bar}$ for Equation (\ref{eq:tlar}) to obtain $\tau _{\rm Lar}$. By comparing $\tau _{\rm Lar}$ to $\tau _{\rm gas}/10$ with $\tau _{\rm gas}$ given by Equation (\ref{eq:tau _gas}) over the grain size distribution, one can determine the value of $a_{\rm max,JB}^{\rm Lar}$.

\subsubsection{Modeling the dependence of Internal Alignment on Grain Magnetic properties}\label{sec:POLARIS_amaxaJ}
The second effect that we take into account is the dependence of internal alignment on the grain magnetic properties. As discussed in the previous section, the current version of POLARIS assumes that all grains have {\it right} IA with $\ahat_{1}$ parallel to \B due to fast internal relaxation. However, as shown in Equation (\ref{eq:tbar}), the rate of internal relaxation by Barnett effect depends sensitively on the grain size, the magnetic susceptibility ($\chi$), and the grain angular momentum (J), which can be faster or slower than the gas randomization. Therefore, it is critically important to determine the range of grain sizes which has fast internal relaxation (see \citealt{Hoang.2022,Hoang_et_al_2022}). Here, we follow the approach proposed in \cite{Hoang_et_al_2022} by considering separately grains aligned at high-J and low-J attractors.

\textbf{(a) Grains at high-\textit{J} attractors}\\
Grains aligned with \B at high-\textit{J} attractors have suprathermal rotation with the grain angular velocity of $\Omega=\Omega_{\rm RAT}$. The precession rate of \J around $\ahat_{1}$ is given by $\omega = \Omega_{\rm RAT} (h-1) \cos\theta$. We take $\theta = 45^{\circ}$ be the average angle between \J and $\ahat_{1}$ for numerical calculations (\citealt{Purcell_1979}, \citealt{Hoang_et_al_2022}). For PM grains, we calculate $\chi_{\rm 2,pm}(\omega)$ by putting $\chi(0)$ (\textbf{which is calculated} in Section \ref{sec:POLARIS_magnetic_susceptibility}) and $\Omega_{\rm RAT}$ (\textbf{which is calculated in} Section \ref{sec:POLARIS_alignment_current}) to Equation (\ref{eq:chi_omega_parmagnetic}), then put $\chi_{\rm 2,pm}(\omega)$ and $J = I_{\parallel}\Omega_{\rm RAT}$ to Equation (\ref{eq:tbar}) to obtain the Barnett relaxation timescale $\tau _{\rm BR}$. Then, by comparing $\tau _{\rm BR}$ with $\tau _{\rm gas}$ over the grain size distribution, one can determine the minimum size ($a_{\rm min,aJ}^{\rm high-J}$) and maximum size ($a_{\rm max,aJ}^{\rm high-J}$) that PM grains have fast internal relaxation at high-\textit{J} attractors. We follow the same method for SPM grains, but $\chi_{\rm 2,spm}(\omega)$ is calculated by putting $\chi_{\rm spm}(0)$ (\textbf{which is calculated in} Section \ref{sec:POLARIS_magnetic_susceptibility}) into Equation (\ref{eq:chi_omega_super_paramagnetic}).

 \textbf{(b) Grains at low-\textit{J} attractors}\\
Grains aligned with \B at low-\textit{J} attractors rotate with thermal angular velocity of $\Omega=\Omega_{d} = J_{\rm d}/I_{\parallel}$ with $J_{\rm d}$ the thermal angular momentum (Section \ref{sec:internal_alignment}). The precession rate of \J around $\ahat_{1}$ is $\omega = \Omega_{\rm d}(h-1)\cos(\theta)$. Following the same flow as grains at high-\textit{J} attractors, we first calculate $\chi_{2}(\omega)$ and put it into Equation (\ref{eq:tbar}) to get $\tau _{\rm BR}$. Then following the condition of $\tau_{\rm BR} \leq \tau_{\rm gas}$, we can determine the range of grains with fast internal relaxation at low-\textit{J} attractors $[a_{\rm min,aJ}^{\rm low-J} - a_{\rm max,aJ}^{\rm low-J}$.

 \begin{figure}
        \includegraphics[width=0.46\textwidth,height=0.5\textheight,keepaspectratio]{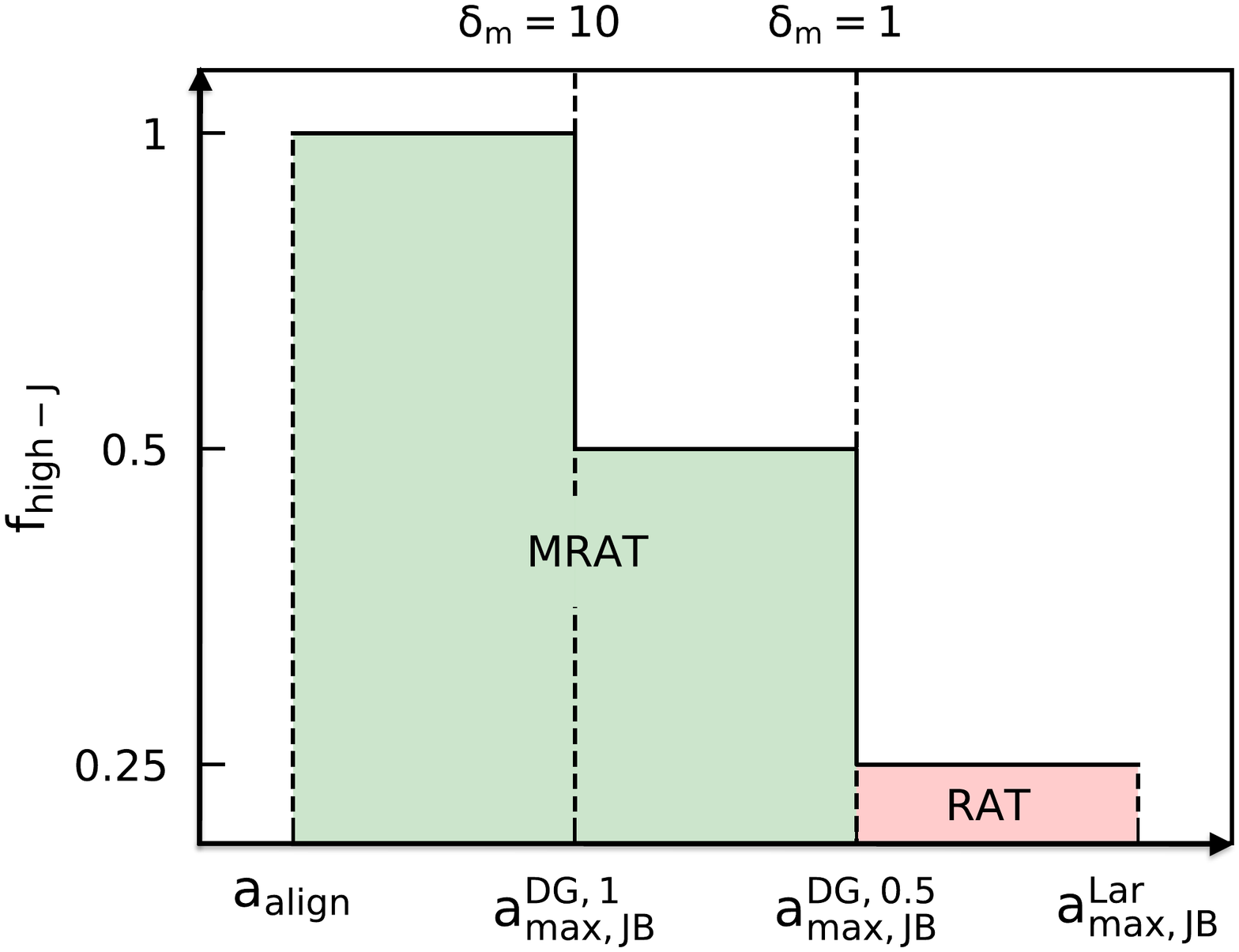}
     \caption{Illustration of the variation of $f_{\rm high-J}$ with alignment grain size within $[a_{\rm align} - a_{\rm max,JB}^{\rm Lar}]$ by MRAT alignment, see Equation (\ref{eq:f_highJ}). Vertical dashed lines mark the location where $\delta_{m}=1, 10$ and the corresponding critical sizes, $a_{\rm max,JB}^{\rm DG,0.5}, a_{\rm max,JB}^{\rm DG,1}$, respectively.}
     \label{fig:fhighJ}
\end{figure}

\subsubsection{Modeling the increase of $f_{\rm high-J}$ with superparamagnetic relaxation}\label{sec:POLARIS_amaxJB_DG}
The last effect that we consider is the increase in the RAT alignment by magnetic relaxation, described by the MRAT model \citep{Hoang_Lazarian_2016a}. The current version of POLARIS treats $f_{\rm high-J}$ as the free parameter (Section \ref{sec:POLARIS_alignment_current}), which does not change with the grain size and grain magnetic properties. However, the value of $f_{\rm high-J}$ is found to increase with the magnetic relaxation, which is described by the magnetic parameter $\delta_{m}$ \citep{Hoang_Lazarian_2016a}. To model the increase of $f_{\rm high-J}$ with $\delta_{m}$ as numerically calculated in \cite{Hoang_Lazarian_2016a}, we introduce the following parametric model:

\bea 
f_{\rm high-J}(\delta_{m}) = \left\{
\begin{array}{l l}    
    0.25 ~ ~  {\rm ~ for~ } \delta_{\rm m} < 1   \\
    0.5 ~ ~  {\rm ~ for~ } 1 \leq \delta_{\rm m} \leq 10 \\
    1    ~ ~ ~ ~  {\rm ~ for~}  \delta_{\rm m} > 10 \\
 
\end{array} \right\},
\label{eq:f_highJ}
\ena
where $\delta_{\rm m}$ depends on the grain size, magnetic susceptibility, and gas density. Above, we have adopted a typical value of $f_{\rm high-J} = 0.25$ for the case $\delta_{\rm m}<1$ at which grains are mainly aligned with \B by RATs. \footnote{The value of $f_{\rm high-J}$ driven by RATs can vary between $\sim 0.2 - 0.7$, depending on grain shapes and radiation fields (\citealt{Herranen_2021}).} For grains with significantly enhanced magnetic relaxation, i.e., $\delta_{\rm m}>10$, we take $f_{\rm high-J}=1$ due to the joint effect of RATs and superparamagnetic relaxation \citep{Hoang_Lazarian_2016a,Lazarian_Hoang_2021}. And we adopt the intermediate value of $f_{\rm high-J} = 0.5$ for grains with $1  \leq \delta_{\rm m}  \leq 10$.

For a given local condition, $f_{\rm high-J}$ is a function of the grain size and magnetic susceptibility because $\delta_{\rm m}$ depends on these parameters (Equation \ref{eq:delta_m}). To obtain the function $f_{\rm high-J}(a)$ for SPM grains, we first calculate $\chi_{\rm 2, spm}(\Omega_{\rm RAT})$ by using Equation (\ref{eq:chi_omega_super_paramagnetic}), then put this parameter to Equation (\ref{eq:tmag}) to obtain superparamagnetic relaxation timescale. Then, by calculating $\delta_{\rm m}$ (Equation \ref{eq:delta_m}) over the grain size distribution and following Equation (\ref{eq:f_highJ}), we can determine the maximum size that grains have $f_{\rm high-J} = 0.5$ (denoted by $a_{\rm max,JB}^{\rm DG,0.5}$) and $f_{\rm high-J} = 1$ (denoted by $a_{\rm max,JB}^{\rm DG,1}$). 

Figure \ref{fig:fhighJ} presents the schematic illustration of the variation of $f_{\rm high-J}$ with grain alignment size within $[a_{\rm align} - a_{\rm max,JB}^{\rm Lar}]$ and critical sizes for grain alignment by the MRAT mechanism. Small SPM grains of size $a<a_{\rm max,JB}^{\rm DG,1}$ can have $f_{\rm high-J} = 1$ because of their faster magnetic relaxation (Equation \ref{eq:tmag}). Larger grains of size $a_{\rm max,JB}^{\rm DG,1} < a < a_{\rm max, JB}^{\rm DG,0.5}$ have smaller $f_{\rm high-J} = 0.5$ because of their slower magnetic dissipation. Large grains of $a>a_{\rm max,JB}^{\rm DG,0.5}$ have $f_{\rm high-J} = 0.25$ due to the negligible contribution of magnetic relaxation to RAT alignment. 

 \begin{table}
     \centering
          \caption{Grain alignment physics in the current and updated version of POLARIS}
     \begin{tabular}{lll }
     \hline
    Effect & Current POLARIS & Update POLARIS \\
     \hline
    Magnetic properties & Paramagnetic & Grains with Iron Inclusions \\
    \hline
    \multicolumn{3}{c}{\textbf{Internal Alignment}}\\
    \hline
    
    Internal relaxation rate & Fast & Fast and Slow \\
    \hline
    \multicolumn{3}{c}{\textbf{External Alignment}}\\
    \hline
    
    Fast Larmor precession & $\tau_{\rm Lar} \leq \tau_{\rm gas}$ & $\tau_{\rm Lar} \leq \tau_{\rm gas}/10$ \\
    Alignment mechanism  & RAT, fixed $f_{\rm high-J}$ & MRAT, $f_{\rm high-J}$ depends on \\
    & & magnetic susceptibility. \\
 
    \hline
     \end{tabular}
 
     \label{tab:physics_old_new}
 \end{table}

\subsubsection{Improved Rayleigh reduction factor}\label{sec:POLARIS-New_R}
Due to the dependence of the internal and external alignment of grains on iron inclusions and local conditions described in the previous section, the Rayleigh reduction factor $R$ also changes. The illustration of our new flow for calculating $R$ is shown in the middle and bottom parts of Figure \ref{fig:Rayleigh} as following:

For grains aligning with \B at high-\textit{J} attractors (right panel), grains of size $a_{\rm min,aJ}^{\rm high-J} < a < a_{\rm max,aJ}^{\rm high-J}$ will have fast internal relaxation and then efficient IA due to suprathermal rotation of grains, giving $Q_{\rm X,high-J} = 1$, as adopted in the current version of POLARIS (Section \ref{sec:POLARIS_alignment_current}). Grains beyond this range have the slow internal relaxation, which induces the inefficient IA. The study of RAT alignment for grains without internal relaxation by \cite{Hoang_Lazarian_2016a} showed that RATs could drive grains to have right IA ($\ahat_{1} \parallel $\J) if they align with \B at high-\textit{J} attractors, but may have right or wrong IA ($\ahat_{1} \perp $\J) for grains at low-\textit{J} attractors. For the latter case, the alignment degree of grains with wrong IA is described by the negative value of $Q_{\rm X} < 0$. The detailed angle distribution $f(\theta)$ for grains with slow internal relaxation is still missing, thus we assume $Q_{\rm X}$ to be a free parameter for grains beyond the range $a_{\rm min,aJ}-a_{\rm max,aJ}$. For the modelling of grains with inefficient IA at high-\textit{J} attractors, we adopt $Q_{\rm X,high-J} = 0.15$ as a typical value in this paper.

\begin{figure*}
\centering
        \includegraphics[width=\textwidth,height=\textheight,keepaspectratio]{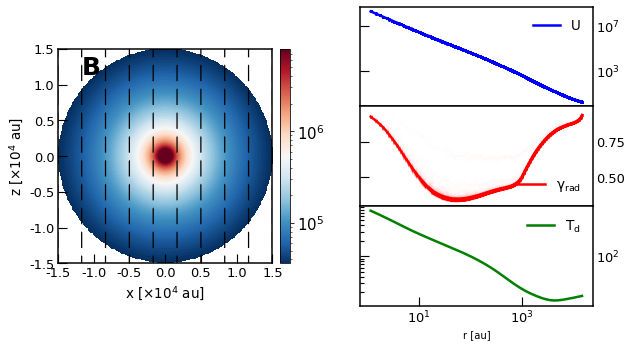}
     \caption{Left panel: Model of a spherical protostellar core of size 15000 au with the embedded low-mass protostar in the center. The gas density is uniform with $n_{\rm H} = 7\times10^{6}\cm^{-3}$ in the central region of size 1000 au (red circle) and decreases with $n_{\rm H} \sim r^{-2}$ in the envelope. The white dashed line shows the magnetic field direction, with the constant magnetic field strength of $B = 134\mu G$. Right panel: Variation of the radiation field strength $U$, the average anisotropic degree $\gamma_{\rm rad}$, and the dust temperature $T_{\rm d}$ along r direction.}
     \label{fig:model}
\end{figure*}

For grains aligned with \B at low$-J$ attractors (left part), grains within $a_{\rm min, aJ}^{\rm low-J} < a < a_{\rm max, aJ}^{\rm low-J}$ have fast internal relaxation but inefficient IA due to thermal fluctuations inside grains. The angle $\theta$ follows the Boltzmann distribution given by Equation (\ref{eq:f_zeta}) and the internal alignment degree $Q_{\rm X,low-J}$ is given by Equation (\ref{eq:QX_lowJ}). Grains beyond this range have slow internal relaxation and inefficient IA with right or wrong IA. We adopt $Q_{\rm X,low-J} = -0.1$ for the case of wrong IA and $Q_{\rm X,low-J} = 0.05$ for grains with right IA to study the effect of grains with slow internal relaxation on synthetic polarizations of protostellar cores. The choice of low $Q_{\rm X} \sim 0$ for grains with slow internal relaxation is to present the weak internal alignment between $\ahat_{1}$ and \J (grains with random orientation has $Q_{\rm X} = 0$).
 
For the alignment between \J and \B, we use the same assumption of the perfect external alignment as in the current version of POLARIS. Then, by plugging the terms $Q_{\rm X}$ and $Q_{\rm J}$ for grains at high and low-\textit{J} attractors into Equation (\ref{eq:R_new}), one can calculate the alignment degree of grains with \B (i.e., $R$), and use it to perform synthetic polarizations as 
described in Section \ref{sec:POLARIS_stoke_parameter}. The summary of new effects that we incorporate into POLARIS is 
given
in Table \ref{tab:physics_old_new}.
 
\section{Modelling results for a protostellar core}\label{sec:grain_alignment}
 
\subsection{Physical model of a protostellar core} \label{sec:model_setup}
We consider a spherical protostellar core and describe it by the spherical grid with $N_{\rm r}\times N_{\vartheta} \times N_{\varphi} = 250\times101\times1$ where $N_{\rm r}, N_{\vartheta}$, and $N_{\varphi}$ are the number of cells along the radial, polar, and azimuthal direction. \footnote{The choice of one cell in the azimuthal direction $\varphi$ is to save the computation time due to the symmetric of gas density around the the protostar (see Appendix \ref{sec:grid_model}).}

The gas density follows the Bonor-Ebert distribution:
\bea 
n_{\rm H} = n_{\rm H,0}\left\{
\begin{array}{l l}
     1~~~~~~~~ ~~~~~~~~~ {\rm for} ~ r \leq R_{\rm center} \\
     (r/r_{\rm center})^{-2}~~ {\rm for} ~ R_{\rm center} < r \leq R_{\rm outer}
\end{array}\right.,
\label{eq:n_H}
\ena
which is 
constant in the central region of $r < R_{\rm center}$ and decreases outward. We take the model of the protostellar core B335 adopted in \cite{Brauer_2016} with  $R_{\rm center} = 1000$ au and $R_{\rm outer} = 15000$ au. We assume the total gas mass in the protostellar core is $M_{\rm gas} = 8M_{\odot}$, which lies in the range of gas mass between $M_{\rm gas} = 4M_{\odot}$ and $\sim 40 M_{\odot}$ for the Bok globule (\citealt{Bok_1977}, \citealt{Leung_1985}, \citealt{Clemens_1991}). The gas density distribution on the x$-$z plane is illustrated in the left panel of Figure \ref{fig:model}, with the constant gas density of  $n_{\rm H} = 7\times 10^{6}\cm^{-3}$ in the central region and $n_{\rm H} = 4\times 10^{4}\cm^{-3}$ at $R_{\rm outer} = 15000$ au. The magnetic field is uniform along the vertical direction $z$ (white dashed line in the left panel of Figure \ref{fig:model}), with the constant magnetic field strength of $B = 134~\mu \rm G$ measured by \cite{Wolf_2004} toward B335.

For the radiation source, we consider a single low-mass protostar in the center of the core and assume it is a blackbody with the temperature of $T_{\rm star} = 6000$ K, the stellar radius of $R_{\rm star} = 2R_{\odot}$, and the total luminosity of $L_{\rm star} = 4.6L_{\odot}$  (\citealt{Brauer_2016}). We also consider ISRF, which is the main source for grain alignment in the outer boundary of the protostellar envelope (\citealt{Hoang_et_al_2021}), with the spectral energy density distribution given by \cite{Mathis_1983}. In contrast to the protostellar source that sents out photons to the grid, photons from the interstellar radiation field (ISRF) 
are ejected randomly from the outer boundary of the envelope inward during the MCRT simulation. For the range of radiation field, we choose the lower limit of $\lambda_{\rm min} = 0.1\mum$ at which all shorter wavelengths are mostly absorbed by the photoionization of hydrogen atoms and the upper limit of $\lambda = 3$ mm at which RATs efficiency is negligible with grain size.

For the dust model, we adopt the composite grains including $67.5\%$ of silicate and $37.5\%$ of graphite (e.g., \citealt{Draine_2021}). This composite dust model is motivated by the astrodust model proposed by \cite{Draine_2021} and is reasonable in protostellar environments when sub-micron grains bounded by ice mantles can collide and stick together to form larger fluffy composite grains (\citealt{Kataoka_2013}, \citealt{Okuzumi}). We assume the uniform dust-to-gas mass ratio of $\eta = 0.01$ in the entire protostellar core, and consider that they follow the standard MRN distribution $dn/da = C a^{-3.5}$  (\citealt{Mathis_1977}) with $C$ the normalization constant derived from $\eta$ (see \citealt{Mathis_1977} and the derivation in Section \ref{sec:stoke_parameter}). We adopt the minimum grain size of $a_{\rm min} = 3.5$ {\AA} and the maximum grain size of $a_{\rm max} = 100\mum$. Our choice of the maximum grain size is motivated by the detection of grains of $a > 10\mum$ in the envelope and the central region of Class 0 YSOs (\citealt{Miotello_2014}, \citealt{Kwon_2009}). Indeed, recent studies about Class II YSOs show that the maximum grain size in this stage should be $a_{\rm max} \leq 100\mum$ (see e.g., \citealt{Dent_2019}, \citealt{Okuzumi_2019}). Thus, the presence of VLGs up to $a_{\rm max} = 100\mum$ may be not common around Class 0/I YSOs. However, \cite{Brauer_2016} showed that the alignment of VLGs allows polarization by dichroic extinction to become the main source of polarization at sub-mm. Therefore, to quantify the efficiency of all available polarization mechanisms around the protostellar core, we choose $a_{\rm max} = 100\mum$. The modelling results for the maximum grain size of $a_{\rm max} = 10\mum$ are shown in Appendix \ref{sec:amax_10um}.

To model the effect of iron inclusions on grain alignment and synthetic polarizations, we consider PM grains with $f_{\rm p} = 0.14285$. For SPM grains, we fix $\phi_{\rm sp} = 0.005$ that corresponds to $\sim 1.67\%$ of iron abundance presented in the form of iron clusters (\citealt{Hoang_Lazarian_2016a}), and vary $N_{\rm cl}$ to describe different magnetic properties of grains. The parameters used in our model are summaried in Table \ref{tab:input_parameter}.

\begin{table}
\centering
\caption{Parameters used in POLARIS}
\begin{tabular}{lll}
\hline
Quantity & Symbol & Value \\   
\hline
\multicolumn{3}{c}{\textbf{Protostellar core model and radiation sources}}\\
\hline
Central boundary         & $R_{\rm center}$     & 1000 au  \\
Envelope boundary        & $R_{\rm outer}$      & 15000 au \\
Total gas mass              & $M_{\rm gas}$        & $8 M_{\odot}$ \\   
Stellar radius              & $R_{\rm star}$       & $2R_{\odot}$ \\ 
Effective temperature       & $T_{\rm star}$       &  6000 K  \\
Stellar luminosity          & $L_{\rm star}$       & $4.6 L_{\odot}$ \\
Magnetic field              & $B$                  & $134\mu \rm G$ \\
\hline
\multicolumn{3}{c}{\textbf{Dust model}}\\
\hline
Grain axial ratio                & $s$                    & 0.5\\
Dust-to-gas mass ratio      & $\eta$               & 0.01\\
Grain size distribution           & $\rm dn/da$              & C$ a^{-3.5}$\\
Minimum grain size          & $a_{\rm min}$        & 3.5\AA  \\ 
Maximum grain size          & $a_{\rm max}$        & $100\mu m$  \\
Fraction of silicate        &                      & $67.5\%$  \\
Fraction of graphite        &                      & $32.5\%$  \\
Iron fraction               & $f_{\rm p}$          & 0.14285\\
Volume filling factor       & $\phi_{\rm sp}$      & 0.005\\
Iron atoms/cluster  & $N_{\rm cl}$       & $10 - 10^{4}$ \\
\hline
\multicolumn{3}{c}{\textbf{Observation parameters}} \\ 
\hline
Distance to observer        &                      & 100 pc  \\
Map size - Resolution       &  Full map            & 30000 au - 120 au \\
                            &  Zoom in 1000 au     & 1000 au - 4 au \\
Wavelengths                  &  \multicolumn{2}{l}{$89, ~250, ~450,~ 870\mum$, 1.3 mm, 2 mm} \\
 
  \hline
    \label{tab:input_parameter}
    \end{tabular}   
\end{table}

\subsection{Results for grain alignment}\label{sec:result_grain_alignment}

\subsubsection{Radiation field distribution}

The right panel of Figure \ref{fig:model} shows the radial distribution of the radiation field strength  $U = u_{\rm rad}/u_{\rm ISRF}$ with $u_{\rm rad}$ the energy density of the radiation spectrum and $u_{\rm ISRF} = 8.64\times10^{-13} ~\rm erg cm^{-3}$ the energy density of ISRF (\citealt{Mathis_1983}), the mean anisotropic degree $\gamma_{\rm rad} = \int \gamma_{\lambda} u_{\rm \lambda} d\lambda / \int u_{\lambda} d\lambda$, and the dust temperature $T_{\rm d}$, from top to bottom, respectively. Generally, $U$ and $T_{\rm d}$ decrease outward due to the dust extinction. However, $T_{\rm d}$ slightly increases in the envelope scale of $r > 10000$ au due to the increasing contribution from ISRF and the drop of gas density in the envelope (Equation \ref{eq:n_H}). The protostellar radiation field is highly anisotropic ($\gamma_{\rm rad} \sim 1$) within 10 au from the protostar, then becomes less anisotropic ($\gamma_{\rm rad}$ decreases to $\sim 0.7$) outward due to the strong scattering of protostellar radiation by dust and the strong emission from hot dust in the dense central region. The radiation field becomes highly anisotropic again ($\gamma_{\rm rad}$ increases to 1) beyond $r > 1000$ au due to the weak interaction between thermal dust emission and dust grains in the envelope. 

\subsubsection{Critical sizes for external alignment via B-RAT ($a_{\rm align}$, $a_{\rm max,JB}^{\rm Lar}$)}\label{sec:align_amaxJB}

\begin{figure}
    \includegraphics[width=0.46\textwidth,height=0.5\textheight,keepaspectratio]{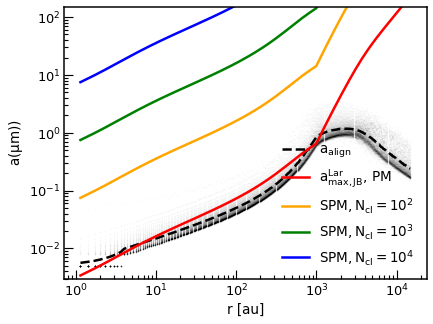}
 
    \caption{Variation of the average minimum $a_{\rm align}$ (black dashed line) and maximum alignment size $a_{\rm max,JB}^{\rm Lar}$ (color lines) as a function of distance $r$ of PM and SPM grains with different $N_{\rm cl}$, assuming $\phi_{\rm sp} = 0.005$. Within 1000 au, the alignment range shifts from sub-micron size near the protostar to micron size at $r \sim 1000$ au due to the reduced RAT alignment efficiency by dust extinction. Beyond 1000 au, the alignment range is broader, i.e., smaller $a_{\rm align}$ and larger $a_{\rm max,JB}^{\rm Lar}$, due to the enhanced RAT alignment by ISRF and the reduced gas randomization in the envelope. More large grains can be aligned with \B with increasing $N_{\rm cl}$ due to the enhanced Larmor precession by iron inclusions. Typically, VLGs can have the magnetic alignment even in the central region with $n_{\rm H} = 1.6\times10^{7}\cm^{-3}$ if they have $N_{\rm cl} \geq 10^{4}$ and $\phi_{\rm sp} \geq 0.005$.}
     \label{fig:align_amaxJB}
\end{figure}

 \begin{figure*}
 \centering
        \includegraphics[width=\textwidth,height=\textheight,keepaspectratio]{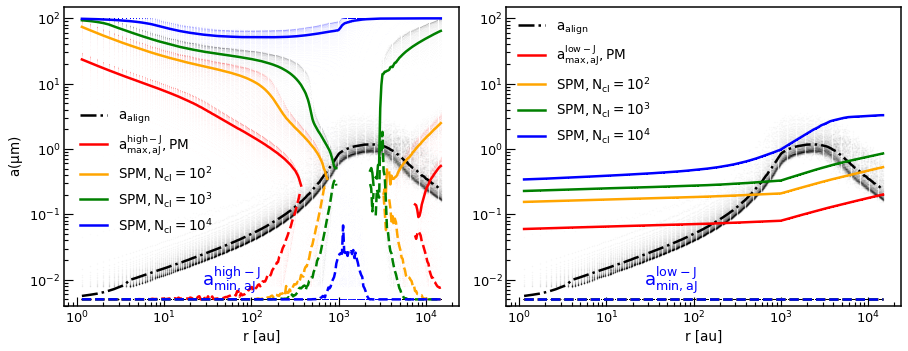}
       
     \caption{Variation of the average minimum and maximum size with fast internal relaxation for PM and SPM grains with different $N_{\rm cl}$. The left panel shows the results for grains aligned with \B at high-\textit{J} attractors ($a_{\rm min,aJ}^{\rm high-J} - a_{\rm max,aJ}^{\rm high-J}$), and the right panel is for grains at low-\textit{J} attractors ($a_{\rm min,aJ}^{\rm low-J} - a_{\rm max,aJ}^{\rm low-J}$). At high-\textit{J}, the range of grains with fast internal relaxation is broader near the protostar and in the outer boundary of the envelope irradiated by ISRF. In contrast, at low-\textit{J}, grains tend to have fast internal relaxation in the envelope due to the reduced gas randomization. However, grains only rotate at thermal rotation, thus, all large grains of $a\geq 1\mum$ cannot have fast internal relaxation.}
     \label{fig:align_amaxaJ}
\end{figure*}

Figure \ref{fig:align_amaxJB} shows the variation of the minimum size of external alignment determined by RATs, $a_{\rm align}$ (black dashed line), and the maximum size of magnetic alignment determined by the Larmor precesssion condition, $a_{\rm max,JB}^{\rm Lar}$ (solid lines) for PM and SPM at different distances to the central protostar $r$. Different values of $N_{\rm cl}$ from 100 to $10^{4}$ are considered. Note that the range $a_{\rm align}-a_{\rm max,JB}^{\rm Lar}$ determines the range of grain sizes which can be aligned with \B by RATs (i.e., B$-$RAT mechanism).

For the minimum size of B-RAT alignment, one can see that $a_{\rm align}$ increases continuously with radial distances due to the decrease of RAT alignment caused by the attenuation of protostellar radiation. The grain alignment size reaches the maximum of $a_{\rm align} \sim 1\mum$ at the boundary between the central region and the envelope at $r\sim 10^{3}$ au, then decreases to $a_{\rm align} \sim 0.2\mum$ at $r = 15000$ au due to the additional contribution from ISRF and the reduced gas density in the envelope (see analytical studies in \citealt{Hoang_et_al_2021}).

For the maximum size of B-RAT alignment, in the central region of $r < 1000$ au with the constant gas density, $a_{\rm max,JB}^{\rm Lar}$ increases outward due to the fast Larmor precession driven by the increased magnetic susceptibility with decreasing grain temperature (Equations \ref{eq:tau _gas} and \ref{eq:tlar_pm}). Beyond $r > 1000$ au, $a_{\rm max,JB}^{\rm Lar}$ increases faster due to the reduced gas randomization in the envelope. 

The maximum size of B-RAT alignment is larger for SPM with higher $N_{\rm cl}$ due to the enhanced Larmor precession by iron inclusions. Consequently, more large (micron-sized) grains are able to align with \B even in the dense central region of protostellar environments. In particular, PM grains cannot be aligned with \B in the central region with $n_{\rm H} \sim 10^{7} \cm^{-3}$ due to its slow Larmor precession, i.e., $a_{\rm max,JB}^{\rm Lar} < a_{\rm align}$. In contrast, VLGs can have magnetic alignment in this region if they are SPM with $N_{\rm cl} \sim 10^{4}$ and $\phi_{\rm sp} = 0.005$.

\subsubsection{Critical sizes with fast internal relaxation ($a_{\rm min,aJ},a_{\rm max,aJ}$)}\label{sec:align_amaxaj}

 \begin{table*}
  \centering
         \caption{Parameters of the grain alignment model}
  \begin{tabular} {lllccll}
  \hline 
  Model name &  $a_{\rm max} (\mum)$ & Magnetic & Slow Larmor & Slow internal & Internal Alignment & B$(\mu G)$\\ 
         &        & properties        &  precession   &  relaxation   &  at low-\textit{J} attractors    &  \\
      \hline
Ideal $^{1}$                  & 100   & $--$          & no        & no        & right IA  ($\ahat_{1} \parallel$ \J)         &  $134$\\
Realistic$^{2}-$rIA
      & 100    & PM, SPM       & yes       & yes       & right IA  & $134$\\

Realistic$-$wIA    & 100   & PM, SPM       & yes       & yes       & wrong IA ($\ahat_{1}\perp \J$)  & $134$\\

Realistic$-$rIA$-$amax  & $[5-150]^{3}$   & PM, SPM       & yes       & yes       & right IA  & $134$\\ 

Realistic$-$wIA$-$amax  & $[5-150]$   & PM, SPM       & yes       & yes       & wrong IA  & $134$\\ 

Realistic$-$rIA$-$amax$-$B    & $[5-150]$   & PM, SPM       & yes       & yes       & right IA  & $[134-1000]$ $^{4}$\\ 

Realistic$-$wIA$-$amax$-$B    & $[5-150]$   & PM, SPM       & yes       & yes       & wrong IA  & $[134-1000]$\\ 

Ideal$-$a10                & 10   & $--$          & no        & no        & right IA              &  $134$\\
Realistic$-$rIA$-$a10 $^{5}$    & 10    & PM, SPM       & yes       & yes       & right IA  & $134$\\
Realistic$-$wIA$-$a10    & 10   & PM, SPM       & yes       & yes       & wrong IA   & $134$\\

  \hline
    \label{tab:model_table}
    \end{tabular}

 \footnotesize{($^1$): In Ideal model, all grains of $a \geq a_{\rm align}$ are considered to align with \B, with $f_{\rm high-J} = 1$.}\\
 \footnotesize{($^2$): In Realistic model, all effects described from Section \ref{sec:POLARIS_amaxJB} to Section \ref{sec:POLARIS_amaxJB_DG} are taken into account, with the maximum alignment size is determined by the condition $\tau_{\rm Lar} \leq \tau_{\rm gas}/10$, the grain with fast internal relaxation is determined by condition $\tau_{\rm BR} \leq \tau_{\rm gas}$, and $f_{\rm high-J}$ varies with $\delta_{\rm m}$ given by Equation (\ref{eq:f_highJ}).}\\
  \footnotesize{($^3$): The maximum grain size $a_{\rm max}$ varies from $5\mum$ to $150\mum$.}\\
   \footnotesize{($^4$): Consider two values of magnetic fields, $B = 1000\mu \rm G$ in the central 1000 au region and $B = 100\mu \rm G$ in the envelope.}\\
      \footnotesize{($^5$): Results for model $...-\rm a10$ are shown in Appendix \ref{sec:amax_10um}.}
\end{table*}

 \begin{figure}[t]
        \includegraphics[width=0.46\textwidth,height=0.5\textheight,keepaspectratio]{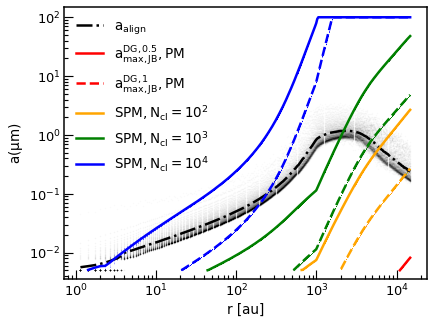}
     \caption{Variation of the maximum size for magnetic alignment by the MRAT mechanism, $a_{\rm max,JB}^{\rm DG,0.5}$ and $a_{\rm max,JB}^{\rm DG,1}$, for PM and SPM grains with different $N_{\rm cl}$. The values of $a_{\rm max,JB}^{\rm DG}$ increase with increasing $N_{\rm cl}$, but decrease toward the central protostar. The range of grains with efficient external alignment ($a\sim [a_{\rm align}-a_{\rm max,JB}^{\rm DG}]$) is more extended in the envelope due to the reduced gas randomization.}
     \label{fig:align_amaxJB_DG}
\end{figure}

The left panel of Figure \ref{fig:align_amaxaJ} shows the variation at different distances $r$ of the minimum size, $a_{\rm min,aJ}^{\rm high-J}$ (dashed lines), and maximum size, $a_{\rm max,aJ}^{\rm high-J}$ (solid lines), of grains having fast internal relaxation by Barnett relaxation for PM and SPM grains  aligned with \B at high-\textit{J} attractors. The range of grains having fast internal relaxation is more extended toward the protostar, i.e., broader range of $a_{\rm min,aJ}^{\rm high-J} - a_{\rm max,aJ}^{\rm high-J}$, due to the faster Barnett relaxation driven by suprathermal rotation of grains by efficient RATs. In the inner region of the envelope of $r \sim 5000$ au, grains tend to have slow internal relaxation due to the reduced Barnett relaxation as a result of the slow rotation of grains in attenuated radiation fields. In the outer envelope of $r > 5000$ au, aligned grains can have fast internal relaxation again due to the increased grain rotational rate by ISRF and the decreased gas randomization. The size range of fast internal relaxation is more extended to large micron-sized grains and also in space for SPM grains with larger $N_{\rm cl}$ due to stronger Barnett relaxation by iron inclusions. For example, all SPM grains of $a\geq a_{\rm align}$ (black dashed line) can have fast internal relaxation at high-\textit{J} attractors and then efficient IA in the entire protostellar core with $N_{\rm cl} \sim 10^{4}$ and $\phi_{\rm sp} = 0.005$.

The right panel of Figure \ref{fig:align_amaxaJ} shows the similar results as the left panel but for grains aligning with \B at low-\textit{J} attractors. In contrast to the complex variation of grains with fast internal relaxation on distances at high-\textit{J} attractors, the size range of grains with fast relaxation at low-\textit{J} attractors simply increases from the central region toward the envelope due to the reduced effect of gas randomization. This simple tendency is because grains at low-J attractors rotate with the thermal angular velocity $\Omega_{T}$ which is not sensitive to the change in the radiation field as grains at high-\textit{J} attractors. The smaller value of $a_{\max,aJ}^{\rm low-J}$ in the central region is caused by the decrease of the magnetic susceptibility with increasing grain temperature near the protostellar source (Equations \ref{eq:chi_0_paramagnetic} and \ref{eq:chi_0_super_paramagnetic} and Figure \ref{fig:model}, right panel). The size range of grains with fast relaxation is extended for SPM grains with larger $N_{\rm cl}$. However, for grains at low-\textit{J}, only small grains of $a<1\mum$ can have fast relaxation, while larger grains of $a>1\mum$ always have slow internal relaxation due to their slow rotation.  

\subsubsection{Critical sizes for magnetically enhanced RAT alignment (MRAT),  $a_{\rm max,JB}^{\rm DG}$}\label{sec:align_amaxJB_DG}

Figure \ref{fig:align_amaxJB_DG} shows our numerical results for the critical sizes at which the magnetic relaxation is efficient in enhancing the RAT alignment (i.e., the MRAT mechanism), $a_{\rm max,JB}^{\rm DG,0.5}$ (solid lines) and $a_{\rm max,JB}^{\rm DG,1}$ (dashed lines), for different magnetic properties of grains. For PM grains (red lines), they are mainly aligned by RATs because of the negligible effect of magnetic relaxation, i.e., $a_{\rm max,JB}^{\rm DG,0.5}\ll a_{\rm align}$. In contrast, SPM grains with high $N_{\rm cl}$ can have efficient external alignment in the envelope by the MRAT mechanism. In particular, the high level of iron inclusions with $N_{\rm cl} = 10^{4}$ and $\phi_{\rm sp} = 0.005$ allows $100\%$ of grains to be aligned with \B at high-\textit{J} attractors beyond $\sim 1000$ au.

However, the high level of iron inclusions cannot significantly increase the efficiency of MRAT alignment in the central region of $r < 1000$ au, i.e., $a_{\rm max,JB}^{\rm DG}$ drops toward the center. The decreased magnetic relaxation is due to the decrease of the magnetic susceptibility with increasing grain temperature toward the protostar (Equations \ref{eq:chi_0_paramagnetic} and \ref{eq:chi_0_super_paramagnetic}). In addition, the increase in gas randomization also results in the reduction of the efficiency of the MRAT mechanism here. Thus, grains in the central region are mainly aligned with \B by RATs with the typical values of $f_{\rm high-J} = 0.25$.

\section{Effect of Iron Inclusions on Polarization Pattern} \label{sec:iron_pol_map}

\begin{table}
\centering
\caption{Optical depth in the central region of $1000$ au scale}
\begin{tabular} {cccccc}
  \hline 
   Wavelength        &  $89\mum$  & $250\mum$   & $450\mum$ & $870\mum$ & 2 mm \\
  \hline
  $\tau_{\lambda}$  &  2.75      & 1.14        & 0.59      & 0.08      & 0.02 \\
  \hline
    \label{tab:model_optical_depth}
    \end{tabular}
\end{table}

Using above critical sizes for grain alignment, we can calculate the new Rayleigh reduction factor $R$ described in Section \ref{sec:POLARIS-New_R} and use it to model synthetic polarization maps at optically thin wavelengths of $870\mum$ and $450\mum$ (Table \ref{tab:model_optical_depth}), which can be observed by ALMA. We place the detector at 100 pc from the protostellar core along $y$ direction. The plane detector has $N_{\rm x}\times N_{\rm z} = 250\times 250$ pixels on x direction and z direction, giving the spatial resolution of 120 au when observing the entire protostellar core of 30000 au and 4 au when zooming into the central region of scale 1000 au (Observation parameters are summarized in Table \ref{tab:input_parameter}). 

We first consider the model Ideal in which all grains larger than $a_{\rm align}$ have the perfect alignment with \B, i.e., $f_{\rm high-J} = 1$. Then, we consider the realistic models that include the new effects of slow Larmor precession, slow internal relaxation, and the enhanced external alignment by the MRAT mechanism as presented in Section \ref{sec:POLARIS+}. As discussed in Section \ref{sec:POLARIS_amaxaJ}, grains with slow internal relaxation at low-\textit{J} attractors can have right or wrong IA. However, the fraction and conditions that drive the grain internal alignment state are still unclear. Therefore, we consider two scenarios, the first scenario with $100\%$ of grains with slow internal relaxation at low-\textit{J} having right IA, denoted as model Realistic-rIA, and the second case with $100\%$ of grains with slow internal relaxation at low-\textit{J} attractors having wrong IA, denoted as Realistic-wIA. The summary of parameters of our models are listed in the first fouth rows in Table \ref{tab:model_table}. 

\begin{figure*}
        \includegraphics[width=\textwidth,height=\textheight,keepaspectratio]{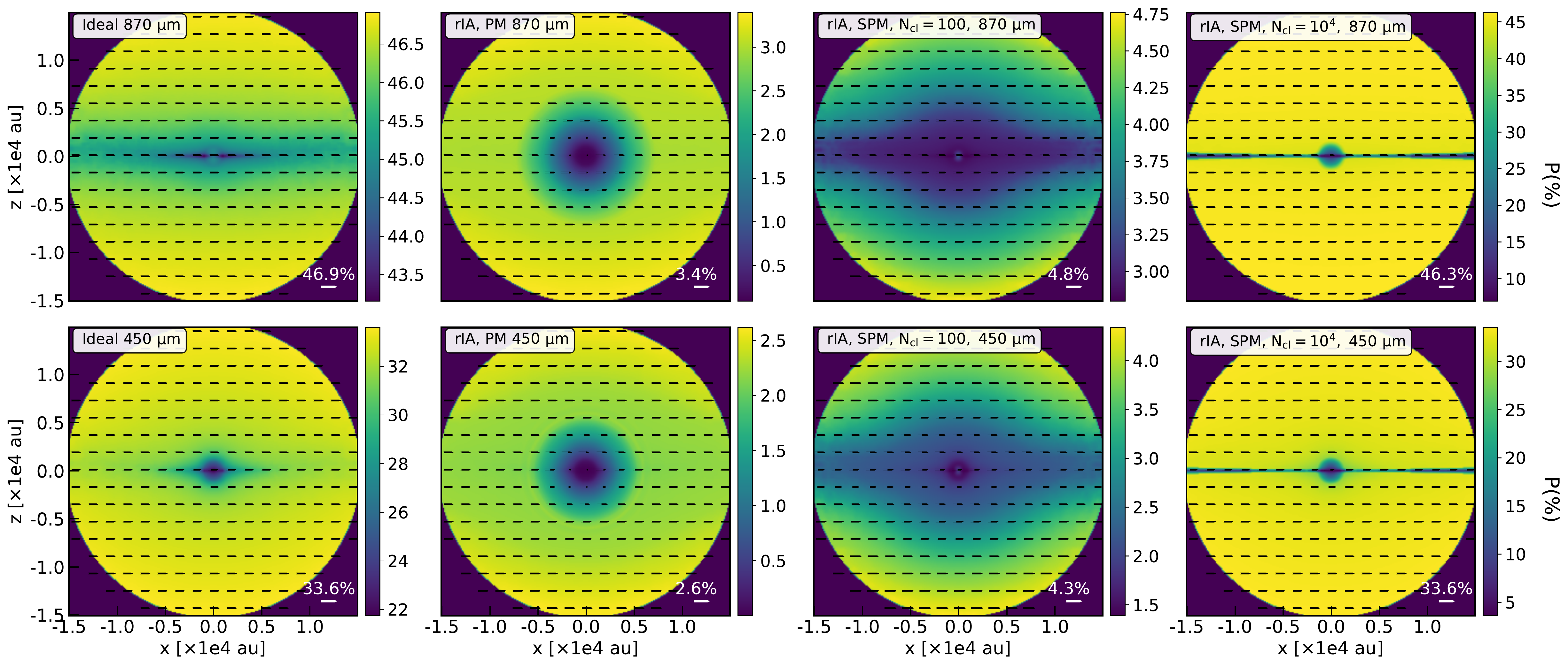}
     \caption{Synthetic polarization maps obtained in the envelope on the x$-$z plane at $870\mum$ (upper panels) and $450\mum$ (lower panels) for model Ideal (first column) and model Realistic-rIA for PM grains (second column) and SPM grains with $N_{\rm cl} = 100$ (third column) and $N_{\rm cl} = 10^{4}$ (fourth column). The color code shows the polarization degree, black segments show  polarization vectors \P scaled with the polarization degree, the magnetic field is along the vertical direction (z axis). Our models of PM and SPM grains show the uniform polarization pattern with \PperpB at $870\mum$ and $450\mum$ because all aligned dust grains have right IA. The polarization degree generally decreases inward due to the reduced grain alignment efficiency in the dense central region. The lower $p$ on the equatorial plane is due to the reduced RATs efficiency, i.e., larger $a_{\rm align}$,  and the reduced amount of grains with perfect magnetic alignment, i.e., smaller $a_{\rm max, JB}^{\rm DG,0.5}$ and $a_{\rm max, JB}^{\rm DG, 1}$} where \textbf{k} $\perp$ \B (see Appendix \ref{sec:alignment_xaxis} for details).
 
     \label{fig:Realistic_rIA_envelope}
\end{figure*}

\begin{figure*}
\centering
    \includegraphics[width=\textwidth,height=\textheight,keepaspectratio]{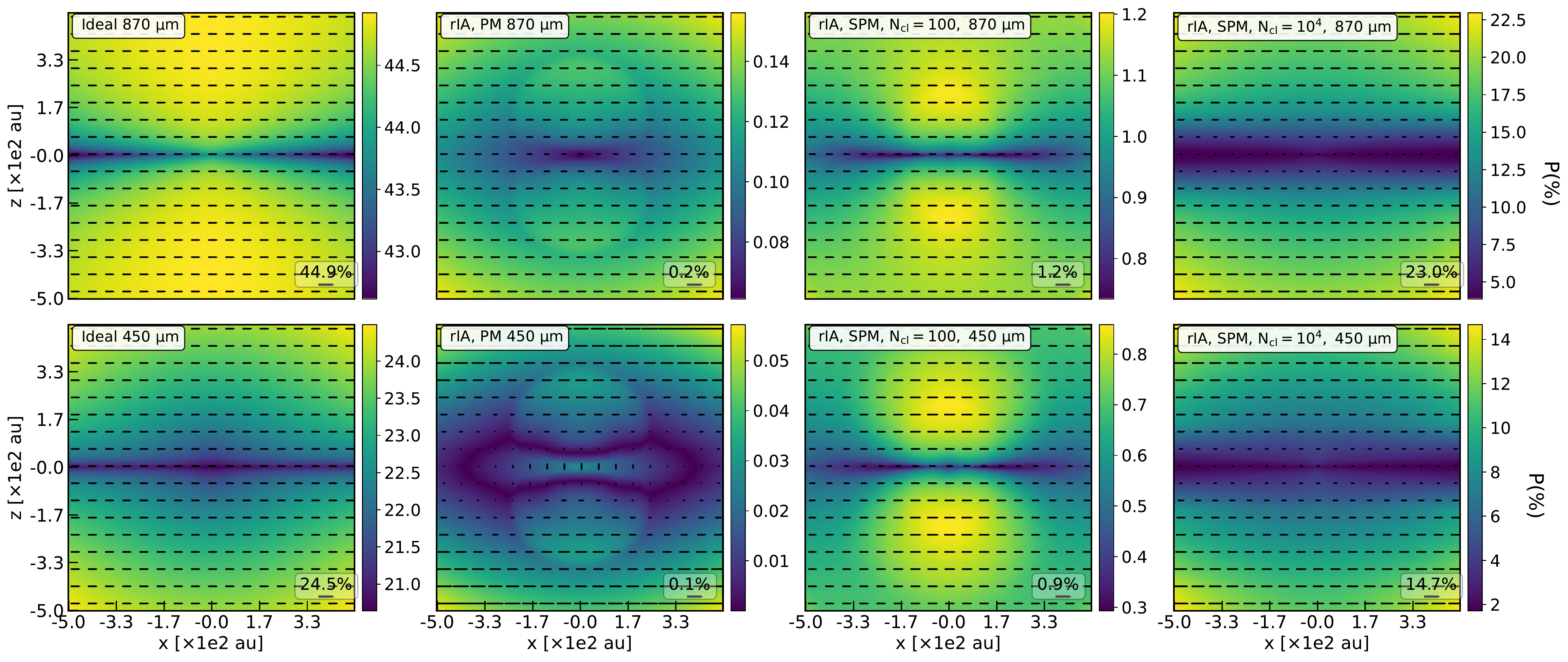}
    \caption{Similar as Figure \ref{fig:Realistic_rIA_envelope} but focusing in the central region of scale 1000 au. The polarization patterns produced by both PM and SPM grains are uniform with \PperpB at both $870\mum$ and $450\mum$ if all grains have right IA. The polarization degree is  slightly smaller on the equatorial plane due to the reduced RAT efficiency when \textbf{k} $\perp$ \B.} 
    
     \label{fig:Realistic_rIA_center}
\end{figure*}

Figures \ref{fig:Realistic_rIA_envelope} and \ref{fig:Realistic_rIA_center} show the comparison of the synthetic polarization pattern obtained from the protostellar envelope and the central region of 1000 au on x$-$z plane at $870\mum$ (upper panels) and $450\mum$ (lower panels) between model Ideal (first column) and model Realistic$-$rIA (second to fourth columns). The color code shows the polarization fraction $p(\%)$, black segments show polarization vectors \P with the length representing the degree of polarization. The magnetic field is along the vertical direction (z-axis). One can clearly see that with the presence of grains with slow internal relaxation, both PM and SPM grains with $N_{\rm cl} = 100$ or $N_{\rm cl} = 10^{4}$ produce the uniform polarization pattern with \PperpB as the Ideal model if they have right IA with $\ahat_{1} \parallel$ \J. The polarization pattern is uniform in the entire protostellar core and does not change with wavelengths. In addition, the polarization degree along the equatorial plane is smaller than ones at higher latitudes. It is a result of the reduced RAT alignment efficiency, i.e., larger $a_{\rm align}$, and the reduced amount of grains having perfect magnetic alignment, i.e., smaller $a_{\rm max,JB}^{\rm DG, 0.5}$ and $a_{\rm max,JB}^{\rm DG,1}$, in the area where the radiation field is perpendicular to B-fields (see Appendix \ref{sec:alignment_xaxis} for the detailed variation of grain alignment on x and z direction).

\begin{figure*}    
\centering
       \includegraphics[width=\textwidth,height=\textheight,keepaspectratio]{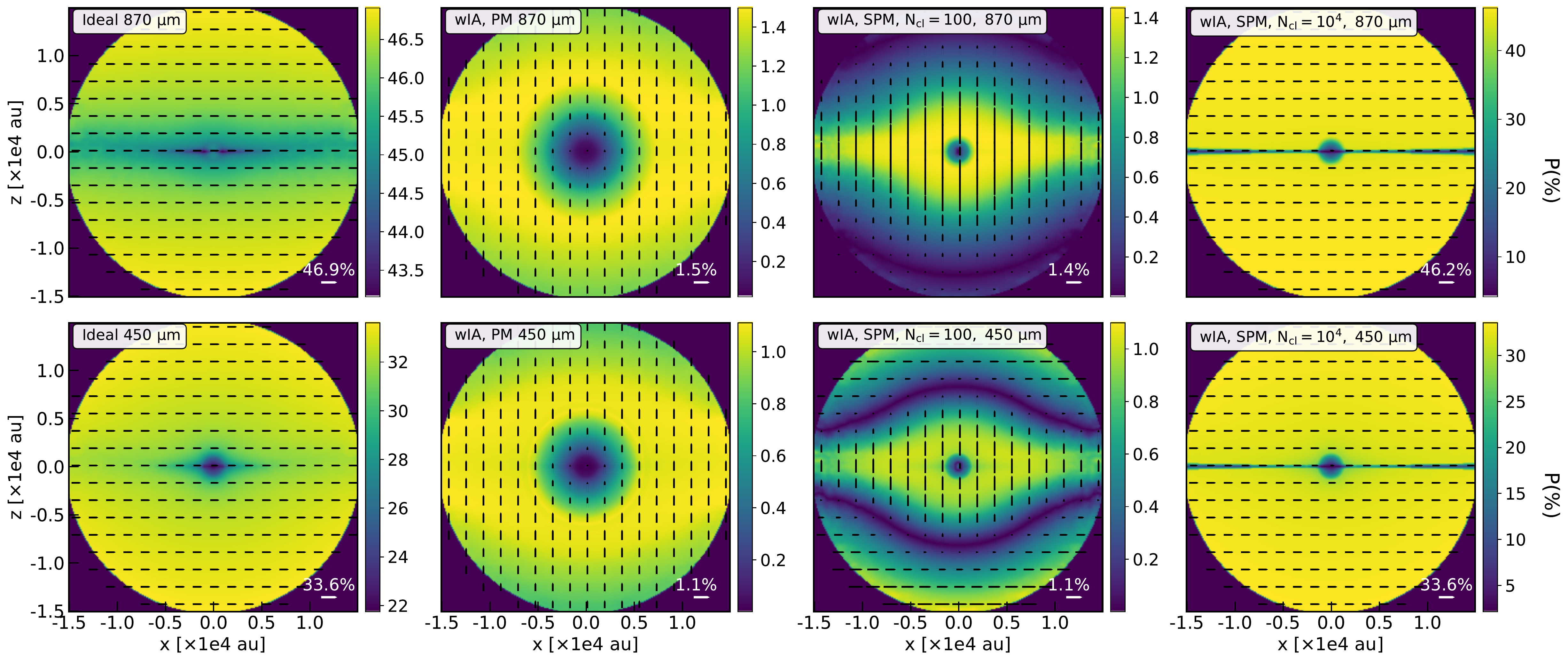}
     \caption{Synthetic polarization maps produced by Ideal model (first column) and model Realistic$-$wIA for PM grains and SPM grains with $N_{\rm cl} = 100$ and $10^{4}$ (second to fourth columns). With the presence of grains with wrong IA by slow internal relaxation, PM grains produce the uniform polarization pattern with \PparaB at both $870\mum$ and $450\mum$ due to the dominant emission from VLGs with wrong IA. SPM grains with $N_{\rm cl} = 100$ produce the uniform polarization pattern with \PparaB at $870\mum$, but the complex one with \PperpB beyond $\sim 5000$ au and \PparaB within 5000 au at $450\mum$. The variation of \P with wavelengths beyond $\sim 5000$ au is driven by the change in IA of the emission source, from VLGs with wrong IA to micron-sized grains with right IA (Figure \ref{fig:align_amaxaJ}, left panel). In contrast, SPM grains with $N_{\rm cl} = 10^{4}$ produce the uniform polarization pattern with \PperpB as model Ideal due to enhanced grain alignment by the MRAT alignment.}
     \label{fig:Realistic_wIA_envelope}
\end{figure*}

\begin{figure*}
\centering
       \includegraphics[width=\textwidth,height=\textheight,keepaspectratio]{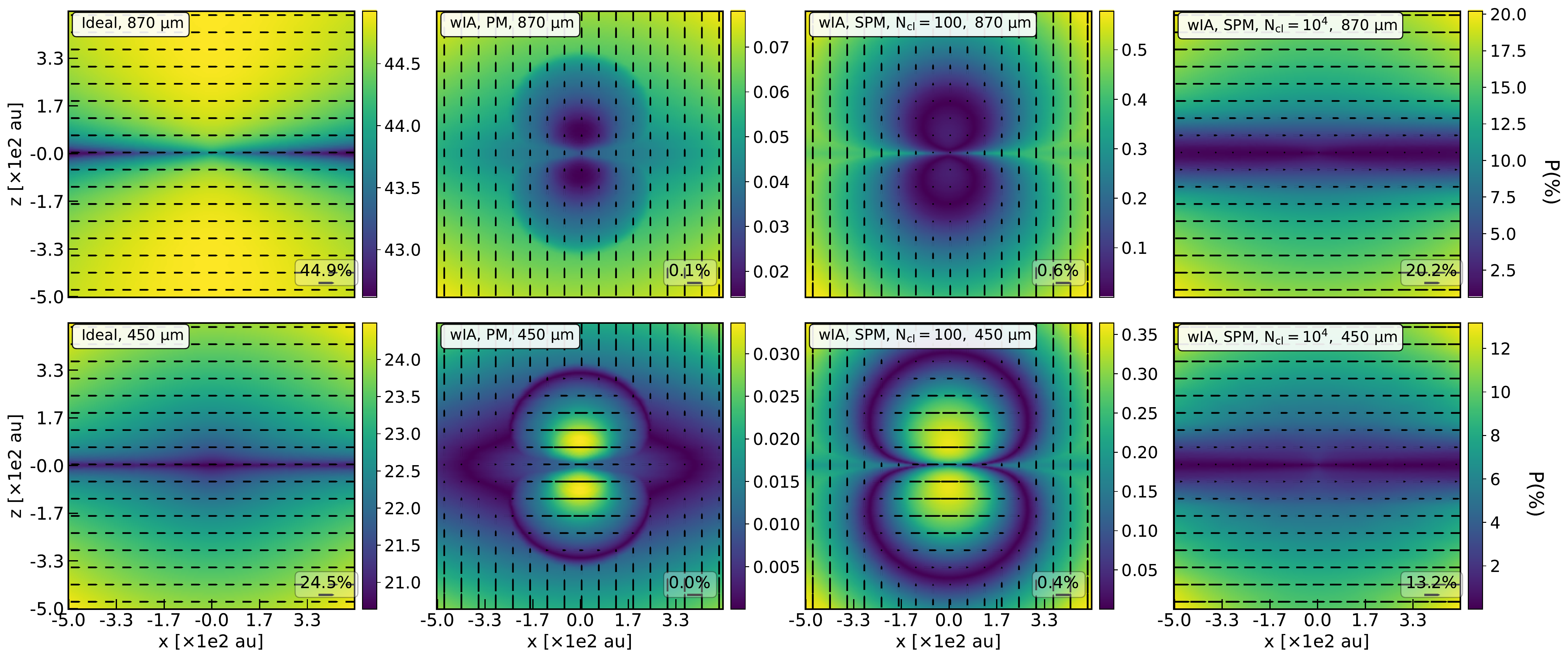}
     \caption{Similar to Figure \ref{fig:Realistic_wIA_envelope} but zoom in into the inner region of 1000 au around the protostar. For PM and SPM with $N_{\rm cl} = 100$, polarization vectors are uniform with \PparaB at $870\mum$ but become complicated with \PparaB beyond 200 au and \PperpB within 200 au at $450\mum$. The polarization flipping in the center is due to the change of the emission source from VLGs with wrong IA to micron-sized grains with right IA. In contrast, SPM grains with $N_{\rm cl} = 10^{4}$ produce the uniform polarization pattern with \PperpB at both $870\mum$ and $450\mum$ as Ideal model because most grains of $a \geq 1\mum$ can have efficient IA even in the central region.}
     \label{fig:Realistic_wIA_center}
\end{figure*}

However, if grains with slow internal relaxation have wrong IA with $\ahat_{1} \perp$ \J (model Realistic$-$wIA), the synthetic polarization pattern becomes much more complicated depending on the level of iron inclusions locked inside dust grains. The results obtained in the envelope for model Realistic$-$wIA is shown in Figure \ref{fig:Realistic_wIA_envelope}.

\begin{figure*}[t]
 \centering
    \includegraphics[width=\textwidth,height=\textheight,keepaspectratio]{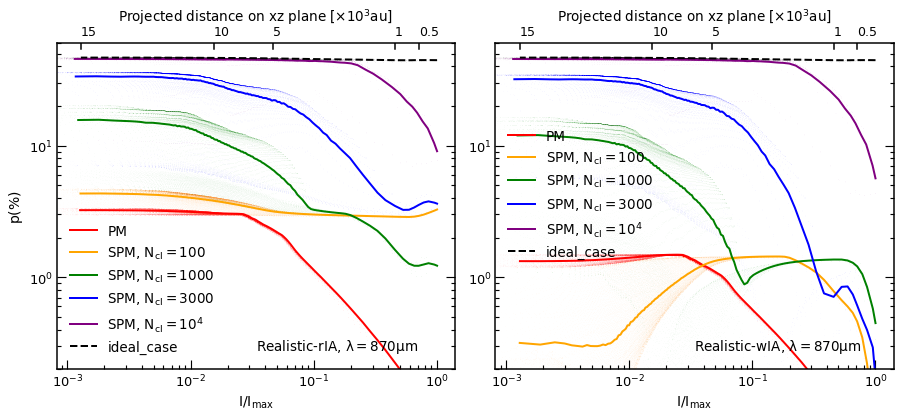}
    \includegraphics[width=\textwidth,height=\textheight,keepaspectratio]{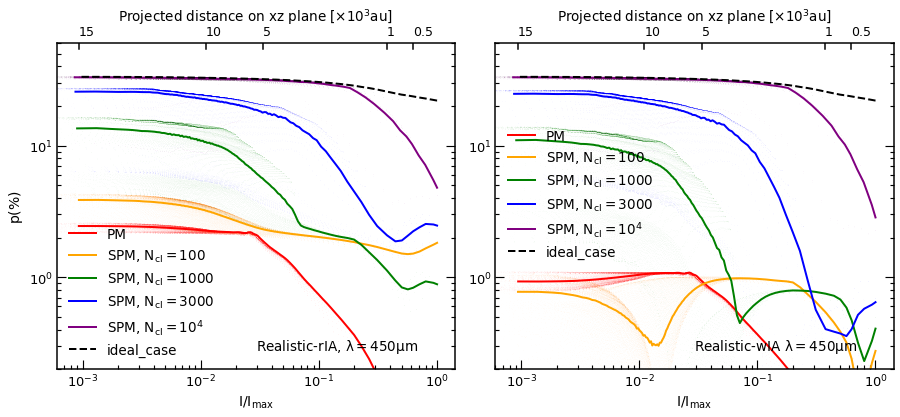}
 
     \caption{Variation of the polarization degree at $870\mum$ (upper panels) and $450\mum$ (lower panels) with the normalized intensity $I/I_{\rm max}$ for different magnetic properties of grains. The result from model Ideal is plotted by the black dashed line for comparison. The left panel shows the results for model Realistic$-$rIA, and the right panel is for model Realistic$-$wIA. In our models, the polarization degree always decreases with increasing $I/I_{\rm max}$ and generally increases with increasing $N_{\rm cl}$. And this feature is not different between the calculations at $870\mum$ and $450\mum$. In addition, if grains with wrong IA exist in the core, one will obtain the "valley-polarization hole" with $p < 1\%$ in $p-I/I_{\rm max}$ curve caused by the cancelling effect of polarized emission from grains with wrong and right IA. }
 \label{fig:iron_P_I}
\end{figure*}

In particular, PM grains (second column) produce the uniform polarization pattern in the entire envelope at $870\mum$ and $450\mum$, but with \PparaB arising from the emission of grains with wrong IA (i.e., most PM grains have slow internal relaxation, Figure \ref{fig:align_amaxaJ}, left panel, red line). For SPM with the moderate level of iron inclusions ($N_{\rm cl} = 100$, third column), the polarization pattern is uniform with \PparaB at $870\mum$ because VLGs in the envelope have wrong IA by slow internal relaxation. At shorter wavelengths of $450\mum$, the polarization pattern becomes complicated with \PperpB beyond $\sim 5000$ au and \PparaB in the inner region. The change of the polarization pattern in the outer envelope is caused by the change in IA of the emission source from VLGs with wrong IA to micron-sized grains with right IA. In contrast, the polarization pattern with \PparaB within 5000 au does not change with wavelengths because most of grains have slow internal relaxation with wrong IA here (Figure \ref{fig:align_amaxaJ}, left panel). 

If SPM grains have the high level of iron inclusion ($N_{\rm cl} = 10^{4}$, fourth column), they can produce the uniform polarization pattern with \PperpB in the entire envelope at both $870\mum$ and $450\mum$ as the Ideal model (first column). Such polarization pattern is achieved due to the efficient IA of all aligned grain in the envelope driven by the fast Barnett relaxation.

Figure \ref{fig:Realistic_wIA_center} shows the similar results as Figure \ref{fig:Realistic_wIA_envelope} but zoom in into 1000 au from the center. For PM and SPM grains with $N_{\rm cl} = 100$ (second and third columns), the polarization pattern is uniform with \PparaB at $870\mum$ due to the emission of VLGs with wrong IA. At $450\mum$, it becomes complicated with \PparaB beyond 300 au and \PperpB within 300 au from the protostar. The change of the polarization pattern in the inner $\sim 300$ au region with wavelengths is due to the change in IA of the emission source from VLGs with wrong IA to micron-sized grains with right IA. Indeed, for PM grains, grains in the central region do not radiate polarized emission because they are not aligned with \B here (Figure \ref{fig:align_amaxJB}, red line). Therefore, the polarization vectors \PperpB in the inner $\sim 200$ au region at $450\mum$ (second column) is originated from the polarized emission of sub-micron grains with efficient IA in the boundary of the envelope (Figure \ref{fig:align_amaxaJ}, left panel, red line). For SPM with $N_{\rm cl} = 100$, they can have the magnetic alignment and also efficient IA in the central region. The polarization signal with \PperpB in the inner $\sim 300$ au region at $450\mum$  (third column) thus comes from both the emission of aligned dust grains with efficient IA within $\sim 400$ au from the protostar and beyond 4000 au in the envelope (Figure \ref{fig:align_amaxaJ}, left panel, orange line). For SPM grains with high $N_{\rm cl} = 10^{4}$, the polarization pattern is uniform with \PperpB at both $870\mum$ and $450\mum$ as the Ideal model because of the efficient grain alignment by the MRAT mechanism.

\section{Effect of Iron Inclusions on $p - I$ relationship}\label{sec:iron_P_I}

In all polarization maps shown in Section \ref{sec:iron_pol_map}, one can see that the polarization degree tends to decrease toward the central region. In this section, we will study in detail the dependence of the polarization degree $p(\%)$ on the magnetic properties of grains.

Figure \ref{fig:iron_P_I} shows the variation of polarization degree $p(\%)$ with normalized intensity $I/I_{\rm max}$ with $I_{\rm max}$ the maximum intensity of thermal emission obtained in the center of the protostellar core. The projected distances to the center on x$-$z plane $d_{\rm proj}$ corresponding to each value of $I/I_{\rm max}$ are shown in the upper x axis. Scattered points represent the values of $p - I/I_{\rm max}$ on the ring of height $d_{\rm resol} = 120$ au and radius of $d_{\rm proj}$, solid color lines show the average polarization degree obtained by our models for different magnetic properties of grains. The result from model Ideal is plotted by the black dashed line for comparison.

The upper left panel of Figure \ref{fig:iron_P_I} shows the results obtained at $870\mum$ for model Realistic$-$rIA. In contrast to the constant $p \sim 45\%$ with distances in Ideal model, the calculated polarization degree for model Realitic$-$rIA decreases continuously with increasing intensity toward the center, and generally is larger for SPM grains with higher fraction of iron inclusions. Particularly, PM grains only produce the constant low $p \sim 3\%$ in the envelope, i.e.,  $d_{\rm proj} > 5000$ au, because they mainly have inefficient IA driven by slow internal relaxation. Moving toward the center, $p$ decreases to negligible values due to the loss of grain alignment within $1000$ au (Figure \ref{fig:align_amaxJB}). In contrast, SPM grains with $N_{\rm cl}$ changing from 100 to $10^{4}$ produce the constant high polarization degree from $p \sim 4\%$ up to  $p \sim 45\%$ in the envelope due to the increase of amount of grains with efficient IA at high-\textit{J} attractors, i.e., higher $f_{\rm high-J}$ and broader range of $a_{\rm min,aJ}^{\rm high-J} - a_{\rm max,aJ}^{\rm high-J}$ (Figures \ref{fig:align_amaxaJ} and \ref{fig:align_amaxJB_DG}). The polarization degree produced by SPM slightly decreases to $p \sim 3 - 10\%$ near the center due to the reduced internal and external alignment by the MRAT mechanism in the dense central region.

The upper right panel of Figure \ref{fig:iron_P_I} shows the variation of $p(\%)$ with $I/I_{\rm max}$ obtained for model Realistic$-$wIA at $870\mum$. Similar as the left panel for model Realistic$-$rIA, one obtains the decrease of $p$ with increasing intensity due to the reduced grain alignment efficiency by the MRAT mechanism in dense environments and the increase of $p$ with increasing levels of iron locked inside dust grains. However, the polarization degree produced by model Realistic$-$wIA is slightly smaller than model Realistic$-$rIA as a result of the cancelling effect between polarized emission from grains with right and wrong IA.

The lower left (for model Realistic$-$rIA) and lower right panels (for model Realistic$-$wIA) of Figure \ref{fig:iron_P_I} show the similar results as the upper panels, but for the wavelength of $450\mum$. Generally, one can obtain the similar reduction of $p$ with $I/I_{\rm max}$ and the rise of $p$ with $N_{\rm cl}$ as the calculation at $870\mum$. However, in contrast to the smooth reduction of $p(450\mum)$ with $I/I_{\rm max}$ produced by model Realistic$-$rIA, in model Realistic$-$wIA, one can see the sudden drop of the polarization degree to $p < 1\%$ at the distance where the polarization degree changes from high value of $p \sim 10-45\%$ in the envelope to low $p \sim 1\%$ in the center, which is termed as the "valley-polarization hole". The presence of the polarization valley is originated from the cancelling effect of the polarization signal radiating from grains with right and wrong IA. The V-shape of the $p-I/I_{\rm max}$ curve corresponds to the gap with $p < 1\%$ between the outer envelope with \PperpB and the inner region with \PparaB shown in the third column of Figure \ref{fig:Realistic_wIA_envelope}. The second change of the polarization pattern from \PparaB beyond $> 300$ au to \PperpB within $\sim 300$ au (Figure \ref{fig:Realistic_wIA_center}, third column) is also featured by the second valley at $d_{\rm proj} < 500$ au (the orange and green lines in the upper right panel). 

\section{Effect of Iron Inclusions on the Polarization by Dichroic Extinction}\label{sec:iron_extinction}

Next, we examine the effect of iron inclusions on the polarization by dichroic extinction of aligned dust grains. \cite{Brauer_2016} suggested that the dichroic extinction by aligned VLGs in protostellar cores can reduce polarized thermal emission and become the main source of polarization at sub-mm wavelengths. We first study the effect of iron inclusions on the polarization pattern observed in optically thick wavelengths of $250\mum$ and $89\mum$ (Table \ref{tab:model_optical_depth}) in Section \ref{sec:iron_extinction_map} and the relationship of $p-I/I_{\rm max}$ in Section \ref{sec:iron_extinction_p_I}.

\subsection{Polarization maps}\label{sec:iron_extinction_map}

\begin{figure*}    
\centering
       \includegraphics[width=\textwidth,height=\textheight,keepaspectratio]{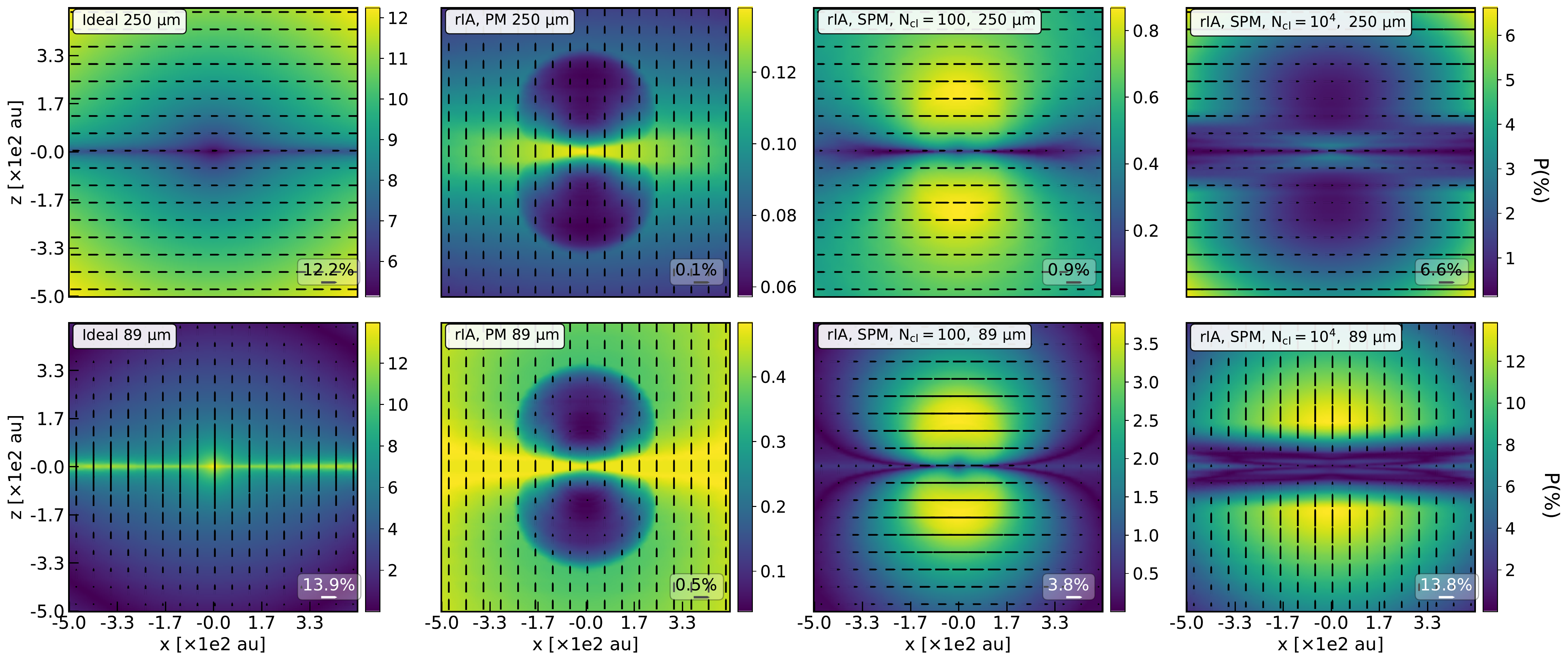}
     \caption{Synthetic polarization maps calculated in the central region of 1000 au at $250\mum$ (upper panels) and $89\mum$ (lower panels) for model Ideal and model Realistic$-$rIA for PM and SPM grains. In the Ideal model and our model for SPM grains with $N_{\rm cl} =10^{4}$, the polarization pattern can rotates $90^{\circ}$ from \PperpB at $250\mum$ to \PparaB at $89\mum$ due to the change of the polarization mechanism from dichroic emission to dichroic extinction. This effect does not happen for SPM grains with low $N_{\rm cl} = 100$ because of the reduced polarization by dichroic extinction caused by the weak alignment of VLGs with \B.}
     \label{fig:Realistic_rIA_center_thick}
 \end{figure*}  
    
\begin{figure*}
        \includegraphics[width=\textwidth,height=\textheight,keepaspectratio]{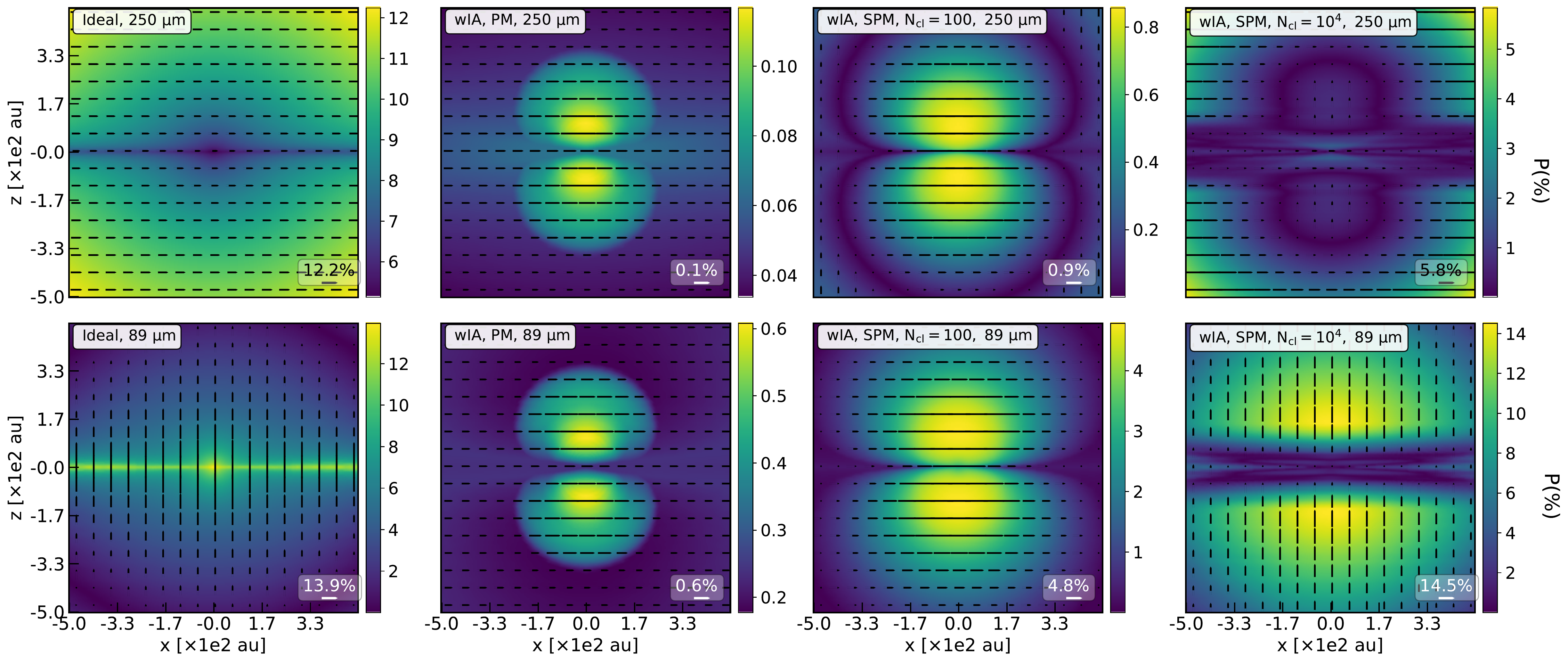}
     \caption{Similar results as Figure \ref{fig:Realistic_rIA_center_thick} but for model Realistic$-$wIA. The $90^{\circ}$ flipping of \P due to the change of the polarization mechanism only happens for SPM grains with high $N_{\rm cl} = 10^{4}$. For PM and SPM grains with $N_{\rm cl} = 100$, dichroic emission is still the main source of polarization at optically thick wavelengths due to the reduced alignment degree of VLGs with \B.}
     \label{fig:Realistic_wIA_center_thick}
 \end{figure*}
 
  \begin{figure*}
 \centering
    \includegraphics[width=\textwidth,height=\textheight,keepaspectratio]{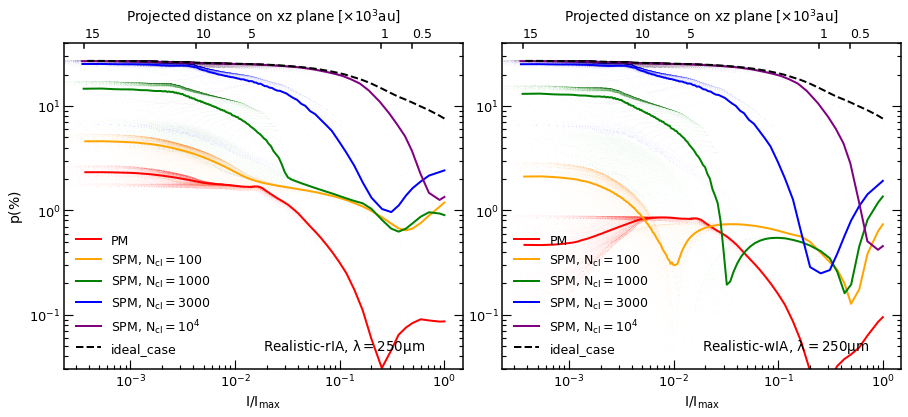}
    \includegraphics[width=\textwidth,height=\textheight,keepaspectratio]{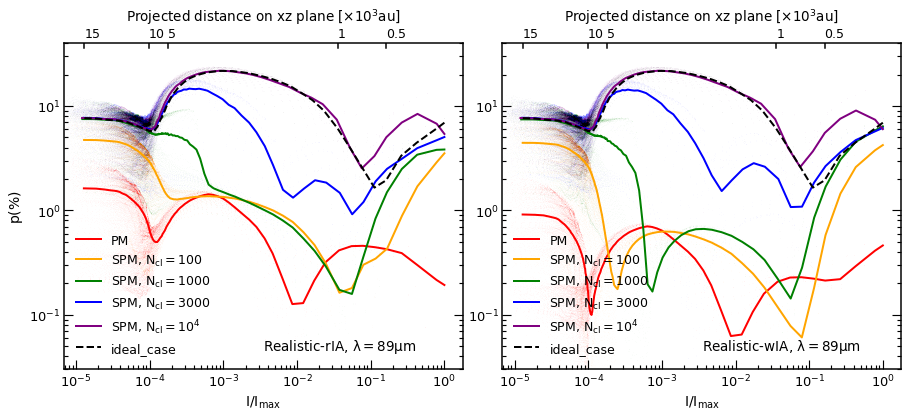}
     \caption{Effect of iron inclusions on the variation of $p$ with $I/I_{\rm max}$ at $250\mum$ (upper panels) and $89\mum$ (lower panels) for model Realistic$-$rIA (left column) and Realistic$-$wIA (right column). At $250\mum$, the polarization degree produced by model Ideal (black dashed line) slightly decreases with increasing intensity due to the extinction effect by aligned VLGs. Our models also show the reduction of $p$ with $I/I_{\rm max}$, but with steeper slope due to the additional effect from the weak grain alignment by the MRAT mechanism in dense environments. The variation of $p$ with $I/I_{\rm max}$ is more complicated at $89\mum$ due to the joint polarization by dichroic emission from micron-sized grains and dichroic extinction by VLGs. But in general, grains with higher $N_{\rm cl}$ produce high $p(\%)$ due to enhanced MRAT efficiency on grains.} 
     \label{fig:iron_P_I_thick}
\end{figure*}

Figure \ref{fig:Realistic_rIA_center_thick} shows the polarization maps obtained in the central region of 1000 au at $250\mum$ (upper panels) and $89\mum$ (lower panels) for model Ideal in the first column and model Realistic$-$rIA for PM and SPM in other columns. In the Ideal model, the polarization pattern changes from \PperpB at $250\mum$ to \PparaB at $89\mum$ due to the change in the polarization mechanism from dichroic emission to dichroic extinction by aligned VLGs. For our model of PM grains (second column), the polarization pattern is uniform with \PparaB at both $250\mum$ and $89\mum$ as a result of the polarization by dichroic extinction by aligned VLGs. The reason for the efficient polarization by dichroic extinction is as follows. For PM, most of aligned grains in the protostellar envelope have the inefficient IA by slow internal relaxation. Thus, VLGs can strongly attenuate polarized thermal emission from micron-sized aligned dust grains, letting dichroic extinction be the main source of polarization at optically thick wavelengths of $\lambda \leq 250\mum$.

In contrast, the polarization pattern produced by SPM grains with $N_{\rm cl} = 100$ (third column) is uniform in the central region with \PperpB at both $250\mum$ and $89\mum$. The independence of \P with wavelengths (\PperpB from $870\mum$ to $89\mum$, Figure \ref{fig:Realistic_wIA_center}) is explained as follows.For SPM grains with 
$N_{\rm cl} = 100$, 
sub-micron and micron-sized aligned grains near the protostar and in the outer envelope can have efficient IA by enhanced Barnett relaxation by iron inclusions. However, VLGs have inefficient IA due to slow internal relaxation due to the their larger sizes (Figure \ref{fig:align_amaxaJ}). Consequently, VLGs cannot efficiently attenuate polarized emission from smaller grains. As a result, dichroic emission is still the main polarization mechanism at optically thick wavelengths of $\lambda \leq 250\mum$. The situation will be different for SPM grains with $N_{\rm cl} = 10^{4}$ (fourth column). The polarization pattern can change from \PperpB at $250\mum$ to \PparaB at $89\mum$ due to the change of the polarization mechanism as model Ideal. The activation of the polarization by dichroic extinction here is due to the enhanced internal and external alignment of VLGs by efficient MRAT alignment.

Figure \ref{fig:Realistic_wIA_center_thick} shows the similar results as Figure \ref{fig:Realistic_rIA_center_thick} but for model Realistic$-$wIA. For PM and SPM grains with $N_{\rm cl} = 100$, one gets the uniform polarization pattern with \PperpB as a result of the dichroic emission of aligned dust grains at both $250\mum$ and $89\mum$ due to the inefficient alignment of VLGs. In contrast, SPM grains with high $N_{\rm cl} = 10^{4}$ produce the change of polarization pattern from \PperpB at $250\mum$ to \PparaB at $89\mum$ due to the change of the polarization mechanism.

Indeed, the polarization mechanism behind the polarization pattern produced by PM grains is different between model Realistic$-$rIA (due to dichroic extinction by aligned VLGs) and model Realistic$-$wIA (due to dichroic emission of micron-sized aligned grains). Such a different polarization mechanism is attributed to the stronger reduced extinction efficiency of VLGs with wrong IA.

\subsection{Intensity-dependent Polarization Degree}\label{sec:iron_extinction_p_I}

The upper panels of Figure \ref{fig:iron_P_I_thick} show the effect of iron inclusions on the mean variation of $p(\%)$ with $I/I_{\rm max}$ calculated at $250\mum$ for model Realistic$-$rIA (left panel) and Realistic$-$wIA (right panel). The result from model Ideal is plotted by the black dashed line for comparison. In the model Ideal, the polarization degree decreases from $p \sim 30\%$ from $d_{\rm proj} \geq 1000$ au to $p \sim 8\%$ in the center as a result of the dichroic extinction by aligned VLGs. For model Realistic$-$rIA (left panel) and Realistic$-$wIA (right panel), $p$ also decreases with increasing $I/I_{\rm max}$, but with a stepper slope than model Ideal due to the additional effect from the inefficient internal and external alignment of grains with \B (for SPM grains) or the loss of grain alignment (for PM grains)  in the central region. For example, the polarization degree produced by SPM grains with $N_{\rm cl} = 10^{4}$ reduces from $p \sim 30\%$ from $d_{\rm proj} \geq 1000$ au to $p \sim 1\%$ in the center. SPM with lower $N_{\rm cl}$ and PM grains produce lower $p$, typically with $p \sim 3 - 20\%$ in the envelope and $p \ll 1\%$ (for PM) and $p \sim 1\%$ (for SPM). And similar to Figure \ref{fig:iron_P_I}, the polarization fraction produced by model Realistic$-$wIA is smaller than model Realistic$-$rIA. One also will obtain the valley-polarization hole with $p < 1\%$ in the transition distance between the envelope with high $p(\%)$ (and \PperpB) and the central region with low $p$ (and \PparaB) resulted from the cancelling effect of polarized emission from aligned dust grains with right and wrong IA.

One interesting feature obtained at $250\mum$ is the slight rise of $p(250\mum)$ with increasing $I/I_{\rm max}$ at $d_{\rm prof} < 1000$ au for SPM grains with low $N_{\rm cl} < 10^{3}$. This feature arises from the strong emission of grains with efficient IA near the protostar (Figure \ref{fig:align_amaxaJ}), which  corresponds to the polarization pattern with \PperpB in the inner 1000 au region in Figures  \ref{fig:Realistic_rIA_center_thick} and \ref{fig:Realistic_wIA_center_thick}.

The lower panels of Figure \ref{fig:iron_P_I_thick} show the similar results as the upper panels but at $89\mum$. The $p-I/I_{\rm max}$ relationship becomes more complicated due to the joint polarization mechanism between dichroic emission and dichroic extinction by aligned dust grains. In detail, for model Ideal (see the black dashed line), the polarization degree is constant at $p \sim 8\%$ at $d_{\rm proj} > 10000$ au , then rises to the peak of $p \sim 20\%$ at $d_{\rm proj} \sim 5000$ au, decreases to $p \sim 10\%$ at $d_{\rm proj} \sim 1000$ au, and rises again to $p \sim 8\%$ in the center. The first rise of $p$ in the envelope is due to the high emission of warm grains at short wavelengths (in the envelope, polarization by dichroic emission is still the main source of polarization due to low gas density).  The following decrease of $p$ with increasing intensity arises from the extinction by aligned VLGs. And the rise of $p$ from  $d_{\rm proj} \leq 500$ au toward the center is due to the increased efficiency of polarization by dichroic extinction in dense regions around the protostar.

For the realistic model of grain alignment (solid lines), the polarization degree generally follows the similar complex dependence with intensity as model Ideal, but with lower $p \sim 1-10\%$ due to the inefficient grain alignment in dense environments. However, except the case of SPM grains with high $N_{\rm cl} = 10^{4}$ that the variation of $p$ with $I/I_{\rm max}$ is controlled by the joint effect of polarization by dichroic emission and dichroic extinction, the decrease and increase of $p$ for SPM with low $N_{\rm cl}\leq 10^{3}$ are  attributed to the change in IA efficiency of the emission source, from efficient IA to inefficient IA by slow internal relaxation in the envelope and the opposite trend for grains in the central region. 
 
\section{Effect of Iron Inclusions on Polarization Spectrum} \label{sec:iron_p_lambda}
\begin{figure*}
        \includegraphics[width=\textwidth,height=\textheight,keepaspectratio]{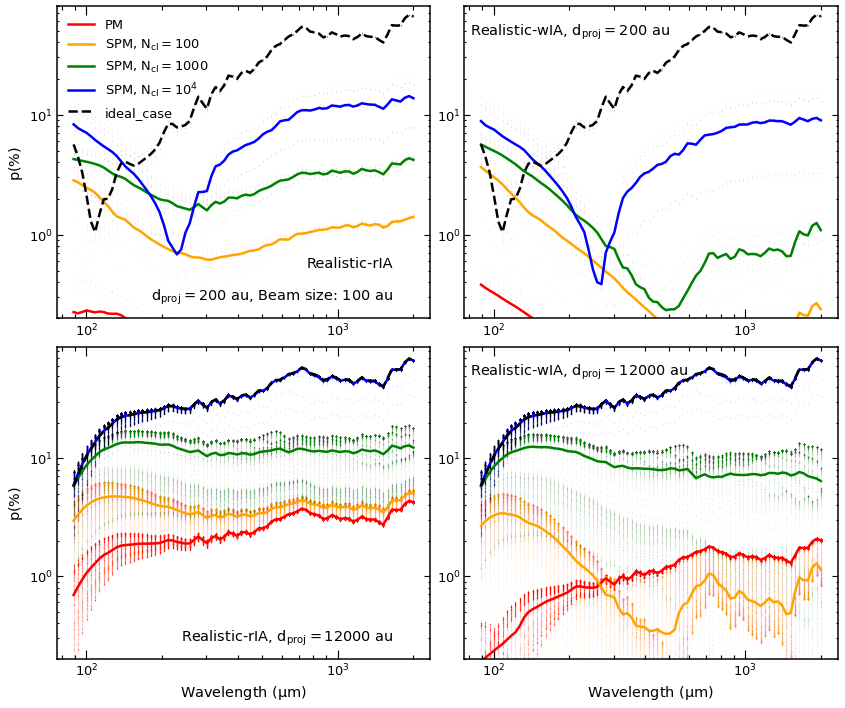}
     \caption{Effect of iron inclusions on the variation of $p$ from $89\mum$ to 2 mm at the projected distance to the protostar of $d_{\rm prof} = 200$ au (upper panels) and $d_{\rm proj} = 12000$ au (lower panels) for model Realistic$-$rIA (left column) and Realistic$-$wIA (right column). At $d_{\rm proj} = 200$ au, $p$ produced by model Ideal (black dashed line) and our models for SPM grains with $N_{\rm cl} = 10^{4}$ decreases continuously from 2 mm to $\lambda \sim 100 - 200\mum$ then rises toward $\lambda = 89\mum$ due to the increased polarization by dichroic extinction. SPM grains with low $N_{\rm cl}  = 100$ also produce the decrease and increase of $p(\%)$ with decreasing wavelengths, but it is due to the enhanced emission of micron-sized grains with efficient IA at shorter wavelengths. At $d_{\rm proj} = 12000$ au, $p(\%)$ slightly decreases with decreasing $\lambda$ due to the weak emission of cold dust grains at shorter wavelengths.}
     \label{fig:iron_P_lambda}
\end{figure*}

Lastly, we study the effect of iron inclusions on the wavelength-dependent polarization degree $p(\lambda)$ (i.e., polarization spectrum), from $89\mum$ to 2 mm. The results calculated in the central region at $d_{\rm proj} = 200$ au are shown in the upper panels of Figure \ref{fig:iron_P_lambda}, and the results in the envelope at $d_{\rm proj} = 12000$ au are shown in the lower panels
The beam size is $120$ au in both cases. 
The left column shows the results for model Realistic$-$rIA and the right column is for model Realistic$-$wIA. The polarization curve for the ideal case is plotted by the black dashed line for comparison.
 
In the ideal case, the polarization degree calculated at $d_{\rm proj} = 200$ au decreases continuously from $p > 30\%$ at 2mm to $p \sim 1\%$ at $\lambda \sim 100\mum$ due to the increased dichroic extinction by aligned VLGs at short wavelengths. At $\lambda < 100\mum$, dichroic extinction by aligned VLGs becomes the main source of polarization, inducing the increase of $p(\%)$ with decreasing wavelengths. 

For our realistic model of grain alignment, the polarization degree calculated at $d_{\rm proj} = 200$ au also decreases then increases with decreasing wavelengths from 2mm to $89\mum$. The polarization degree produced by SPM grains with larger $N_{\rm cl}$ is higher, but it is always smaller than model Ideal with $p \leq 10\%$. However, for SPM with high $N_{\rm cl}= 10^{4}$, the curvature of the polarization curve is controlled by the joint action of polarization by dichroic emission (at $\lambda > 200$) and dichroic extinction (at $\lambda < 200\mum$) as model Ideal. For SPM with lower $N_{\rm cl}\leq 10^{3}$, the decrease and increase of $p$ with decreasing $\lambda$ is due to the change in IA of the polarized emission source, from grains with inefficient IA at $\lambda > 600\mum$ to grains with efficient IA at $\lambda < 600\mum$. If grains with slow internal relaxation have wrong IA (upper right panel), the V-shape of the polarization curve is clearer corresponding to the polarization flipping in the central region from \PparaB at $870\mum$ (Figure \ref{fig:Realistic_wIA_center}, upper center panel) and 2 mm (Appendix \ref{sec:mm_a100}, Figure \ref{fig:Realistic_wIA_mm}, lower center panel) to \PperpB at $\leq 450\mum$ (Figure \ref{fig:Realistic_wIA_center}, lower center panel and Figure \ref{fig:Realistic_wIA_center_thick}). The valley-polarization hole on $p(\lambda)$ caused by the change in IA of the emission source appears at optically thin wavelengths $\lambda \sim 400-600\mum$),  while the valley-polarization hole caused by the change of the polarization mechanism  appears at optically thick wavelengths of $\lambda < 300\mum$.

Toward the envelope (lower panels), the polarization degree produced by both model Ideal, Realistic$-$rIA, and Realistic$-$wIA just slightly decreases with decreasing wavelengths as a result of the weak emission of cold dust grains at short wavelengths. The polarization degree obtained in the envelope is generally higher than one in the central region (upper panels) because of increasing grain alignment efficiency by the MRAT mechanism with decreasing gas density. SPM grains with higher $N_{\rm cl}$ produces higher $p(\%)$, typically with $p \sim 1-10\%$ in the envelope for $N_{\rm cl} \leq 10^{3}$ and $p \sim 10-45\%$ for $N_{\rm cl} = 10^{4}$. 

\section{Discussion} \label{sec:discussion} 
\subsection{Effects of Iron Inclusions on Grain Alignment in Protostellar Cores}\label{sec:align_state}

The alignment of dust grains with \B (i.e., magnetic alignment) induces the polarization of thermal dust emission and background starlight. Dust polarization is a powerful tool for studying magnetic fields in astrophysics and understanding the role of \B-fields in star and planet formation. However, grain alignment (both internal and external) depends on the local conditions and grain properties. In the diffuse ISM and MCs, PM grains can have efficient internal alignment (IA) with the grain longest axis perpendicular to its angular momentum due to fast internal relaxation \citep{Lazarian_2007,Hoang.2022}. Moreover, the grain angular momentum can be efficiently aligned with \B by RATs due to the fast Larmor precession compared to the grain randomization by gas collisions \citep{Hoang_Lazarian_2014,Hoang.2022}. Thus, thermal dust polarization is always perpendicular to \B (see \citealt{Tram_Hoang_2022} for a review). The existence of iron clusters (SPM) embedded in dust grains can enhance the degree of RAT alignment due to the joint effect of enhanced magnetic relaxation and RATs (a mechanism called MRAT), resulting in perfect grain alignment (\citealt{Hoang_Lazarian_2016a}). Consequently, dust polarization is a robust tool for tracing \B-fields in the diffuse ISM and MCs.

In dense environments as starless cores, the RAT alignment efficiency is significantly reduced due to the strong attenuation of ISRF by dust extinction and stronger gas randomization (\citealt{Hoang_et_al_2021}). In protostellar cores with an embedded protostar, dust grains near the protostar are predicted to be more efficiently aligned with \B due to the contribution of protostellar radiation (\citealt{Hoang_et_al_2021}). However, the gas randomization is also enhanced due to the high gas density, which can significantly suppress the alignment of grains with magnetic fields, as shown in \cite{Hoang.2022}. Thus, the pressing question is whether dust polarization can reliably trace \B-fields in protostellar environments (\citealt{Hoang_Lazarian_2016a,Yang_2021}). 

Using the RAT paradigm in which grains can be aligned at low-J and high-J attractors, numerical calculations by \cite{Hoang_et_al_2022} show the crucial importance of iron inclusions on both internal and external alignment of grains with \B in protostellar environments. Specifically, the authors show that PM grains cannot be aligned with \B, and only SPM grains can have the magnetic alignment due to the enhanced Larmor precession. Iron inclusions also help grains that rotate suprathermally have fast internal relaxation (then efficient IA) due to enhanced Barnett relaxation. However, they do not help grains at low-\textit{J} attractors due to their slow rotation. 
 
Our detailed modeling of grain alignment using the RAT paradigm with radiative transfer with POLARIS confirms the numerical calculations of grain alignment in protostellar cores from previous studies \cite{Hoang.2022,Hoang_et_al_2022}. Particularly, in Section \ref{sec:result_grain_alignment}, we show that PM grains only can be aligned with \B in the protostellar envelope due to the reduced gas randomization, but most of them have slow internal relaxation at both low and high-\textit{J} attractors. In contrast, SPM grains of micron-sizes can have the magnetic alignment and a fraction of them can have efficient IA in the dense central region with $n_{\rm H} > 10^{7}\cm^{-3}$ due to the enhanced Barnett relaxation and Larmor precession by iron inclusions (Figures \ref{fig:align_amaxJB} and the left panel of \ref{fig:align_amaxaJ}). However, we found that even with a high level of iron inclusions, micron-sized grains aligned with \B at low-\textit{J} attractors cannot have the fast internal relaxation due to its thermal rotation (the right panel of Figure \ref{fig:align_amaxaJ}). Consequently, they may align with their longest axis perpendicular or parallel with \B, which causes the ambiguity on the orientation between \P and \B.

The external alignment between SPM grains with \B can be enhanced by the joint effect of RATs and enhanced magnetic relaxation, as predicted by the MRAT mechanism \citep{Hoang_Lazarian_2016a}. We found that this effect is very efficient in increasing the external alignment of SPM grains with a high fraction of iron inclusions in the protostellar envelope. It allows all SPM grains to have perfect alignment ($f_{\rm high-J} = 1$) with \B, regardless of the difference on the grain sizes and their orientation in the envelope. However, in the central region, the efficiency of magnetic relaxation is negligible due to the reduction of the grain magnetic susceptibility with dust temperature in hot environments near the protostar (Figure \ref{fig:align_amaxJB_DG}) and the enhancement of gas randomization. As a result, both PM and SPM grains near the protostar are only aligned with \B by RATs, with the fraction of grains at high-\textit{J} attractors varying from $0.25 - 0.7$, depending on the grain shape, grain size, etc (\citealt{Herranen_2021}).
 
\subsection{Dependence of dust polarization pattern at optically thin wavelengths on Iron inclusions}\label{sec:iron_polarization_pattern}

As discussed in Section \ref{sec:align_state}, large (micron-sized) grains in protostellar cores tend to have inefficient IA due to slow internal relaxation, even when they rotate suprathermally at high-J attractors. Different from grains with fast internal relaxation that always have right IA ($\ahat_{1} \parallel$ \J), the slow internal relaxation produces the uncertainty in the alignment direction between $\ahat_{1}$ and \J. \cite{Hoang_Lazarian_2009} showed that grains with slow internal relaxation can still have right IA at high-\textit{J} attractors due to RATs, but grains at low-\textit{J} attractors may have right IA or wrong IA ($\ahat_{1} \perp$ \J). A detailed study for the conditions for the right or wrong IA is not yet available. 

In Section \ref{sec:iron_pol_map}, we show that if grains with slow internal relaxation have right IA (model Realistic$-$rIA), the polarization patterns are always uniform  with \PperpB in the entire protostellar core from mm to sub-mm wavelengths, regardless of the grain magnetic properties (Figures \ref{fig:Realistic_rIA_envelope}, \ref{fig:Realistic_rIA_center}, and \ref{fig:Realistic_rIA_mm}). In contrast, if grains with slow internal relaxation have wrong IA (model Realistic$-$wIA), the polarization patterns will depend on the magnetic properties of grains. Particularly, PM grains produce the uniform polarization pattern with \PparaB at both mm and sub-mm. SPM grains with a moderate level of iron inclusions produce the uniform pattern with \PparaB at 2mm (Figure \ref{fig:Realistic_wIA_mm}) and $870\mum$ (Figure \ref{fig:Realistic_wIA_envelope}), but the complex ones at $450\mum$ with the polarization pattern changing from \PperpB to \PparaB from the outer to the inner envelope (Figure \ref{fig:Realistic_wIA_envelope} and \P changing from \PparaB to \PperpB from the outer to the inner region of the central region (Figure \ref{fig:Realistic_wIA_center}). SPM grains with a high level of iron inclusions simply produce the uniform polarization pattern, with \PperpB due to the efficient alignment of large grains by MRAT alignment (Figures \ref{fig:Realistic_wIA_envelope}, \ref{fig:Realistic_wIA_center}, and \ref{fig:Realistic_wIA_mm}).

In summary, in contrast to the coherent polarization pattern with \PperpB at sub-mm/mm observed from the diffuse ISM and MCs, the polarization patterns observed toward protostellar cores are diverse due to the existence of large grains with slow internal relaxation. It strongly depends on the magnetic properties of grains and the nature of internal alignment between $\ahat_{1}$ and \J. This complicated feature causes the confusion on the orientation of \P produced by thermal emission from magnetically aligned dust grains, especially when the level of iron locked inside dust grains is unknown. However, we found that the abundance of embedded iron is revealed through the polarization degree, which will be discussed in the next section. It gives us a hint for constraining the orientation between \P and \B, and this issue will be discussed in detail in Section \ref{sec:rotate_P?}.

\subsection{Dependence of polarization degree on iron inclusions and maximum grain size}\label{sec:iron_polarization_degree}

 \begin{figure*}     
    \includegraphics[width=\textwidth,height=\textheight,keepaspectratio]{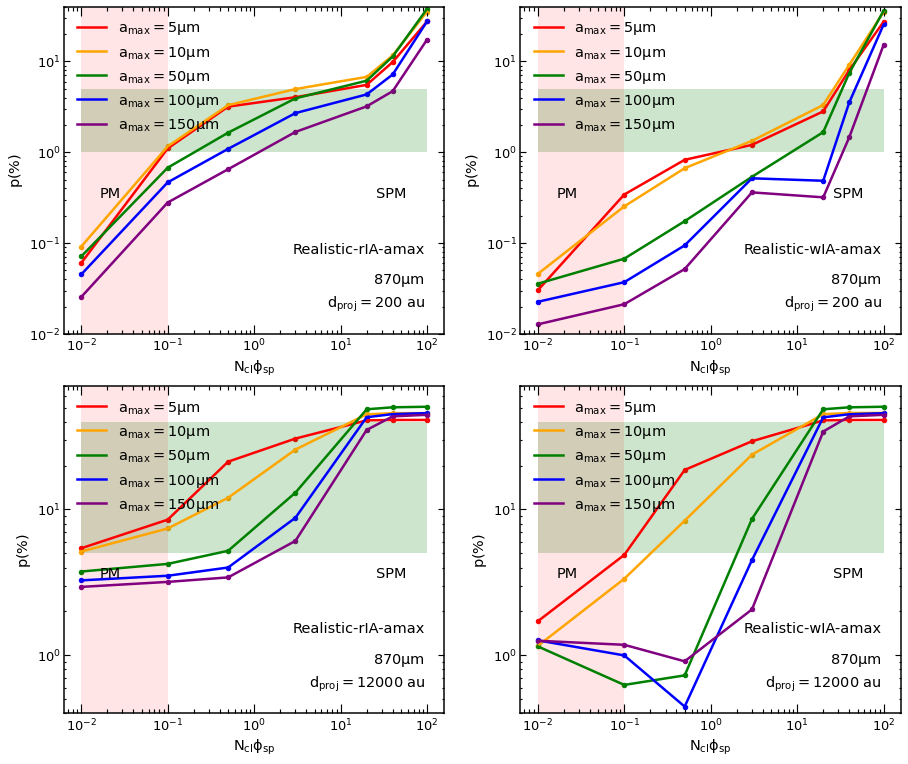}
     \caption{Dependence of $p$ calculated at $d_{\rm proj} = 200$ au (upper panels) and $d_{\rm proj} = 12000$ au (lower panels) as the function of iron fractions locked inside dust grains $N_{\rm cl} \phi_{\rm sp}$ from model Realistic$-$rIA$-$amax (left column) and model Realistic$-$wIA$-$amax (right column), with different grain size from $a_{\rm max} = 5\mum$ to $a_{\rm max} = 150\mum$. The red shaded area marks the value produced by PM grains, and the green shaded area marks the polarization degree observed toward protostellar cores by ALMA. The polarization degree increases with increasing the levels of iron inclusions locked inside grains, and one will obtain higher $p(\%)$ if grains with slow internal relaxation have right IA. The high amount of iron is required to reproduce observed polarization fraction of $p > 1\%$ in the central region and $p \sim 5-40\%$ in the envelope of Class 0/I YSOs. }
 \label{fig:P_Ncl}
\end{figure*}

\begin{figure*}     
    \includegraphics[width=\textwidth,height=\textheight,keepaspectratio]{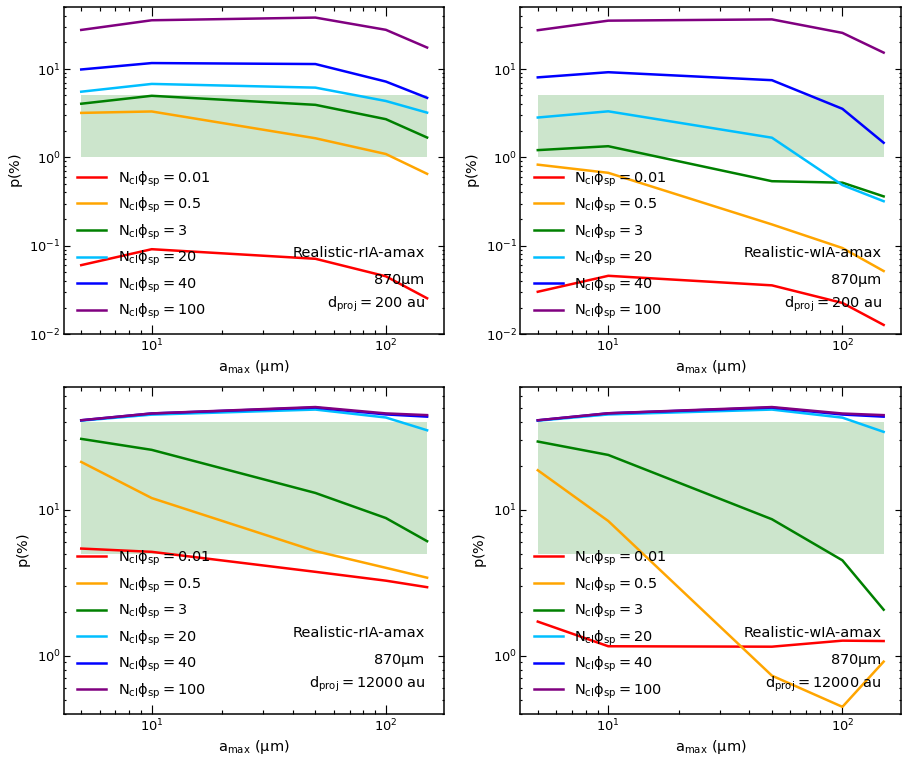}
 
     \caption{Similar as Figure \ref{fig:P_Ncl} but for the variation of $p(\%)$ with maximum grain size $a_{\rm max}$. The presence of large grains generally reduces the fraction of polarization. Therefore, they must contain a high level of iron inclusions in order to produce the detected level of $p \geq 1\%$ in the central region and $p \geq 10\%$ in the envelope. }
 \label{fig:P_amax}
\end{figure*}

In Section \ref{sec:iron_P_I}, we showed that the calculated polarization degree $p(\%)$ is higher for higher values of $N_{\rm cl}$, assuming a constant volume filling factor of iron clusters, $\phi_{\rm sp}$. We found that in particularly, PM grains only can produce low $p \sim 1\%$ in the envelope and negligible polarization signal in the inner 1000 au region where grains are not aligned with \B. In constrast, iron inclusions can lead VLGs to achieve perfect alignment in the envelope and produce the maximum polarization degree up to $p \sim  40\%$ here. However, they cannot strongly enhance RAT alignment in the central region due to the stronger gas randomization. As a result, the polarization degree decreases from the envelope toward the protostar. But in contrast to the negligible $p < 1\%$ in the central region produced by PM grains, SPM grains produce $p > 1\%$ because micron-sized grains can be aligned with \B here. The dependence of $p$ on iron inclusions are nearly similar for both model with and without grains with wrong IA.

In our modelling till now, we have fixed the volume filling factor of iron clusters $\phi_{\rm sp} = 0.005$ that corresponds to $1.67\%$ (\citealt{Hoang_Lazarian_2016a}) iron abundance present in the form of embedded iron clusters, and vary $N_{\rm cl}$ to describe different magnetic properties of grains. Indeed, \cite{Jenkins_2009} showed that $\sim 90\%$ of iron can be locked inside grains, that corresponds to $\phi_{\rm sp} \sim 0.3$ if all iron are in the form of clusters. That means with a higher $\phi_{\rm sp}$, one only needs a small number of iron atoms per each cluster to achieve the high magnetic susceptibility that can drive most grains to be perfectly aligned with B as  our studied case with $N_{\rm cl} = 10^{4}$ and $\phi_{\rm sp} = 0.005$ (Equation \ref{eq:chi_0_super_paramagnetic}).

To generalize the dependence of the polarization degree on the level of iron inclusions, which is described by a term $N_{\rm cl}\phi_{\rm sp}$, we adopt models Realistic$-$rIA$-$amax and Realistic$-$wIA$-$amax with different maximum grain sizes from $a_{\rm max} = 5\mum$ to $a_{\rm max} = 150\mum$ (Table \ref{tab:model_parameter}) and plot the results in Figure \ref{fig:P_Ncl}. The left column shows the results calculated by model Realistic$-$rIA$-$amax, and the right column is for model Realistic$-$wIA$-$amax. The upper panels show the results calculated in the central region at $d_{\rm proj} = 200$ au, and the lower panels are for the envelope at $d_{\rm proj} = 12000$ au. The minimum $N_{\rm cl}\phi_{\rm sp} = 0.01$ corresponds to the magnetic susceptibility of PM (Equation \ref{eq:chi_0_paramagnetic}), which is marked by the red box. The maximum value $N_{\rm cl}\phi_{\rm sp} = 100$ is for SPM grains with the magnetic susceptibility enhanced by a factor of $\sim 10^{4}$ compared to PM. One can clearly see that grains with higher level of iron inclusions radiate stronger polarized thermal emission with higher $p(\%)$. In the central region, iron inclusions help grains to be stably aligned with \B and produce the detected level of polarization of $p > 1\%$. In the envelope, high levels of iron inclusions lead grains to achieve perfect alignment with \B and produce very high polarization up to $p \sim 40\%$, as predicted by the MRAT mechanism (\citealt{Hoang_Lazarian_2016a}). With the same amount of iron inclusions, the model with grains having wrong IA produce lower $p$ than model without grains having wrong IA as a result of the reduced dust polarization due to the co-existence of grains with right and wrong IA.
 
Figure \ref{fig:P_amax} shows the variation of $p(\%)$ with $a_{\rm max}$ calculated by the model Realistic$-$rIA$-$amax (left column) and model Realistic$-$wIA$-$amax (right column) at $d_{\rm proj} = 200$ au (upper panels) and $d_{\rm proj} = 12000$ au (lower panels). For a fixed level of iron inclusions, the polarization degree decreases with increasing maximum grain size because large grains tend to have inefficient alignment with \B-fields in dense environments. Thus, higher levels of iron inclusions are required to present inside micron-sized grains to help them efficiently align with \B and produce the detected fraction of polarization of $p > 1\%$ in both the central region and the envelope. For example, with $a_{\rm max} = 10\mum$, grains need $N_{\rm cl}\phi_{\rm sp} \sim 3$ to produce $p \sim 1\%$ at $d_{\rm proj} = 200$ au, but if they grow to $a_{\rm max} = 100\mum$, the level of iron inclusions must increase to $N_{\rm cl}\phi_{\rm sp} \sim 30$ to produce $p \sim 1\%$ in the central region. The value of $N_{\rm cl}\phi_{\rm sp}$ must increase more to produce $p \sim 1\%$ if grains with slow internal relaxation have wrong IA.

\begin{figure*}
\centering
    \includegraphics[width=\textwidth,height=\textheight,keepaspectratio]{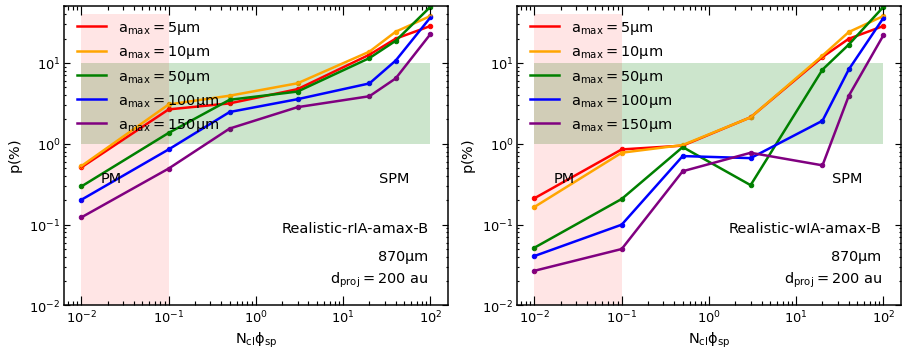}
     \caption{Dependence of $p$ with $N_{\rm cl}\phi_{\rm sp}$ for different values of $a_{\rm max}$ for 
     models with stronger \B-fields in the central region
     Realistic$-$rIA$-$amax$-$B (left column) and model Realistic$-$wIA$-$amax$-$B (right column) calculated in the central region at $d_{\rm proj} = 200$ au. Similar as the left panel of Figure \ref{fig:P_Ncl}, only SPM grains can reproduce the observed degree of polarization of $p > 1\%$, and higher iron fractions are required to align grains with \B if grains grow to VLGs. The polarization degree increases with increasing \B-field strength for low $N_{\rm cl}\phi_{\rm sp} \leq 20$ as a result of the enhanced Larmor precession. However, stronger \B-fields does not help to increase $p(\%)$ for grains with high $N_{\rm cl}\phi_{\rm sp}$ because the MRAT alignment is not the major grain alignment mechanism in the dense central region.} 
 \label{fig:P_Ncl_B}
\end{figure*}

Besides the fraction of iron locked inside grains, magnetic field strength also affects on the alignment of grains with \B. The stronger magnetic field strength enhances the Larmor precession and the magnetic relaxation (Section \ref{sec:mag_properties}), allowing more large grains to be efficiently aligned with \B in the densest central region. In our modeling, we have adopted the uniform $B = 134\mum$ in the entire protostellar core to emphasize the effects of grain alignment physics on synthetic dust polarization. However, the magnetic field strength is amplified with increasing gas density inward of the protostar as a result of the freezing of B-fields. Observations toward Class 0/I YSO L1448 IRS 2 by \cite{Kwon_2019} reveal the highly order hourglass-shape magnetic field in the protostellar system, and they estimated the magnetic field strength in the inner 100 au region of $B \sim 600\mu\rm G$. \cite{Myers_2021} also found the positive correlation between the magnetic field strength measured by the David Chandrasekhar Fermi (DCF) method (\citealt{Chandra_1953}) and the gas density of $B_{\rm pos} \sim n_{\rm H}^{0.66\pm 0.05}$ in the sample of 17 low-mass prestellar and protostellar cores. To understand how the magnetic field strength affects the magnetic alignment in the densest central region, we increase the magnetic field strength in the central 1000 au region into $B = 1$ mG while keeping the same $B = 134\mu G$ at $r > 1000$ au, denoted as model Realistic$-$rIA$-$amax$-$B and Realistic$-$wIA$-$amax$-$B (Table \ref{tab:model_table}). The variation of $p(\%)$ with $N_{\rm cl}\phi_{\rm sp}$ in the central region at $d_{\rm proj} = 200$ au is shown in Figure \ref{fig:P_Ncl_B}.

The left panel of Figure \ref{fig:P_Ncl_B} shows the results for model Realistic$-$rIA$-$amax$-$B and the right panel is for model Realistic$-$wIA$-$amax$-$B. Comparing to the upper panels of Figure \ref{fig:P_Ncl}, one gets higher $p(\%)$ for all values of $N_{\rm cl}\phi_{\rm sp}$. For low $N_{\rm cl}\phi_{\rm sp} < 20$, the difference is clearer as a result of the extend of the range of grain alignment to larger sizes, i.e., higher $a_{\rm max,JB}^{\rm Lar}$, due to the enhanced Larmor precession by stronger magnetic field strength. For high $N_{\rm cl}\phi_{\rm sp} > 20$ in which most large micron-sized grains are aligned with \B, the increase of $p(\%)$ is not clear because the stronger \B-field strength cannot enhance the magnetic relaxation over the strong gas randomization here. It means that grains within few hundreds of au around the protostar are mainly aligned with \B by RATs (instead of MRAT alignment as grains with a high level of iron inclusions in the envelope).
 
\subsection{Implications for the high polarization degree observed by ALMA toward protostellar envelopes}\label{sec:highP}

In dense environments such as protostellar cores, one usually expects to observe a much lower dust polarization degree than in the diffuse ISM due to the inefficient alignment of grains with magnetic fields, according to the RAT theory (\citealt{Lazarian_Hoang_2007}). However, SMA and ALMA observations toward Class 0/I YSOs reveal a high degree of dust polarization up to $p \sim 10-40\%$ in the envelope \citep{Hull_2014,Cox_2018,Galametz_2019}, much higher than the maximum values of $p \sim 20\%$ observed in the diffuse ISM (\citealt{Planck_2015}). The polarization degree observed in the central region is low but still detectable of $p > 1\%$. Although the RAT alignment theory expects the rise of $p$ toward the protostar due to increasing radiation field (\citealt{Hoang_et_al_2021}), the reduction of polarization degree toward the central region is still possible if the randomization effect by gas-grain collisions is more effective than the grain alignment by RATs. However, the numerical modeling with POLARIS by \cite{Gouellec_2020} showed that the RAT alignment mechanism implemented in the current version of POLARIS produces a much lower polarization degree than observations toward both low and high-mass protostars by ALMA. They found that ALMA data only can be successfully reproduced by the model in which all grains are perfectly aligned with \B. 

Theoretical studies in \cite{Hoang.2022} and \cite{Hoang_et_al_2022} and our detailed modelling of grain alignment by the MRAT mechanism in Section \ref{sec:result_grain_alignment} clearly show that large micron-sized grains can achieve the efficient magnetic alignment in protostellar environments if they contain a high level of iron inclusions. This effect is more prominent in the envelope, where VLGs can have perfect alignment and produce the high polarization degree up to $p \sim 40\%$. Thus, the detection of high $p \sim 40\%$ toward Class 0/I YSOs by ALMA and the implication of perfect grain alignment by \cite{Gouellec_2020} can be successfully reproduced by the MRAT mechanism. Moreover, this finding indicates that dust grains in protostellar environments must contain iron inclusions. 
  
\subsection{Toward constraining iron abundance embedded in dust using dust polarization}\label{sec:iron_level}
Iron is among the most abundant element in the Universe. Observations reveal that more than $90\%$ of iron abundance are missing from the gas \citep{Jenkins_2009}, which implies that iron must be dominantly present inside dust grains \citep{Dwek_2016}. Yet, observational study of iron in dust is rather difficult. \cite{Hoang_Lazarian_2016a} first showed that one can constrain the level of iron in dust using sub(mm) polarization data. \cite{Lazarian_Hoang_2019} obtained the lower and upper limit for magnetic susceptibility using the ALMA polarization patterns toward protostellar disks.  

A synthetic modeling of dust polarization by \cite{Lam_2021}, which assumes the perfect alignment for all grains with Larmor precession faster than the gas randomization, found that the magnetic susceptibility of grains of $a \geq 1\mum$ must be increased by a factor of 20 compared to PM to reproduce $p \geq 1\%$ in the inner 100 au of a dozen of Class 0/I YSOs observed by ALMA \citep{Cox_2018}. Moreover, a synthetic modelling with POLARIS by \cite{Valdivia_2019}, which does not consider the grain magnetic properites, concluded that grains must grow to $a \sim 10-50\mum$ to reproduce $p \geq 1\%$ observed in the samples of Class 0 YSOs by \cite{Hull_2014}. Recently, numerical studies of grain alignment in protostellar cores and protoplanetery disks by \cite{Hoang.2022} and \cite{Hoang_et_al_2022} showed that large micron-sized grains must contain the high level of iron inclusions to be efficiently aligned with \B in such dense environments and produce polarized thermal emission.

In Section \ref{sec:iron_polarization_degree}, we show the dependence of $p$ with the level of iron inclusions $N_{\rm cl}\phi_{\rm sp}$ in Figure \ref{fig:P_Ncl}. The green shaded area in the left and right panel of the figure shows the polarization degree observed by ALMA, JVLA, SMA, etc, toward Class 0/I YSOs (\citealt{Hull_2014}, \citealt{Ko_2020}, \citealt{Gouellec_2020}, \citealt{Cox_2018}). It is very clear that grains must be SPM with the magnetic susceptibility enhanced by a factor of $\geq 5$ compared with PM grains to reproduce $p > 1\%$ in the central region and $p \sim 5 - 40\%$ in the envelope. Using our modeling results of polarization degree and observed polarization, we can infer again the level of iron inclusions locked inside dust grains. For example, to reproduce $p \sim 1\%$ at $d_{\rm proj} = 200$ au with $a_{\rm max} = 10\mum$, one needs $N_{\rm cl}\phi_{\rm sp} \geq 1$. It corresponds to $N_{\rm cl} = 200$ if $\phi_{\rm sp} = 0.005$ (or $1.67\%$ of iron abundance locked inside grains under the form of cluster), or $N_{\rm cl} = 10$ if $\phi_{\rm sp} = 0.1$ (or $30\%$ of iron abundance locked inside grains). If grains grow to VLGs with $a_{\rm max} = 100\mum$, they need more embedded iron of $N_{\rm c}\phi_{\rm sp} \geq 20$ to be aligned with \B in the central region. It corresponds to $N_{\rm cl} = 4000$ and $N_{\rm cl} =20$ if $\phi_{\rm sp} = 0.005$ and $\phi_{\rm sp} = 0.1$, respectively.

Note that the main goal of our present study is to demonstrate the key effects of grain alignment physics on synthetic dust polarization in protostellar environments, thus we adopt the simple spherical symmetry for the protostellar core and uniform B-fields along the $z$ direction. We also perform the synthetic modeling assuming that the magnetic field only lies in the plane of the sky to understand how the grain alignment efficiency affects properties of polarized dust emission. In realistic situations, magnetic fields in protostellar cores usually have the hourglass shape driven by gravitational collapse toward protostars, which are theoretically predicted by \cite{Fiedler_Mouschovias_1993}, \cite{Allen_2003}, and widely confirmed by observations, e.g., \cite{Gigart_2006}, \cite{Rao_2009}, \cite{Hull_2014}. The complex \B-field structure along the line of sight combined with the tangling of \B-fields by turbulence results in a smaller polarization degree than obtained by our modeling with uniform \B-fields. In addition, the gas density in the central region of our model is $n_{\rm H} = 10^{7}\cm^{-3}$, which is much lower than the measured gas density in protostellar disks of Class 0/I YSOs of $n_{\rm H} \sim 10^{8} - 10^{9}\cm^{-3}$ (e.g.,  \citealt{Sandell_1991}, \citealt{Takahashi_2012}) and the gas density from MHD simulations of collapsing protostellar cores (\citealt{Lam_2019}). The magnetic field strength is also correlated with the gas density instead of being uniform as our assumption. Thus, the higher gas density will decrease the efficiency of grain alignment, but higher B-field strength would help to enhance MRAT alignment efficiency in the central region. Detailed synthetic observations of non-ideal MHD simulations of collapsing protostellar cores are required to accurately understand how grains around protostars get aligned with \B and how much iron inclusions are needed to reproduce the observed polarization level of $p > 1\%$ from ALMA, JVLA, SMA, etc, observations. Such a study also allows us to better interpret observational data and clarify the role of B-fields the star formation process. We will present the results from this study in our followup paper.

\subsection{Effect of iron inclusions on polarization by dichroic extinction }\label{sec:dichroic_extinction}

In protostellar environments, the polarization by dichroic extinction is expected to be activated at sub-mm (\citealt{Brauer_2016}) due to the presence of VLGs (\citealt{Kwon_2009}, \citealt{Testi_2014}, \citealt{Miotello_2014}). VLGs can strongly attenuate polarized thermal emission at sub-mm, and this feature is suggested to be the origin of the polarization hole (\citealt{Brauer_2016}, \citealt{Liu_et_al_2016}, \citealt{Ko_2020}, \citealt{Liu_2021}). Polarization by dichroic extinction is suggested to become the main source of polarization at optically thick wavelengths and cause the $90^{\circ}$ flipping of the polarization pattern between mm and sub-mm wavelengths observed in IRAS4A (\citealt{Ko_2020}) and OMC-3/MMS 6 (\citealt{Liu_2021}).

However, the study by \cite{Brauer_2016} and the discussion in \cite{Ko_2020} and \cite{Liu_2021} assumed that VLGs are efficiently aligned with \B. In Section \ref{sec:result_grain_alignment}, we show that this situation only can happen if grains are SPM with a high level of iron inclusions (fourth column of Figures \ref{fig:Realistic_rIA_center_thick} and \ref{fig:Realistic_wIA_center_thick}). If it is not the case, VLGs are not efficiently aligned with \B, then, they less attenuate polarized thermal emission and cannot become the main source of polarization at optically thick wavelengths. Thus, one will not obtain any change of the polarization pattern between optically thin and optically thick wavelengths (Figures \ref{fig:Realistic_rIA_center_thick} and \ref{fig:Realistic_wIA_center_thick}). 

Furthermore, even if VLGs are efficiently aligned \B in protostellar environments, the extinction by aligned VLGs is not efficient in reducing polarized thermal emission in the central region compared to the effect of the inefficient grain alignment on the polarization degree (Section \ref{sec:iron_extinction_p_I}). \cite{Brauer_2016} showed that the efficiency of polarization by dichroic extinction increases with increasing gas density. But the high gas density makes the magnetic alignment of VLGs more difficult, which mimics the effect of polarization by dichroic extinction in dense environments. Thus, simply attributing the extinction by aligned dust grains to any anomalous features observed in dense environments without examining in detail the level of iron inclusions may cause the wrong interpretation of dust polarization.
 
\subsection{Implications for the origin of polarization holes and the role of iron inclusions}\label{sec:polarization_hole}
Polarization observations toward Class 0 YSOs usually report the reduction of the polarization degree at sub-mm/mm wavelengths toward the central region of protostellar cores. For example, the existence of polarization hole is widely reported on 30 star-forming cores observed at 1.3mm by CARMA (\citealt{Hull_2014}), 10 Class 0 YSOs observed at $870\mum$ by ALMA (\citealt{Cox_2018}), 6 objects in Bok globules observed at $870\mum$ by SCUBA (\citealt{Henning_2001}, \citealt{Wolf_2003}). Multi-wavelength observations toward NGC 1333 IRAS4A, OMC-3/MMS 6, and NGC 2071 also show the similar feature from sub-mm to mm wavelengths (\citealt{Ko_2020}, \citealt{Liu_2021}, \citealt{Lapo_2022}).
 
Many scenarios are proposed to explain the polarization hole toward protostars, including the inefficient grain alignment by RATs (\citealt{Hoang_Lazarian_2016a,Hoang_et_al_2021}), the dichroic extinction by aligned VLGs  (\citealt{Brauer_2016}), or the inclination angle and the geometrical effect of magnetic fields due to the self-collapse protostellar cores (\citealt{Tazaki_2017}).  

Polarization by dichroic extinction suggested by \cite{Brauer_2016} is the promising explanation of the polarization hole. But as we discuss in Section \ref{sec:dichroic_extinction}, dichroic extinction by aligned grains only plays a minor role on reducing polarized thermal emission around protostars. We show that the reduced internal alignment and external alignment by the MRAT mechanism in the dense central region is the more effective mechanism producing the polarization hole in protostellar cores (Section \ref{sec:iron_extinction_p_I}). Furthermore, this mechanism does not require the presence of aligned VLGs in very dense environments. Thus, we suggest that this scenario as a main mechanism for explaining the origin of polarization holes in different class 0/I YSOs.
 
We note that the inefficient grain alignment in dense environments depends on the magnetic properties of grains. In particular, the polarization hole will be due to the loss of grain alignment in dense central region if grains are PM. In contrast, for SPM grains, it is because of the increase of grains with inefficient IA by slow internal relaxation and the reduced efficiency of MRAT alignment with increasing gas randomization. Since the magnetic properties of grains can be inferred from the observed polarization degree (Section \ref{sec:iron_level}), one then can know in detail the behind mechanism which is responsible for the reduction of $p$ toward the central region of protostellar cores.
 
\subsection{On the $90^{\circ}$ rotation of polarization vectors}\label{sec:polarization_flipping}
Another interesting feature observed toward Class 0 YSOs is the change of the polarization pattern with wavelengths. Particularly, the polarization patterns in the inner 100 au region of Class 0 YSO NGC 1333 IRAS4A (\citealt{Ko_2020}) and OMC-3/MMS (\citealt{Liu_2021}) change $90^{\circ}$ from mm observed by JVLA to sub-mm observed by ALMA. Another case is the MC NGC2071 which shows the $90^{\circ}$ difference of \P in the central region between $870\mum$ observed by JCMT and $158\mum$ and $214\mum$ observed by SOFIA.

\cite{Lapo_2022} argued that the twist of the polarization pattern observed in NGC 2071 may be due to the presence of grains with wrong IA suggested by \cite{Hoang.2022}. In Figure \ref{fig:Realistic_wIA_envelope}, we show that if SPM grains have a moderate level of iron inclusions and have wrong IA by slow internal relaxation, the polarization pattern at the outer boundary of the envelope can change from \PparaB at $870\mum$ (see also the map at mm in Figure \ref{fig:Realistic_wIA_mm}) to \PperpB at $450\mum$ due to the change in IA of the emission source. In our calculations, this feature happens at optically thin wavelengths. The flipping of \P between $214\mum$ and $870\mum$ observed in the MC NGC 2071 also appears in optically thin central region. Thus, we suppose that this anomalous feature be due to the change in IA of the emission source, in which JCMT and SOFIA measure polarized emission from large grains with wrong IA with \PparaB and from smaller grains with right IA with \PperpB, respectively. Indeed, the gas density and the maximum grain size of the MC NGC 2071 are different from our model adopted for the protostellar core. However, we expect the similar results for dense regions inside MCs because the inefficient IA is naturally produced if the gas randomization is stronger than the internal relaxation.

For the case of IRAS4A and OMC-3/MMS 6, \cite{Ko_2020} and \cite{Liu_2021} found that the cores of two objects are optically thick at sub-mm measured by ALMA and optically thin at mm measured by JVLA. Therefore, they assign the polarization flipping in the inner 100 au region to the change of the polarization mechanism from dichroic emission to dichroic extinction. However, as we show in Section \ref{sec:iron_extinction_map} and discuss in Section \ref{sec:dichroic_extinction}, the change of the polarization pattern from optically thin to optically thick wavelengths in protostellar environments only can happen if VLGs contain a very high level of iron inclusions (Figures \ref{fig:Realistic_rIA_center_thick} and \ref{fig:Realistic_wIA_center_thick}). In our calculations, the inner 1000 au region becomes optically thick at $250\mum$, and VLGs must have $N_{\rm cl}\phi_{\rm sp} = 50$ ($\sim 5000$ larger than PM) to produce the polarization flipping between $250\mum$ and $89\mum$. For IRAS4A and OMC-3/MMS 6, \cite{Ko_2020} and \cite{Liu_2021} found that inner 100 au region becomes optically thick at $\leq 1$ mm, indicating that these objects are very dense. Thus, to activate the effect of dichroic extinction at $\leq 1$ mm, grains in IRAS4A and OMC-3/MMS 6 must have much higher iron inclusions than our studied case. However, the amount of iron locked inside grains still be constrained by the crystalline structure of grains (\citealt{Yang_2021}). Thus, one cannot increase amount of iron too much to make the scenario proposed by \cite{Ko_2020} and \cite{Liu_2021} come true. In case that polarization by dichroic extinction is not effective, we suppose that the change in IA of the emission source from inefficient IA at mm to efficient IA at sub-mm to be the reason of the $90^{\circ}$ flipping of \P with wavelengths. However, IRAS4A and OMC-3/MMS 6 have higher luminosity and higher gas density    (\citealt{Sandell_1991}, \citealt{Takahashi_2012}) than our studied parameters. The strong radiation field can allow more large grains near the source to be aligned with \B with efficient IA (Figure \ref{fig:align_amaxaJ}, left panel) and activate polarization by dicroic extinction at sub-mm. However, the strong radiation field also can disrupt large grains by RAdiative Torque Disruption (RAT-D) mechansim proposed by \cite{Hoang_et_al_2019} (\citealt{Hoang_et_al_2021}). If it is the case, the polarization by dichroic extinction cannot cause the change of \P with wavelengths. The detail study of grain alignment by the MRAT mechanism and grain disruption by RAT-D around high-luminosity protostar is required to accuracy figure out the physics behind the twist of the polarization pattern of IRAS4A and OMC-3/MMS 6.

One interesting feature is that if the change of the polarization pattern with wavelengths is caused by the change in IA of the emission, it implies that grains with slow internal relaxation in the studied objects have wrong IA. Until now, the detailed study about the dynamics of grains with slow internal relaxation is still missing. Thus, it is necessary to study basic properties and effects of environments on establishing the alignment direction of grains with slow internal relaxation. Knowing this information allows us to accurately model grain alignment in dense environments, which is the key for better analyzing dust polarization.

\subsection{From polarization vectors to the magnetic field direction}\label{sec:rotate_P?}
Here we discuss the conditions for which the polarization vectors can be rotated by $90^{\circ}$ to infer the magnetic field direction in dense environments. As discussed in Section \ref{sec:iron_polarization_pattern}, the orientation of \P with \B strongly depends on the level of iron locked inside grains and the internal alignment direction. Furthermore, multi-wavelength observations from mm to sub-mm should be done to accurately answer the question whether we should rotate \P by $90^{\circ}$ or not to infer \B.

In particular, if the polarization pattern does not change from mm to sub-mm, it can be classified to three cases: SPM grains with the high level of iron inclusions, SPM grains with a moderate level of iron inclusions which have right IA with slow internal relaxation, and PM grains. For the first and second cases, grains always align with their longest axis perpendicular to \B and thus emit polarized thermal radiation with \PperpB. This case is accompanied by a high polarization degree of $p > 5\%$ in the envelope and $p > 1\%$ in the central region. With this case, we can confidently rotate \P by $90^{\circ}$ to reconstruct \B-fields morphology on the plane of sky. For the last case, \P can be perpendicular or parallel to \B, but the expected polarization degree is rather low of $p \sim 1\%$ in the envelope and $p \ll 1\%$ in the central region. If it is the case, we should keep \P because we don't exactly know whether the grain longest axis is perpendicular to parallel to \B, or dust polarization does not trace the magnetic field direction. Until now, all observations show the high polarization level of $p> 5\%$ in protostellar envelopes, which requires dust grains to be SPM rather than PM. Therefore, one can infer B-fields in the envelope by rotating the polarization orientation by $90^{\circ}$.

If the polarization pattern changes with wavelengths, it can originate from the change in IA of the emission source (Figures \ref{fig:Realistic_wIA_envelope} and \ref{fig:Realistic_rIA_center}), or due to the change of the polarization mechanism. The latter case happens at optically thick wavelengths, but it requires the additional information of iron fractions locked inside dust grains before giving the conclusion as discussed in Section \ref{sec:polarization_flipping}. Given this case, the polarization signal at optically thin wavelengths is due to polarization by dichroic emission while signal at optically thick wavelengths is due to polarization by dichroic extinction. Thus, we can rotate \P at optically thin wavelengths by $90^{\circ}$ to infer \B, and keep \P at optically thick wavelengths. If the origin is the former scenario, large grains tend to have wrong IA due to slow internal relaxation and produce \PparaB, while small grains can have efficient IA and produce \PperpB. For this case, we should keep \P at long wavelengths and rotate \P at shorter wavelengths by $90^{\circ}$ to reconstruct magnetic fields structure.

In conclusion, we suggest that multi-wavelength polarimetric observations toward protostellar cores are required to accurately probe the dust, gas, and magnetic properties around protostars. A combined analysis of multi-wavelength polarization patterns and polarization degrees from synthetic observations with real observations will help us to first constrain amount of iron locked inside grains, then the grain alignment efficiency in studied environments, and lastly magnetic field geometry.

\section{Conclusions}\label{sec:summary}
In our study, we have incorporated the magnetic properties of dust grains and the advanced grain alignment within the RAT paradigm into the POLARIS code. Then, we used the updated version of POLARIS to study the effects of iron inclusions on grain alignment and synthetic polarimetric observations of the protostellar core. Our findings are summarized as follows:
\begin{enumerate}

\item  Paramagnetic (PM) grains are not aligned with \B in the central region with high gas density of $n_{\rm H} \geq 10^{7}\cm^{-3}$ due to the grain randomization by gas collisions that occurs faster than the Larmor precession. They can be aligned with \B in the envelope but always have inefficient IA due to slow internal relaxation. In contrast, superparamagnetic (SPM) grains can have the magnetic alignment with efficient IA in the protostellar core due to the enhanced Barnett relaxation and Larmor precession by iron inclusions. The grain alignment efficiency increases with the increasing 
level of iron inclusions locked inside dust grains. However, in the protostellar central region, large grains of $a\geq 1\mum$ aligned with \B at low-\textit{J} attractors always have slow internal relaxation and can align their longest axis perpendicular or parallel with \B.
 
\item  The polarization patterns produced by both PM and SPM grains are uniform with \PperpB at optically thin wavelengths when grains with slow internal relaxation have right IA. In contrast, when grains with slow internal relaxation have wrong IA, the polarization pattern produced by PM grains will be uniform in space but with \PparaB. SPM grains with a moderate level of iron inclusions produce the uniform pattern with \PparaB at mm, but they produce complex ones at sub-mm with \PperpB in the boundary of the envelope and in the inner $\sim 300$ au region, and \PparaB in the middle region of the core. The change in polarization pattern with the radial distance and wavelengths at optically thin wavelengths is due to the change in the internal alignment (IA) of the emission source. For SPM grains with a high level of iron inclusions, the polarization pattern is uniform, with \PperpB from mm to sub-mm wavelengths because of the efficient grain alignment by the MRAT mechanism.

\item The polarization degree increases with increasing the abundance of iron present in dust, $N_{\rm cl}\phi_{\rm sp}$, but always decreases in the central region near the protostar. PM grains produce a low polarization of $p \sim 1\%$ in the envelope because of their inefficient grain alignment and a negligible polarization level in the inner 1000 au because they are not aligned with \B there. In contrast, SPM grains with a high level of iron inclusions can achieve perfect alignment by the MRAT mechanism and produce a very high polarization degree up to $p \sim 40\%$ in the envelope. It successfully explains the detection of high $p \sim 40\%$ by ALMA toward Class 0 YSOs. The polarization fraction produced by SPM grains slightly decreases to $p < 10\%$ in the central region due to the reduced MRAT alignment efficiency in very dense environments.
 
\item  The polarization by dichroic extinction at optically thick sub-mm wavelengths is less efficient than previously thought due to the weak alignment of VLGs in protostellar environments. We found that the inefficient grain alignment in dense regions induces much stronger reduction of $p(\%)$ with increasing intensity than the extinction effect by VLGs. Furthermore, dichroic extinction cannot become the main source of polarization at sub-mm and cannot induce the polarization flipping from \PperpB at optically thin wavelengths to \PparaB at optically thick wavelengths, excepts for SPM grains with very high levels of iron inclusions. The estimated amount of iron inclusions locked inside dust is required to accurately quantify the efficiency of grain alignment and interpret polarimetric data observed at optically thick wavelengths.

\item  We found that if grains with wrong IA exist in protostellar cores, the change in IA of the polarized emission source can cause the $90^{\circ}$ flipping of the polarization pattern between the envelope and the central region, and between mm and sub-mm. This feature happens at optically thin wavelengths, and be a potential mechanism for the variation of \P with distances and wavelengths. In addition, it also helps us to study the dynamic of grains with slow internal relaxation. 
 
\item The polarization degree increases with increasing iron abundance in the dust but decreases with increasing the maximum grain size. VLGs must be SPM with the magnetic susceptibility enhanced by a factor of $\geq 10^{3}$ compared to PM to reproduce the observed level of polarization of $p \geq 1\%$ in the central region and high $p \sim 5-40\%$ in the envelope. Thus, it opens the new tool for tracing the level of iron inclusions locked inside dust grains via dust polarimetry. Knowing the magnetic properties of grains in dense environments is very important for probing the reability of studying magnetic fields by using dust polarization. 

\item We find that, in addition to the increase of gas randomization toward protostars, the decrease of grain magnetic susceptibility with increasing dust temperature can also decrease the efficiency of external alignment (via the Larmor precession) and internal alignment (via Barnett relaxation). Therefore, although the RAT alignment increases toward the protostar, the decrease of grain magnetic susceptibility and increase of gas randomization leads to the rapid decrease of $p$ toward the protostar. This is a new effect that can reproduce the polarization hole in the central region.

\end{enumerate}

\section*{Acknowledgements}
We thank Stefan Reissl for help with the Polaris code during the initial stage of this work. We thank the members of Vietnam Astrophysics Research Network (VARNET) for various discussion and comments. T.H. is supported by the National Research Foundation of Korea (NRF) grant funded by the Korea government (MSIT (No. 2019R1A2C1087045). J.-G.K acknowledges support from the EACOA Fellowship awarded by the East Asia Core Observatories Association.
 
\section*{Data Availability}
The data underlying this article will be shared on reasonable request to the corresponding author.



\bibliographystyle{mnras}
\bibliography{main.bib} 




\appendix
 
\section{Polarized radiative transfer equation for the Stokes parameters}\label{sec:stoke_parameter}
The polarization state of radiation can be described by the Stokes vector 
\textbf{S} $ = (I ~ Q ~ U ~ V)^{\rm T}$
where Stokes $I$ presents the total intensity, Stokes $Q$ and $U$ present the linear polarization, and Stokes $V$ presents the circular polarization. The radiative transfer equation of the Stokes vector through a dusty environment is given by (\citealt{Martin_1974}, \citealt{Reissl_2016}):
\begin{gather}
    \frac{d}{ds} 
    \begin{pmatrix}
     I\\
     Q\\
     U\\
     V
    \end{pmatrix}
=
-\begin{pmatrix}
    \alpha_{\rm I} && \alpha_{\rm Q} && 0 && 0 \\
    \alpha_{\rm Q} && \alpha_{\rm I} && 0 && 0 \\
    0 && 0 && \alpha_{\rm I} && \kappa_{\rm Q}  \\
    0 && 0 && -\kappa_{\rm Q} && \alpha_{\rm I} \\
\end{pmatrix}
\begin{pmatrix}
 I\\
 Q\\
 U\\
 V
\end{pmatrix}
+ 
\begin{pmatrix}
 j_{\rm I} \\
 j_{\rm Q} \\
 0 \\
0
 \label{eq:polarized_radiative_transfer}
\end{pmatrix}
\end{gather}
where $\alpha_{\rm I}$ is the extinction coefficient over the grain size distribution:
\bea 
\alpha_{\rm I} = \int_{a_{\rm min}}^{a_{\rm max}} C_{\rm ext}(a,\lambda) (dn/da) da,
\label{eq:alpha_I}
\ena
$\alpha_{\rm Q}$ is the linear polarization coefficient due to extinction:
\bea 
\alpha_{\rm Q} = \int_{a_{\rm align}}^{a_{\rm max,JB}^{\rm Lar}} C_{\rm pol}^{\rm ext}(a,\lambda) (dn/da) da,
\label{eq:alpha_Q}
\ena
$\kappa_{\rm Q}$ is the circular polarization coefficient:
\bea 
\kappa_{\rm Q} = \int_{a_{\rm align}}^{a_{\rm max,JB}^{\rm Lar}} C_{\rm circ}(a,\lambda) (dn/da) da,
\ena
where $C_{\rm ext}(a,\lambda)$ is the extinction cross section, $C_{\rm ext}^{\rm pol}(a,\lambda)$ is the linear polarization cross section, and $C_{\rm circ}(a,\lambda)$ is the circular polarization cross section of the grain of size $a$ and radiation wavelength $\lambda$.

Above, $j_{\rm I}$ and $j_{\rm Q}$ are the thermal dust emissivity from all dust grains and aligned dust grains, respectively. For dust grains at thermal equilibrium of temperature $T_{d}$, they are given by
\bea 
j_{\rm I} =  \int_{a_{\rm min}}^{a_{\rm max}} C_{\rm abs}(a,\lambda) B_{\lambda}(T_{\rm d}(a)) (dn/da) da,
\ena
and
\bea 
j_{\rm Q} =   \int_{a_{\rm align}}^{a_{\rm max,JB}^{\rm Lar}} C_{\rm pol}^{\rm abs}(a,\lambda) B_{\lambda}(T_{\rm d}(a))  (dn/da) da,
\ena
where $B_{\rm \lambda}(T_{\rm d}(a))$ is the Planck function at the grain temperature $T_{\rm d}(a)$, $C_{\rm abs}(a,\lambda)$ is the absorption cross section, and $C_{\rm pol}^{\rm abs}(a,\lambda)$ is the polarization cross section by absorption of the grain of size $a$.
 
The extinction cross section $C_{\rm ext}$ in Equation (\ref{eq:alpha_I}) is calculated as :
\bea 
C_{\rm ext} = \frac{C_{\rm ext,x} + C_{\rm ext, y}}{2},
\label{eq:C_ext}
\ena 
and the linear polarization cross section $C_{\rm pol}^{\rm ext}$ in Equation (\ref{eq:alpha_Q}) is given by:
\bea 
C_{\rm pol}^{\rm ext} = \frac{C_{\rm ext, x} - C_{\rm ext, y}}{2}.
\label{eq:C_ext_pol}
\ena

$C_{\rm ext,x}$ and $C_{\rm ext,y}$ are the extinction cross section of grain of size for the electric field \textbf{E} oscillating along x and y axis on the plane of sky, which is:

\bea
C_{\rm ext,x} = \langle C_{\rm ext} \rangle + \frac{1}{3} ~R~ (C_{\rm ext, \parallel} - C_{\rm ext,\perp}),
\label{eq:C_ext_x}
\ena

and 
\bea 
C_{\rm ext,y} = \langle C_{\rm ext} \rangle + \frac{1}{3} ~R ~(C_{\rm ext, \parallel} - C_{\rm ext, \perp})
(1 - 3\sin^{2}\psi)
\label{eq:C_ext_y}
\ena 

The term $\langle C_{\rm ext} \rangle$ is the extinction cross section over grain orientation in the grain frame, which is:
\bea 
\langle C_{\rm ext} \rangle = \frac{2C_{\rm ext, \parallel} + C_{\rm ext, \perp}}{3},
\ena

where $C_{\rm ext, \parallel}$ and $C_{\rm ext, \perp}$ are the extinction cross section produced when the electric field \textbf{E} oscillating along the major and minor axis of the grain, respectively. 

Putting all terms above into Equations (\ref{eq:C_ext}) and (\ref{eq:C_ext_pol}), one obtains

\bea 
C_{\rm ext} &=& \frac{(4+3R\cos^{2}\psi-R) C_{\rm ext,\|}}{6} \nonumber \\
&+& \frac{(2 - 3\cos^{2}\psi + R)C_{\rm ext, \perp}}{6},
\ena
and $C_{\rm pol}^{\rm ext}$ is:
\bea 
C_{\rm pol}^{\rm ext} = \frac{(C_{\rm ext,\|} - C_{\rm ext,\perp}) R \sin^{2}\psi }{2},
\ena
where $R$ is the Rayleigh reduction factor described in Section \ref{sec:POLARIS_alignment_current}.

The values of $C_{\rm abs}$, $C_{\rm pol}^{\rm abs}$, and  $C_{\rm circ}$ have the same 
form with $C_{\rm ext}$ and $C_{\rm pol}^{\rm ext}$.

The term $dn/da$ in Equation (\ref{eq:alpha_I}) is the grain size distribution. Here we assume the typical power law of $dn/da  = C a^{-3.5}$ as in the diffuse ISM (the MRN distribution, \citealt{Mathis_1977}) where $C$ the normalization constant is determined by the gas-to-dust mass ratio:
 
\bea 
\eta = \frac{\int_{a_{\rm min}}^{a_{\rm max}} m_{\rm d}(a) n_{\rm H} C a^{-3.5}  da}{n_{\rm H} m_{\rm H}},
\ena
where 
$m_{\rm d}(a) = 4\pi \rho s a^{3}/3$ 
is the mass of the grain of size $a$ and axial ratio $s$.  

\section{Validity of the one-cell assumption in the azimuthal direction}
\begin{figure*}
 \centering

    \includegraphics[width=0.45\textwidth,height=0.5\textheight,keepaspectratio]{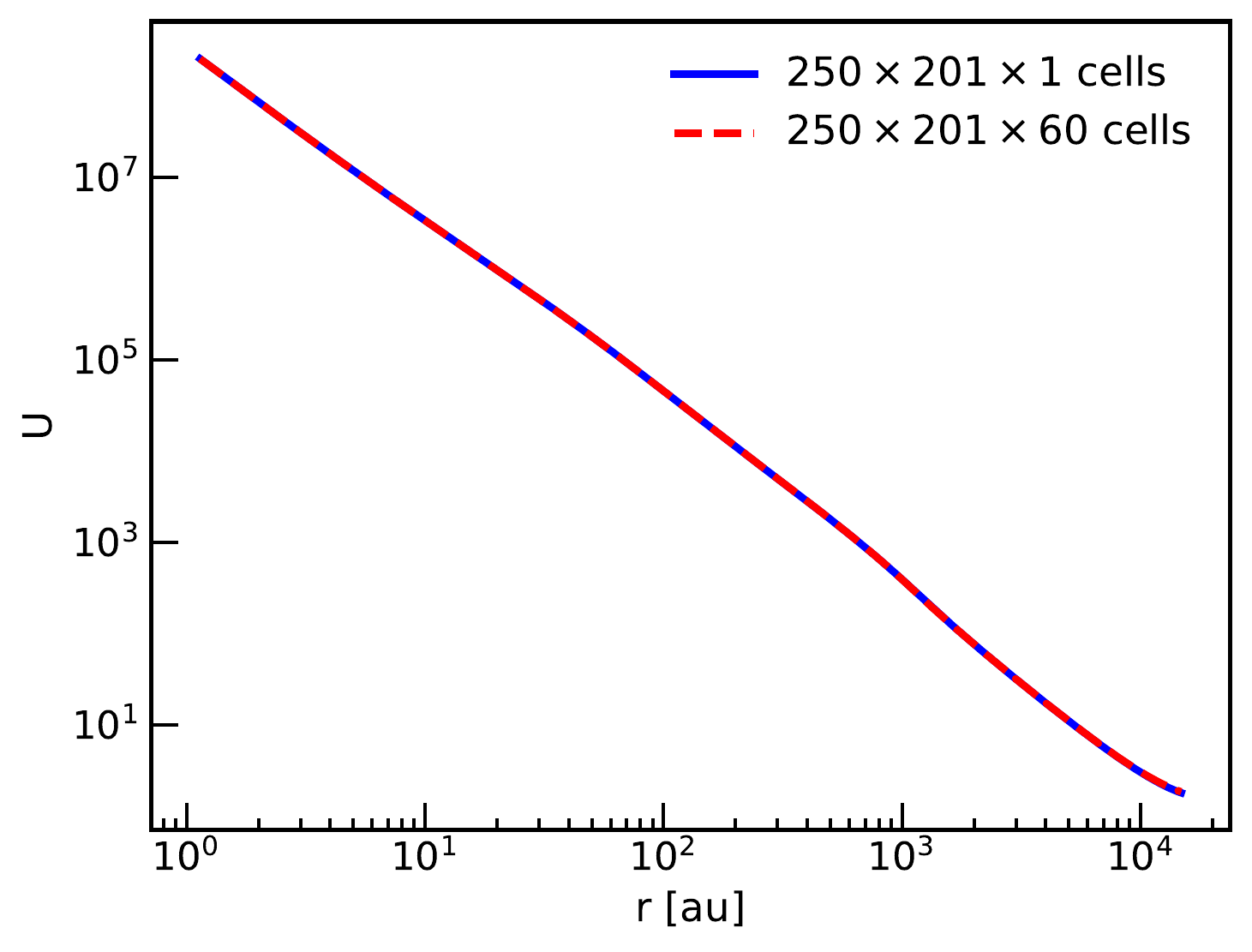}
    \includegraphics[width=0.45\textwidth,height=0.5\textheight,keepaspectratio]{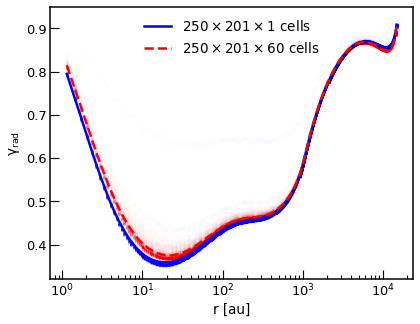}
                
    \caption{Variation of the mean radiation field strength U (left panel) and the anisotropic degree $\gamma_{\rm rad}$ with distances for the spherical grid model with $N_{\rm r}\times N_{\vartheta} \times N_{\varphi} = 250\times101\times1$ and $N_{\rm r}\times N_{\vartheta} \times N_{\varphi} = 250\times101\times60$, respectively. The grid model with one cell along the azimuthal direction gives a similar average $U$ and $\gamma_{\rm rad}$ as the model with 60 cells on the azimuthal direction due to the symmetry of the gas density distribution inside the spherical core.} 
\label{fig:anisotropic_degree}
\end{figure*}

Figure \ref{fig:anisotropic_degree} shows the comparison between the mean radiation field strength $U$ (left panel) and the anisotropic degree  $\gamma_{\rm rad}$ (right panel) as a function of distances to the central protostar using a spherical grid cell with $N_{\rm r}\times N_{\vartheta} \times N_{\varphi} = 250\times101\times1$ and $N_{\rm r}\times N_{\vartheta} \times N_{\varphi} = 250\times101\times60$, respectively. One can clearly see that $U$ and $\gamma_{\rm rad}$ calculated from the model with $N_{\rm varphi} = 1$ are nearly similar to the results from the model with $N_{\rm varphi} = 60$ due to the 
spherical symmetry of the gas density model. Thus, using one cell along the azimuthal direction does not affect 
results of our adopted model.

\section{Grain alignment along x and z direction} \label{sec:alignment_xaxis}
\begin{figure*}
 \centering
    \includegraphics[width=\textwidth,height=\textheight,keepaspectratio]{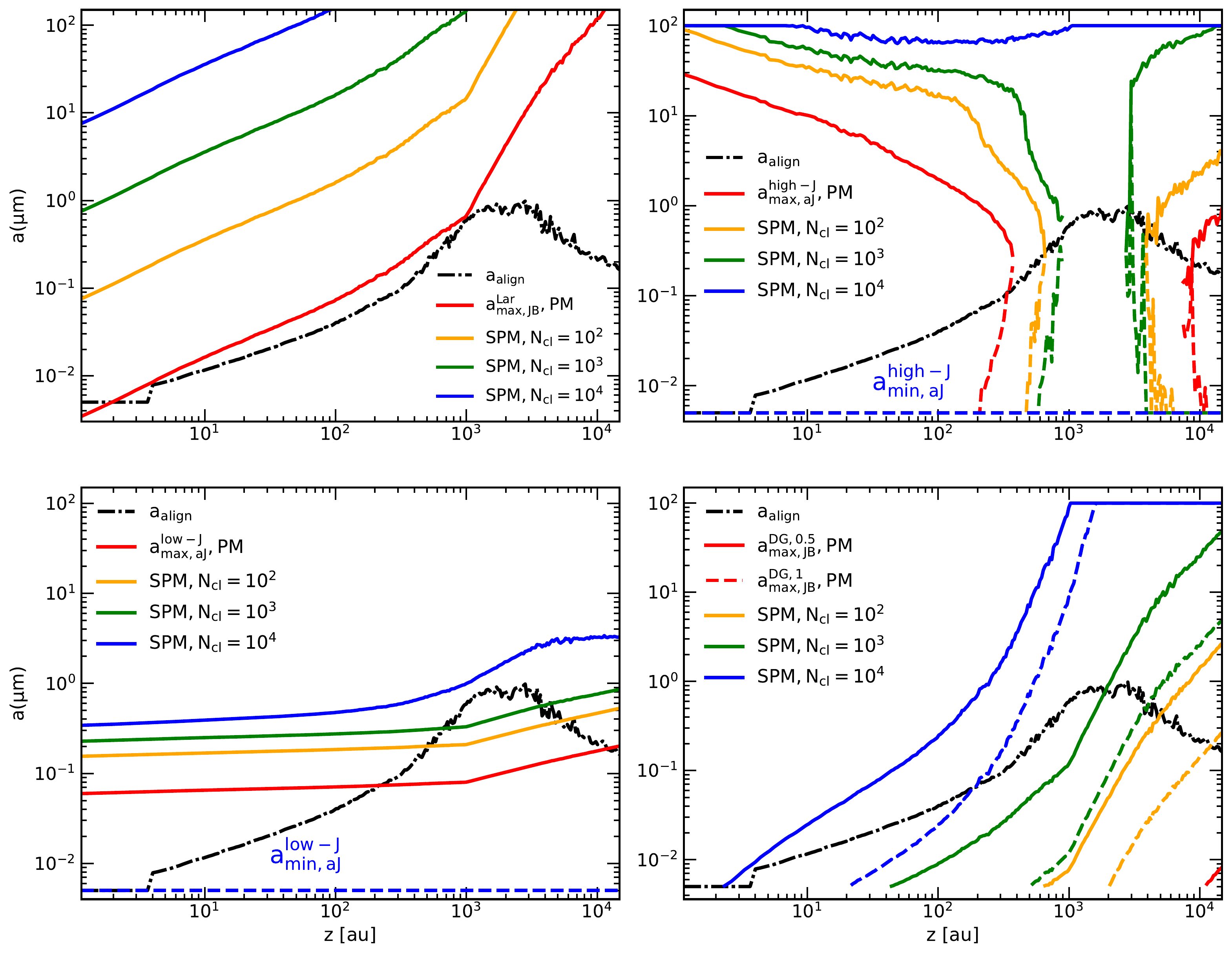}
                
    \caption{Variation along the $z$ direction of the range of aligned dust grains  $a_{\rm align}-a_{\rm max,JB}^{\rm Lar}$ (upper left panel), grains with fast internal relaxation at high and low-\textit{J} attractors $a_{\rm min,aJ} - a_{\rm max,aJ}$ (upper right and lower left panels), and the maximum size that grains are aligned with \B by the MRAT mechanism, $a_{\rm max,JB}^{\rm DG,0.5}$, and $a_{\rm max,JB}^{\rm DG,1}$. The results are for PM and SPM grains with different $N_{\rm cl}$. Generally, large grains can be aligned with \B and have fast internal relaxation in the envelope due to the reduced gas randomization. Iron inclusions enhance the internal and external alignment and help more large grains in the envelope to have perfect alignment with B by MRAT alignment.}

\label{fig:alignment_zaxis}
\end{figure*}

\begin{figure*}
 \centering
    \includegraphics[width=\textwidth,height=\textheight,keepaspectratio]{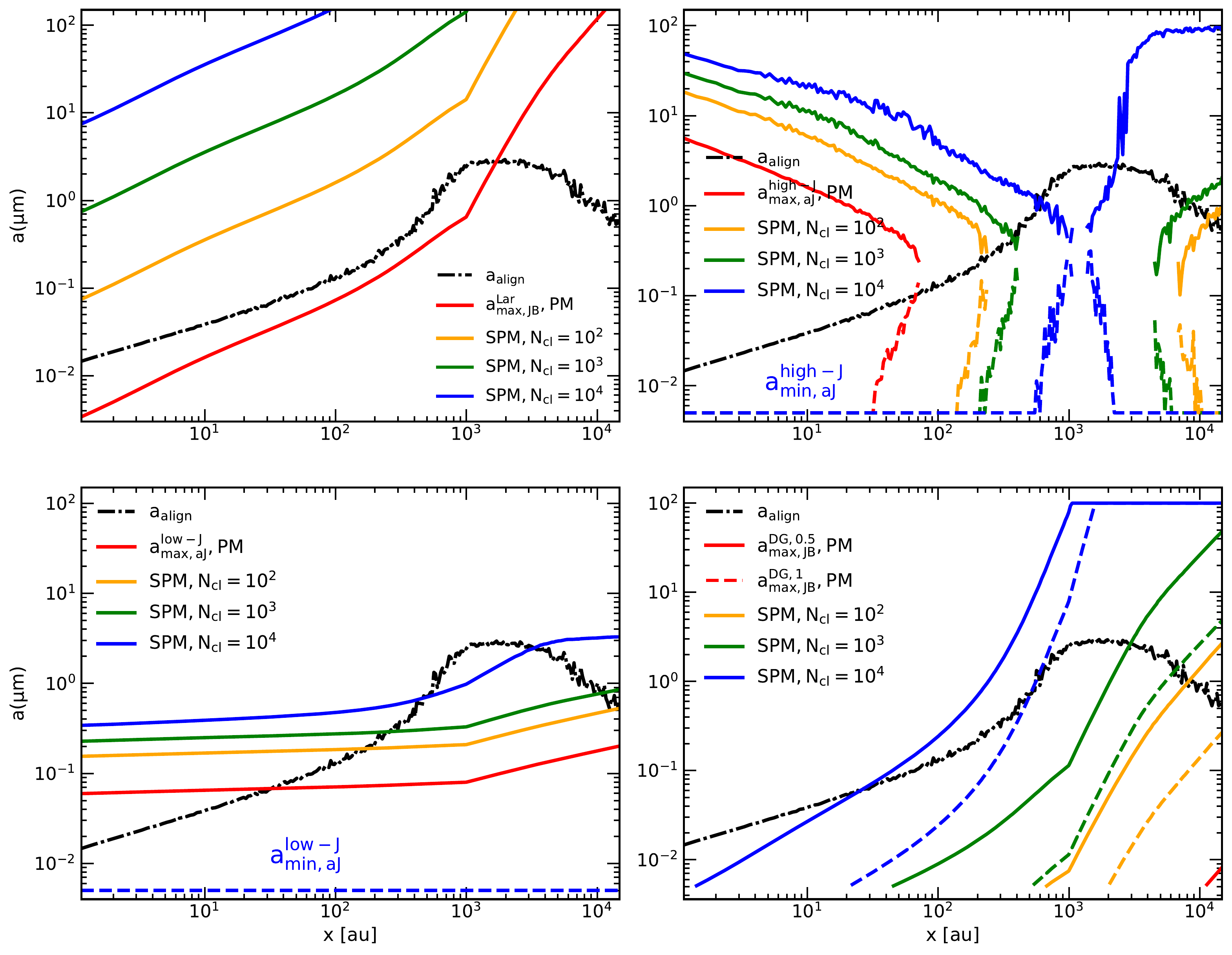}
                
    \caption{Similar as Figure \ref{fig:alignment_zaxis} but for grains on the x direction. The variation of grain alignment along $x$ is similar to along the $z$-direction, but alignment efficiency is weaker, i.e., narrow alignment range of $a_{\rm align} - a_{\rm max, JB}^{\rm Lar}$ (upper left panel) and narrow range of grains with fast internal relaxation at high-\textit{J} attractors $a_{\rm min,aJ}^{\rm high-J} - a_{\rm max,aJ}^{\rm high-J}$ (upper right panel), due to the reduced RATs efficiency in the area where \k $\perp$ \B.}
\label{fig:alignment_xaxis}
\end{figure*}

Figures \ref{fig:alignment_zaxis} and \ref{fig:alignment_xaxis} shows in detail the grain alignment of PM and SPM grains along the z and x direction, respectively (see the gas density map on x-z plane in the left panel of Figure \ref{fig:model}). The upper left panel shows the minimum and maximum grain alignment size $a_{\rm align}$ and $a_{\rm max,JB}^{\rm Lar}$. The upper right and lower left panels show the minimum and maximum size for grains having fast internal relaxation at high-\textit{J} and low-\textit{J} attractors, respectively. The lower right panel shows the maximum size for grains having $f_{\rm high-J} = 0.5$ and $f_{\rm high-J} = 1$. In general, more large grains can be aligned with \B (i.e., larger range $a_{\rm align} - a_{\rm max,JB}^{\rm Lar}$), and have fast internal relaxation at high-\textit{J} attractors (i.e., larger range $a_{\rm min,aJ}^{\rm high-J} - a_{\rm max,aJ}^{\rm high-J}$), in the envelope due to the reduced gas randomization. Iron inclusions allow more grains to have perfect magnetic alignment (i.e., $f_{\rm high-J} = 1$) there due to the enhanced MRAT alignment.

However, one can see that the alignment efficiency of grains on the $x$ direction is slightly smaller than grains on the $z$ direction. It is because grains are spun up by RATs weaker in the area where the radiation field is perpendicular with B-field (Equation \ref{eq:omega_rat}). As a result, the range of aligned dust grains having fast internal relaxation at high-\textit{J} attraction reduces, i.e., larger $a_{\rm align}$ and larger $a_{\rm max,aJ}^{\rm high-J}$, inducing lower $p(\%)$ on the equatorial as shown in the polarization maps in Section \ref{sec:iron_pol_map}.

\section{Polarization map at millimeter wavelengths for $a_{\rm max} = 100\mum$}\label{sec:mm_a100}
 
\begin{figure*}
 \centering
    \includegraphics[width=\textwidth,height=\textheight,keepaspectratio]{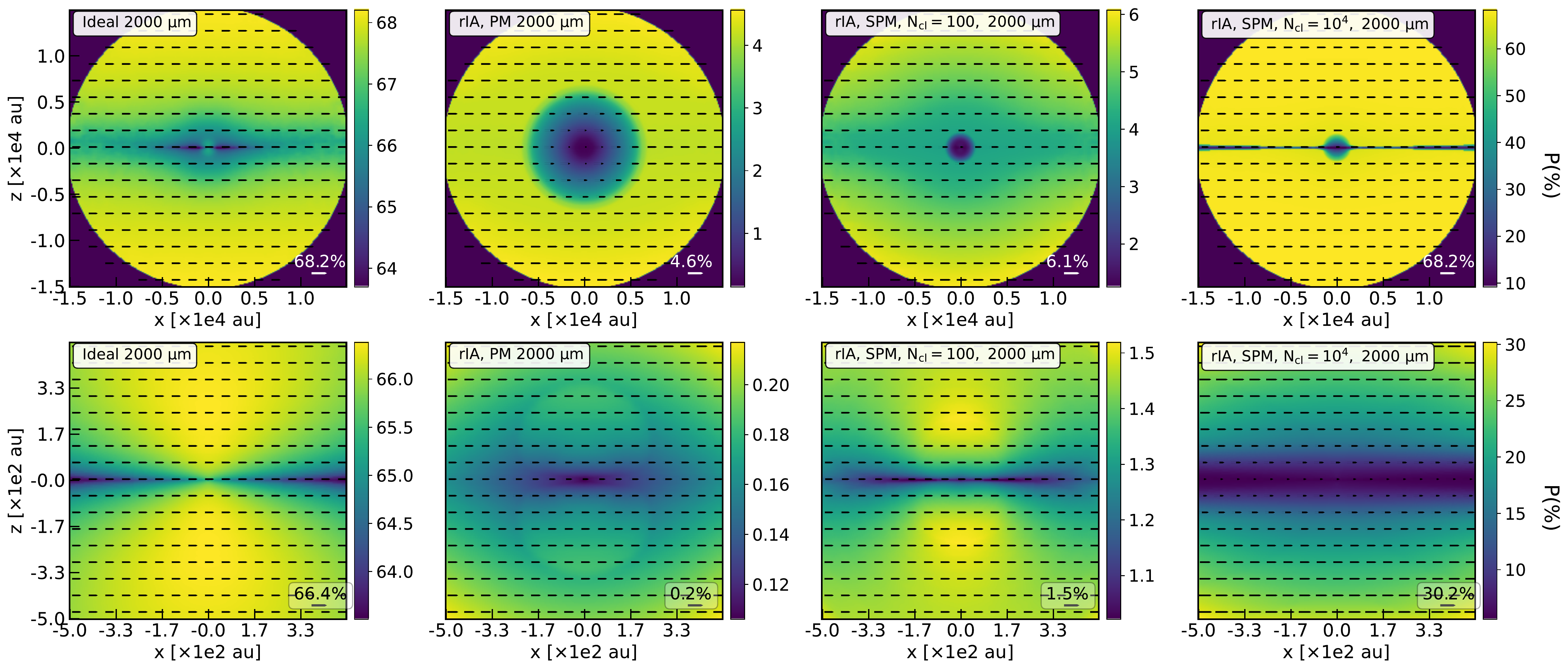}
                
    \caption{Polarization maps calculated at 2 mm for model Ideal (first column) and model Realistic$-$rIA. The upper panels show the results in the envelope, and the lower panels are for the inner 1000 au around the protostar. The polarization patterns for both PM and SPM grains are uniform with \PperpB from the envelope to the center when grains with slow internal alignment have right IA. The polarization degree in our model decreases inward due to the inefficient alignment of grains with \B.}
\label{fig:Realistic_rIA_mm}
\end{figure*}

Figure \ref{fig:Realistic_rIA_mm} shows the polarization patterns calculated at 2mm by model Ideal (first column) and model Realistic$-$rIA. The upper panels show the results in the envelope, and the lower panels show the results for the inner region of 1000 au around the protostar. The polarization patterns produced by both PM and SPM are uniform with \PperpB from the envelope to the central region, assuming that grains with slow internal relaxation have right IA.

\begin{figure*}
 \centering
    \includegraphics[width=\textwidth,height=\textheight,keepaspectratio]{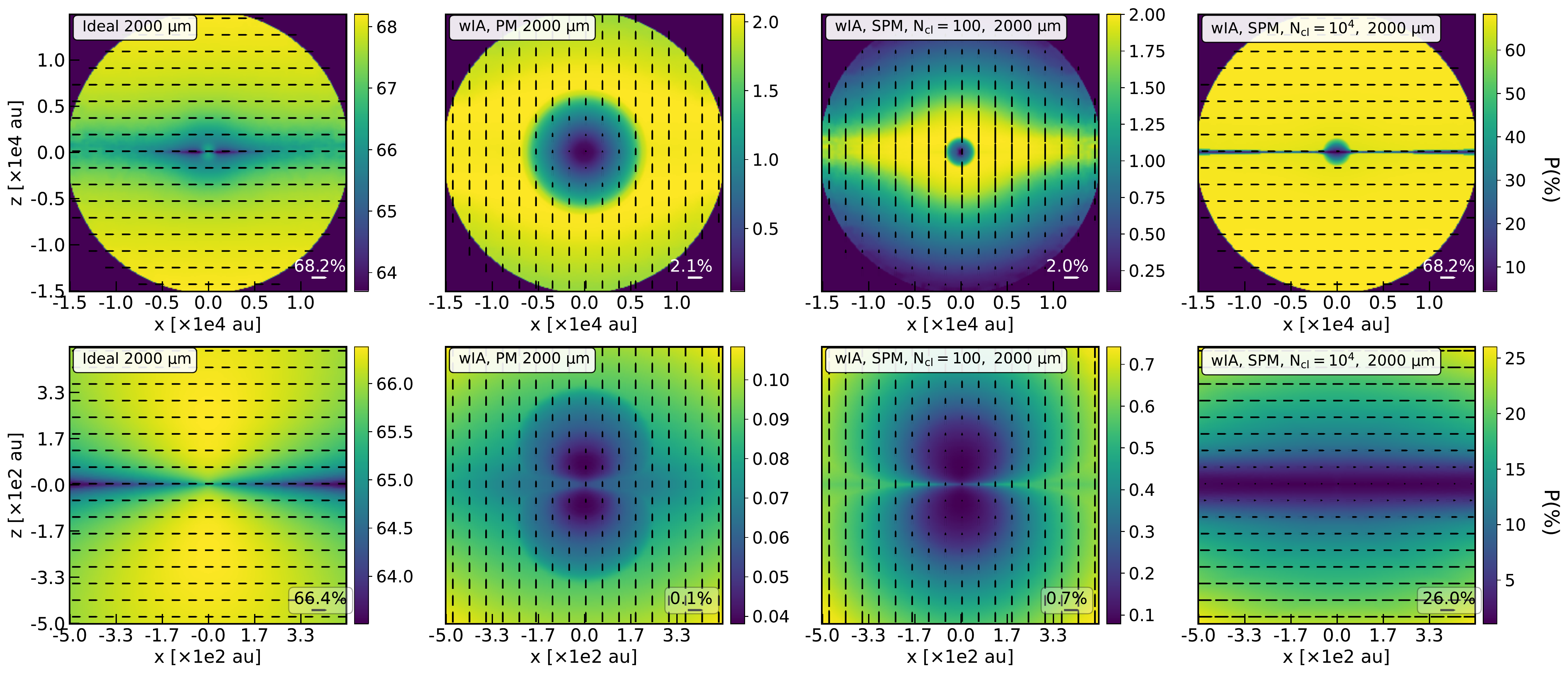}
    \caption{Similar as Figure \ref{fig:Realistic_rIA_mm} but for model Realistic$-$wIA. At mm, PM and SPM grains with low $N_{\rm cl} = 100$ produce the uniform polarization pattern with \PparaB in the entire protostellar core due to the emission of grains with wrong IA. In contrast, SPM grains with high $N_{\rm cl} = 10^{4}$ produce polarization vectors \PperpB as model Ideal.}
\label{fig:Realistic_wIA_mm}
\end{figure*}

Figure \ref{fig:Realistic_wIA_mm} shows the similar results as Figure \ref{fig:Realistic_rIA_mm} but for model Realistic$-$wIA. Similar as the results at $870\mum$ (Figures \ref{fig:Realistic_wIA_envelope} and \ref{fig:Realistic_wIA_center}), PM and SPM grains with $N_{\rm cl} = 100$ produce the uniform polarizaton pattern with \PparaB from the envelope to the center due to the emission of VLGs with wrong IA at mm wavelengths. In contrast, SPM grains with $N_{\rm cl} = 10^{4}$ produce the uniform pattern with \PperpB in the entire protostellar core because of the efficient alignment of VLGs by the MRAT mechanism with \B.
  
\section{Polarization results for $a_{\rm max}=10\mum$}\label{sec:amax_10um}
\subsection{Iron Inclusions and Polarization Pattern}

 \begin{figure*}
 \centering
    \includegraphics[width=\textwidth,height=0.9\textheight,keepaspectratio]{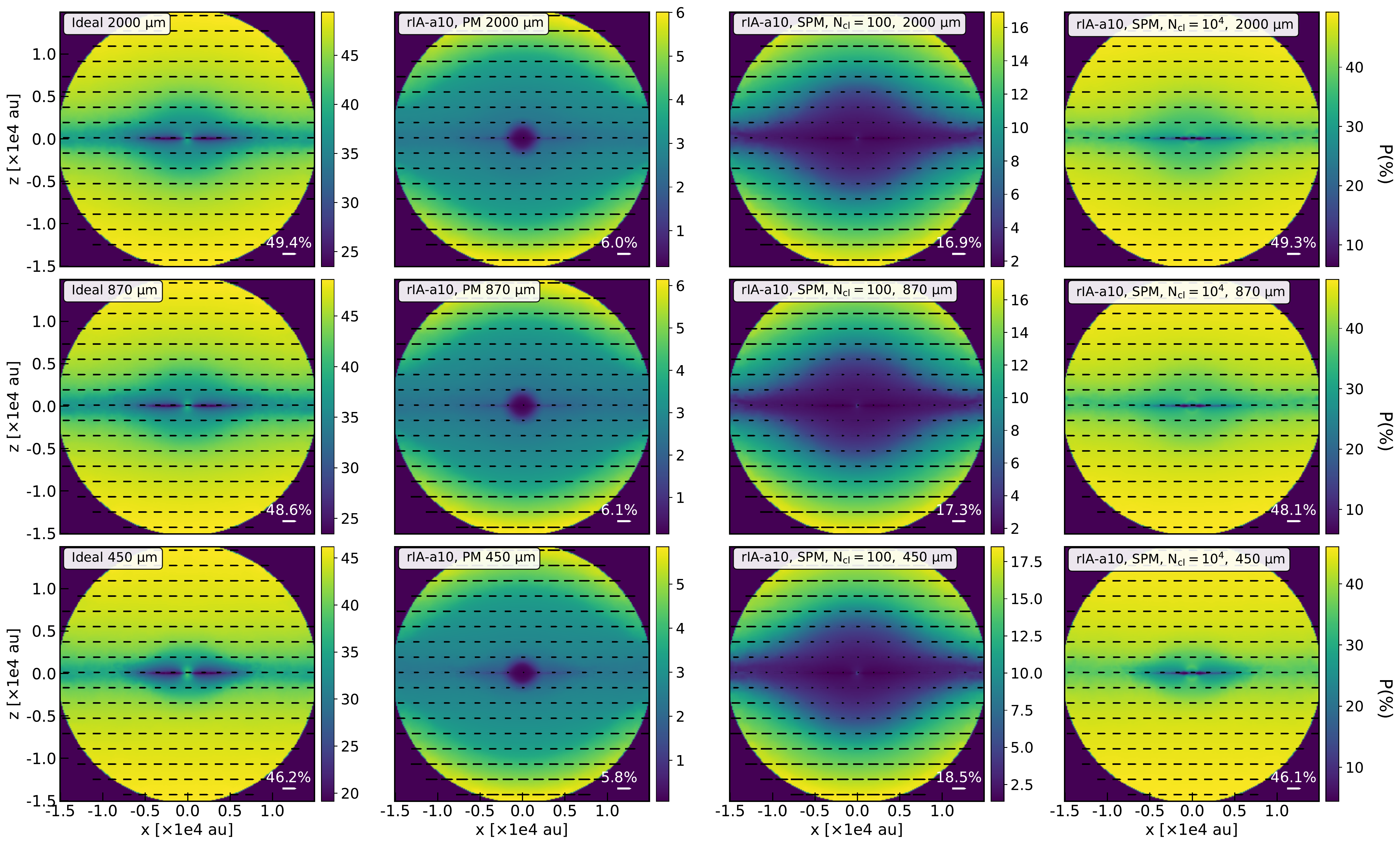}  
    \includegraphics[width=\textwidth,height=0.9\textheight,keepaspectratio]{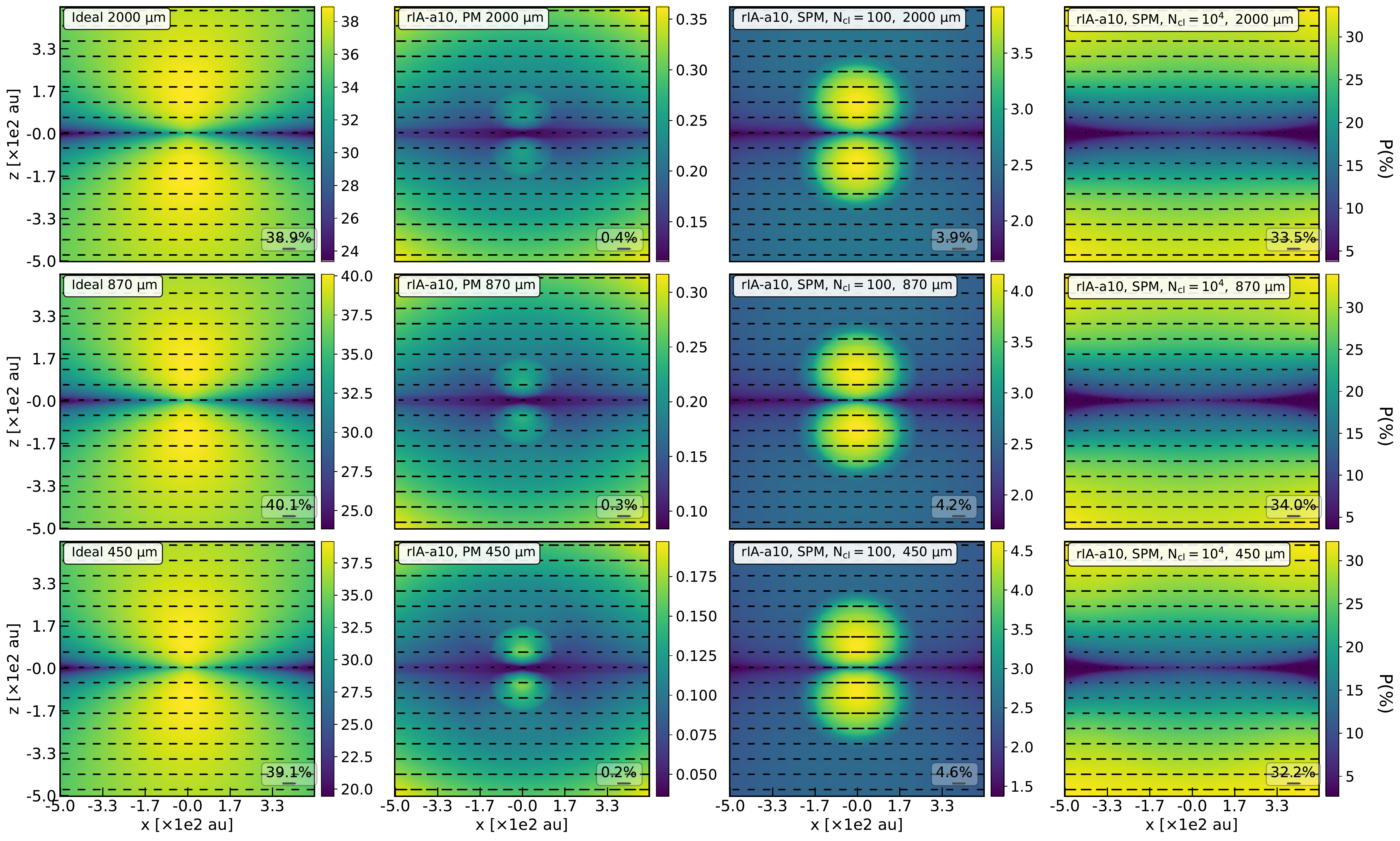}  
    \caption{Polarization maps calculated in the envelope (first to third rows) and the inner region of 1000 au (fourth to sixth rows) for model Ideal$-$a10 (first column) and model Realistic$-$rIA$-$a10 (second to fourth columns) (Table \ref{tab:model_table}). For model that all grains have right IA, the polarization pattern is uniform in space with \PperpB from 2mm to $450\mum$.}
\label{fig:Realistic_rIA_a10}
\end{figure*}
 
\begin{figure*}
 \centering
    \includegraphics[width=\textwidth,height=0.9\textheight,keepaspectratio]{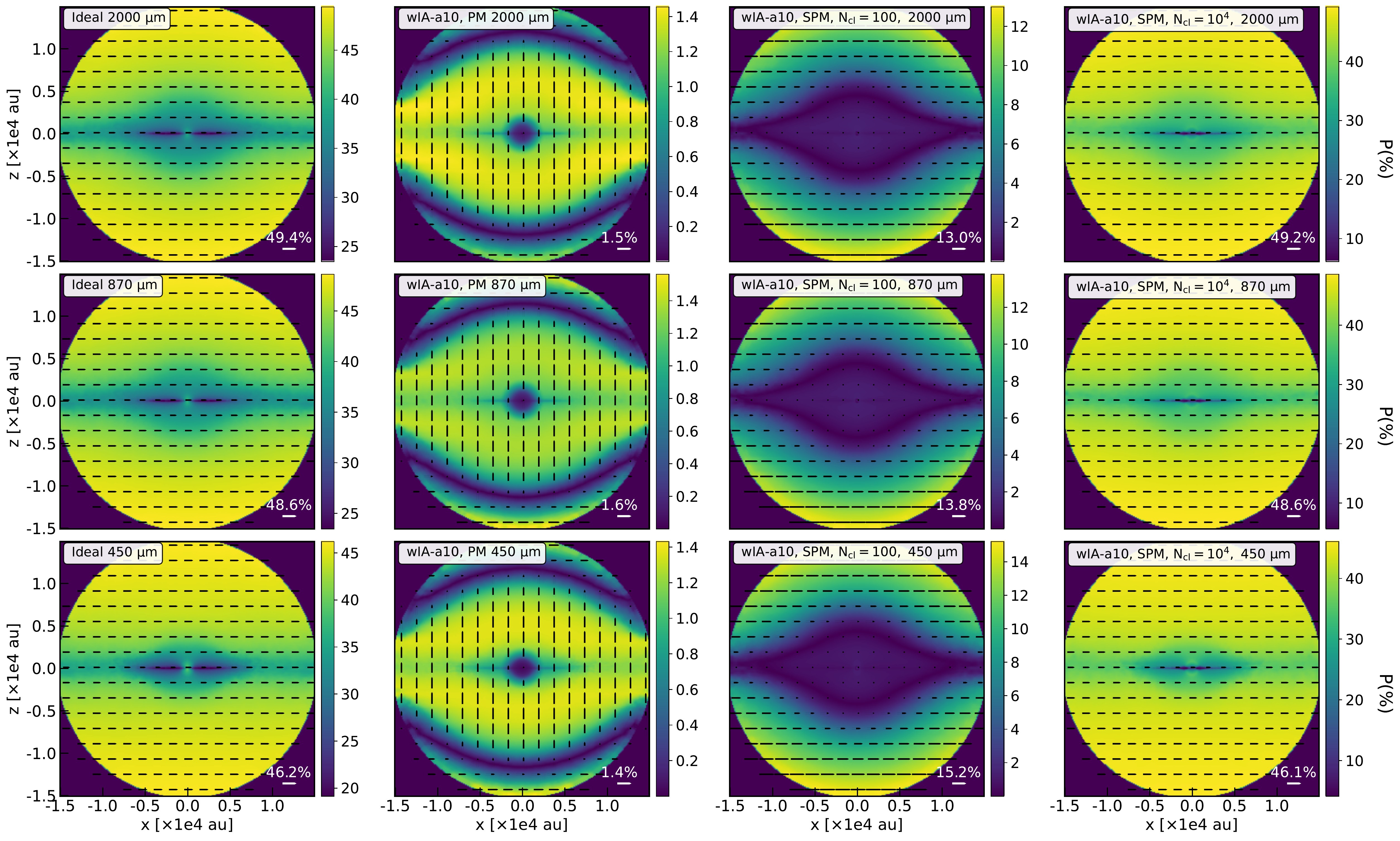}
    \includegraphics[width=\textwidth,height=0.9\textheight,keepaspectratio]{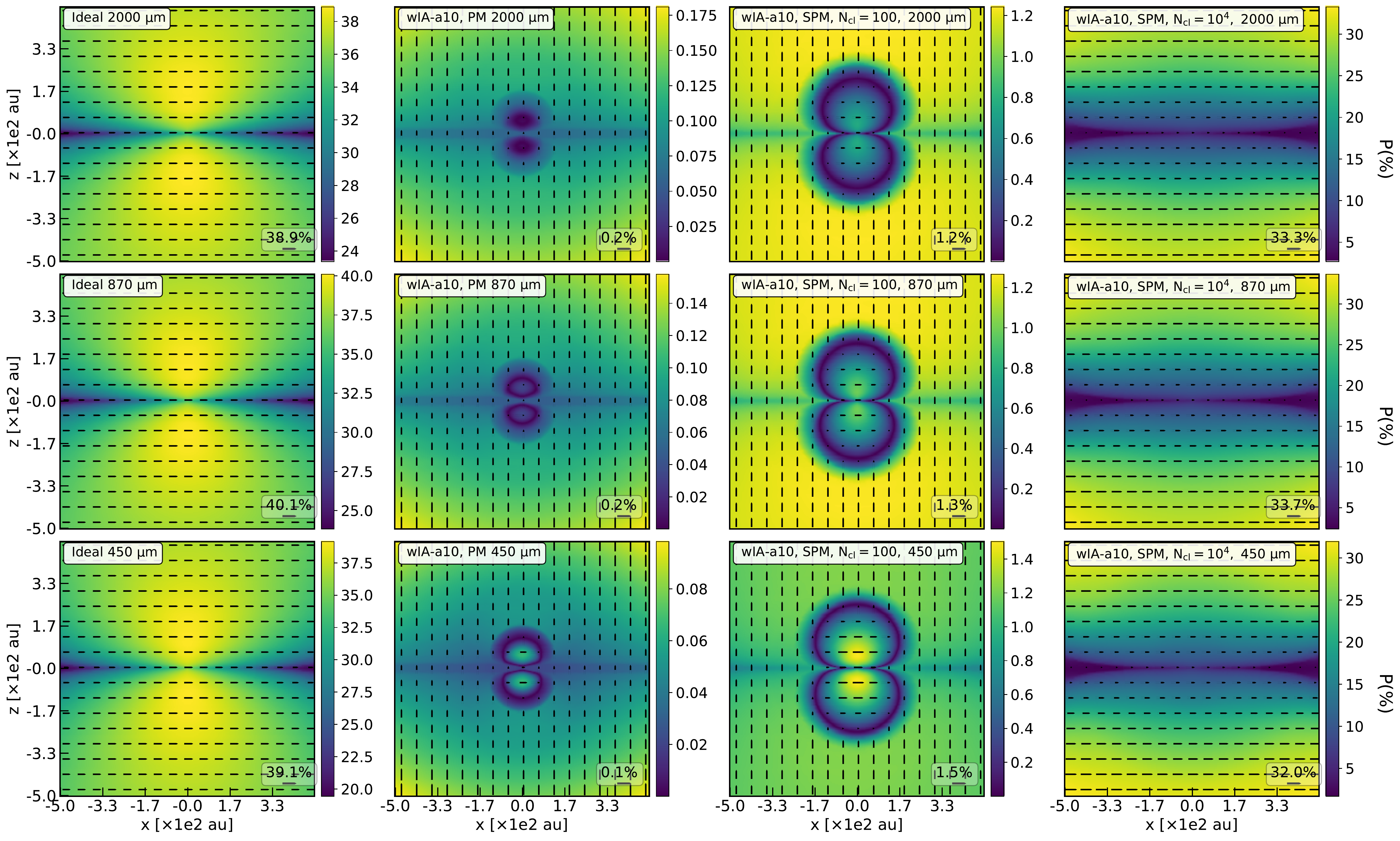} 
      
    \caption{Similar as Figure \ref{fig:Realistic_rIA_a10} but for model Realistic$-$wIA$-$a10. In the presence of grains with wrong IA, at fixed wavelengths, the polarization pattern produced by PM and SPM grains with low $N_{\rm cl}$ changes from \PperpB in the outer boundary of the envelope to \PparaB in the inner region and flips again to \PperpB around protostar. This complicated polarization pattern is due to the complex distribution of grains with right and wrong IA in the protostellar core (e.g., Figure \ref{fig:align_amaxaJ}). For SPM grains with high $N_{\rm cl} = 10^{4}$, \P is uniform with \PperpB. In addition, the polarization patterns in the entire protostellar core do not change with wavelengths for $a_{\rm max} = 10\mum$.}
\label{fig:Realistic_wIA_a10}
\end{figure*}

 \begin{figure*}
 \centering
    \includegraphics[width=\textwidth,height=\textheight,keepaspectratio]{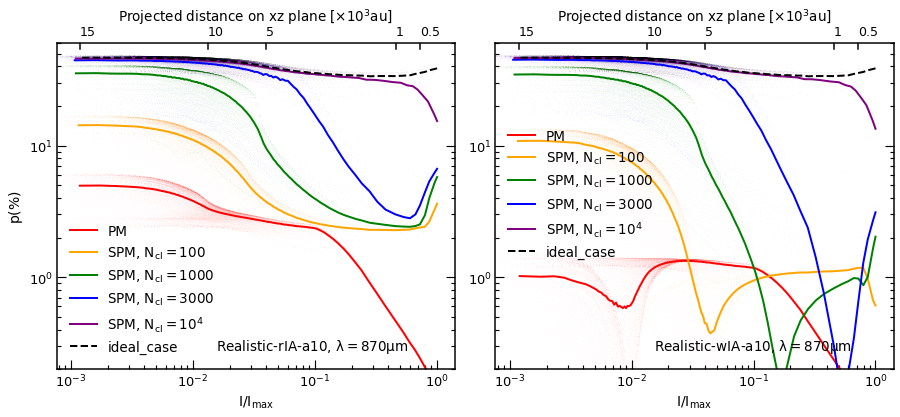} 
    \includegraphics[width=\textwidth,height=\textheight,keepaspectratio]{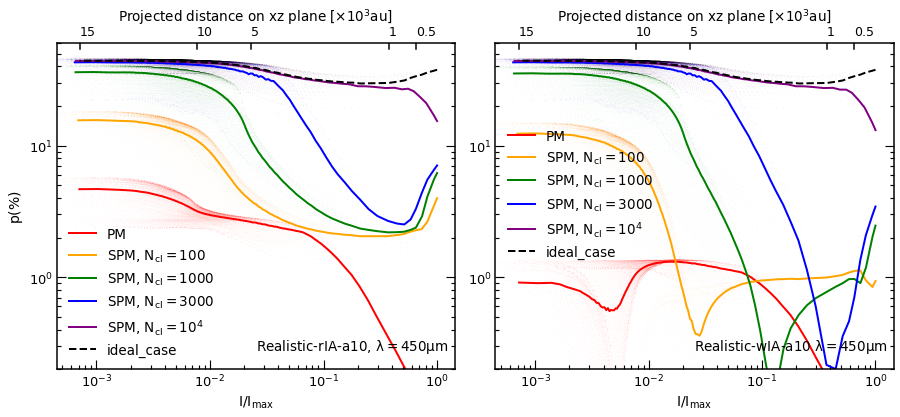} 
    \caption{Effect of iron inclusions on the variation of $p$ with $I/I_{\rm max}$ at $870\mum$ (upper panels) and $450\mum$ (lower panels) for model Realistic$-$rIA$-$a10 (left column) and Realistic$-$wIA$-$a10 (right column). The polarization degree from our models always decreases with increasing intensity due to the reduced grain alignment efficiency in dense environments, and generally increases with increasing $N_{\rm cl}$. If grains with wrong IA are present in the core, one will obtain the "valley-polarization hole" on $p-I/I_{\rm max}$ curve caused by the cancelling effect of polarized emission from grains with right and wrong IA .}
\label{fig:iron_P_I_a10}
\end{figure*}

  \begin{figure*}
 \centering
    \includegraphics[width=\textwidth,height=\textheight,keepaspectratio]{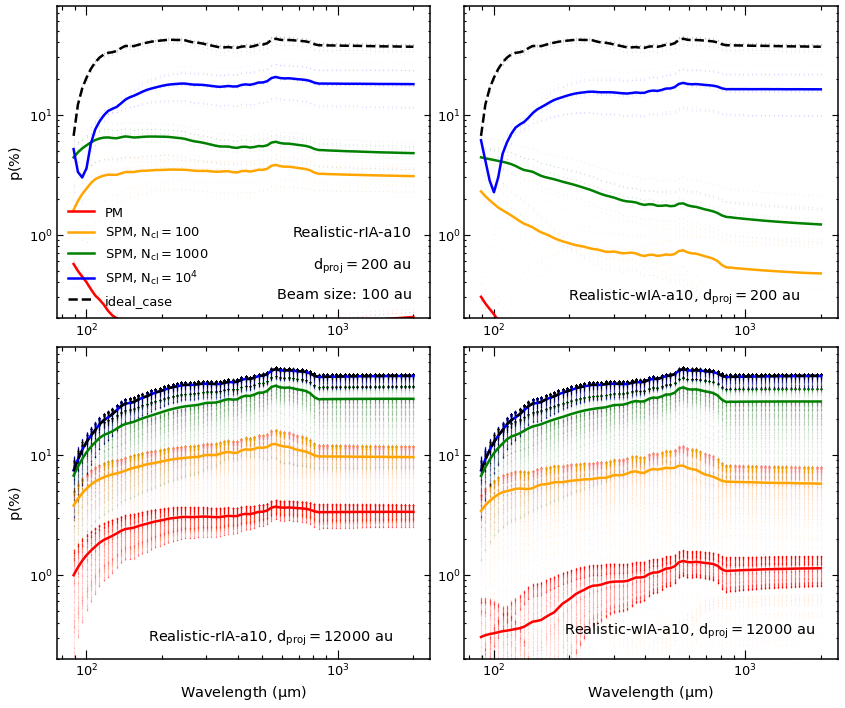}
    \caption{Effect of iron inclusions on the polarization curve $p(\lambda)$ from 2 mm to $89\mum$ calculated at $d_{\rm proj} = 200$ au (upper panels) and $d_{\rm proj} = 2000$ au (lower panels) for model Realistic$-$rIA$-$a10 (left column) and Realistic$-$wIA$-$a10 (right column). With $a_{\rm max} = 10\mum$, the polarization degree is nearly independent with mm/sub-mm wavelengths in both the central region and the envelope (corresponds to the independence of the polarization patterns with wavelengths in Figures \ref{fig:Realistic_rIA_a10} and \ref{fig:Realistic_wIA_a10}).}
\label{fig:iron_P_wave_a10}
\end{figure*}

Figure \ref{fig:Realistic_rIA_a10} shows the polarization maps obtained from the outer envelope (first to third rows) to the central region (fourth to sixth rows) by model Realistic$-$rIA$-$a10 (Table \ref{tab:model_table}). From the top to bottom, the observed wavelengths decrease from 2mm to $870\mum$ and $450\mum$, while from the left to right, the results are for model Ideal$-$a10 (first column) and our models for PM and SPM grains with $N_{\rm cl} = 100$ and $N_{\rm cl} = 10^{4}$, assuming $\phi_{\rm sp} = 0.005$. When all aligned dust grains have right IA, the polarization pattern (black segment) is uniform with \PperpB from the envelope to the central region and does not change with wavelengths. The polarization degree (color code) decreases toward the center due to the inefficient magnetic alignment of grains in dense core. 
 
Figure \ref{fig:Realistic_wIA_a10} shows similar results as Figure \ref{fig:Realistic_rIA_a10} but for model Realistic$-$wIA$-$a10. For PM grains (left column), the polarization pattern is quite uniform with \PparaB within 10000 au where most grains have wrong IA due to slow internal relaxation. However, beyond $\sim 10000$ au, \P flips $90^{\circ}$ to \PperpB because a small amount of micron-sized grains can have efficient IA by ISRF here (Figure \ref{fig:align_amaxaJ}, left panel). For SPM grains with $N_{\rm cl} = 100$ (center column), the polarization pattern changes from \PperpB beyond 5000 au (first three rows)  to \PparaB in the inner region and changes back to \PperpB within 200 au from the protostar (fourth to sixth rows) as a result of the complex distribution of grains with wrong and right IA from the center to the envelope (Figure \ref{fig:align_amaxaJ}, left panel). In contrast, SPM grains with $N_{\rm cl} = 10^{4}$ (right column) produce much more simple polarization pattern, with \PperpB in the entire protostellar core due to their efficient alignment with \B-fields.

One can see that that in the absence of VLGs, the polarization vectors do not rotate from \PparaB at mm to \PperpB at sub-mm as in the case of model Realistic$-$wIA for SPM grains with $N_{\rm cl} =100$ and $a_{\rm max} = 100\mum$ (Figures \ref{fig:Realistic_wIA_envelope} and \ref{fig:Realistic_wIA_center}). It is because for the model with $a_{\rm max} = 10\mum$, most all SPM grains have efficient IA and thus produce \PperpB with wavelengths. In contrast, with the model of $a_{\rm max} = 100\mum$, the main source for therma dust emission at mm (VLGs) arises from VLGs that have wrong IA due to slow internal relaxation, while the source of dust emission at sub-mm arises from small (micron-size) grains that have efficient IA. Consequently, the polarization pattern changes with wavelengths due to the change in IA configuration of the emitting dust grains. This is an interesting feature that (may) be classified as an imprint of the grain growth process.

\subsection{Iron Inclusions and Intensity-dependent Polarization Degree}
Figure \ref{fig:iron_P_I_a10} shows the effect of iron inclusions on the variation of $p(\%)$ with $I/I_{\rm max}$ at $870\mum$ (upper panels) and $50\mum$ (lower panels) for model Realistic$-$rIA$-$a10 (left column) and model Realistic$-$wIA$-$a10 (right column). The result for the ideal model of grain alignment is also plotted by the black dashed line for comparison. Similar as the results with $a_{\rm max} = 100\mum$, $p$ always decreases with increasing $I/I_{\rm max}$ due to the decreased grain alignment efficiency in dense regions. The polarization degree produced by SPM grains with higher $N_{\rm cl}$ generally shows higher values thanks to the enhanced grain alignment by the MRAT mechanism. In addition, if aligned dust grains have wrong IA, one will obtain the "valley-polarization hole" caused by the cancelling effect of polarized emission due to the co-existence of grains with right and wrong IA (left column).

\subsection{Iron Inclusions and Wavelength-dependent Polarization Degree}
 
Figure \ref{fig:iron_P_wave_a10} shows the effect of iron inclusions on the polarization curve $p(\lambda)$ observed at $d_{\rm proj} = 200$ au (upper panels) and $d_{\rm proj} = 12000$ au (lower panels) for model Realistic$-$rIA$-$a10 (left columns) and Realistic$-$wIA$-$a10 (right columns). Without the presence of VLGs (Section \ref{sec:iron_p_lambda}), the polarization degree is nearly independent with wavelengths from $89\mum$ to 2mm in both the envelope and the central region. The slight rise of $p(\%)$ from 2mm to $89\mum$ produced by SPM grains with $N_{\rm cl} < 1000$ for model Realistic$-$wIA$-$a10 is due to the increased emission of micron-sized grains with efficient IA at short wavelengths.


\bsp	
\label{lastpage}
\end{document}